\documentclass[12pt]{article}

\usepackage{amsfonts}
\usepackage{amsmath}
\usepackage[pdftex]{hyperref}
\usepackage{cite}
\usepackage{epsfig}
\usepackage{latexsym}
\usepackage{paralist}
\usepackage{fancyhdr}
\usepackage{graphicx}
\numberwithin{equation}{section}
\usepackage[vcentermath]{youngtab}
\usepackage{etex}
\usepackage{braket}
\usepackage{float}

\def\spa#1{\phantom{\fbox{\rule[-#1cm]{0cm}{0cm}}}}

\def\be{\begin{equation}}
\def\ee{\end{equation}}
\def\bea{\begin{eqnarray}}
\def\eea{\end{eqnarray}}

\def\half{{1\over 2}}
\def\Tr{\mbox{Tr}}

\def\nn{\nonumber}

\renewcommand{\thefootnote}{\fnsymbol{footnote}}

\def\mut{\widetilde{\mu}}
\def\nut{\widetilde{\nu}}

\def\Rt{\widetilde{R}}
\def\Wt{\widetilde{W}}

\def\str{\mathop{\mathrm{str}}\nolimits}
\def\diag{\mathop{\mathrm{diag}}\nolimits}

\let\ev=\bracket

\def\bbC{{\mathbb{C}}}

\def\bbZ{{\mathbb{Z}}}

\def\Nt{{\widetilde{N}}}
\def\cN{{\cal N}}
\def\cO{{\cal O}}
\def\cW{{\cal W}}
\def\sign{\mathop{\mathrm{sign}}\nolimits}
\def\tr{\mathop{\mathrm{tr}}\nolimits}

\begin{document}

\hfuzz=100pt
\title{{\Large \bf{ABJ Wilson loops and Seiberg Duality}}}
\date{}
\author{ Shinji Hirano$^a$, 
Keita Nii$^b$
 and Masaki Shigemori$^{c,d}$
}
\date{}

\maketitle

\thispagestyle{fancy}
\rhead{YITP-14-49}
\cfoot{}
\renewcommand{\headrulewidth}{0.0pt}

\vspace*{-1cm}
\begin{center}
$^{a}${{\it School of Physics and Center for Theoretical Physics}}
\\ {{\it University of the Witwatersrand}}
\\ {{\it WITS 2050, Johannesburg, South Africa}}
  \spa{0.5} \\
$^b${{\it Department of Physics}}
\\ {{\it Nagoya University, Nagoya 464-8602, Japan}}
\spa{0.5}  \\
$^c${{\it Yukawa Institute for Theoretical Physics}}
\\ {{\it Kyoto University, Kyoto 606-8502, Japan}}
\spa{0.5}  \\
$^d${{\it Hakubi Center, Kyoto University, Kyoto 606-8501, Japan}}

\end{center}

\begin{abstract}
We study supersymmetric Wilson loops in the ${\cal N} = 6$
supersymmetric $U(N_1)_k\times U(N_2)_{-k}$ Chern-Simons-matter (CSM)
theory, the ABJ theory, at finite $N_1$, $N_2$ and $k$. This generalizes
our previous study on the ABJ partition function.  First computing the
Wilson loops in the $U(N_1) \times U(N_2)$ lens space matrix model
exactly, we perform an analytic continuation, $N_2$ to $-N_2$, to obtain
the Wilson loops in the ABJ theory that is given in terms of a formal
series and only valid in perturbation theory. Via a Sommerfeld-Watson
type transform, we provide a nonperturbative completion that renders the
formal series well-defined at all couplings. This is given by ${\rm
min}(N_1,N_2)$-dimensional integrals that generalize the ``mirror
description'' of the partition function of the ABJM theory.  Using our
results, we find the maps between the Wilson loops in the original and
Seiberg dual theories and prove the duality. In our approach we can
explicitly see how the perturbative and nonperturbative contributions to
the Wilson loops are exchanged under the duality. The duality maps are
further supported by a heuristic yet very useful argument based on the
brane configuration as well as an alternative derivation based on that
of Kapustin and Willett.
\end{abstract}

\renewcommand{\thefootnote}{\arabic{footnote}}
\setcounter{footnote}{0}

\newpage


\section{Introduction}


Duality is one of the most fascinating phenomena in quantum field
theory.  It provides an alternative, often non-perturbative,
understanding of the theory that is not accessible in the original
description.  Seiberg duality \cite{Seiberg:1994pq}, under which weakly
coupled gauge theory is mapped into strongly coupled one with different
rank, and vice versa, is a prominent example.  Although the mapping of local operators
under Seiberg duality is known to be rather simple
\cite{Seiberg:1994pq}, the transformation properties of non-local
operators, such as Wilson loops, are more non-trivial and less studied.

By its strong-weak nature, any checks of Seiberg duality must
involve a non-perturbative approach.  The localization method
\cite{Witten:1988ze, Pestun:2007rz, Marino:2011nm} is a powerful
technique applicable to supersymmetric field theory and reduces the
infinite dimensional path integral of quantum field theory to a finite
integral which can be often regarded as a matrix model integral.  This
method allows for exact computation of quantities such as partition
function and Wilson loops at strong coupling, and is an ideal tool for
studying non-perturbative physics such as Seiberg duality.

Among various supersymmetric gauge theories calculable by the
localization method, we focus on the three-dimensional $\mathcal{N}=6$
supersymmetric $U(N_1)_k \times U(N_2)_{-k} $ Chern-Simons-matter (CSM)
theory, known as the ABJ theory \cite{Aharony:2008gk}, where
$k\in\bbZ_{\neq 0}$ is the Chern-Simons level.  In the special case with
$N_1=N_2$, this theory is called the ABJM theory \cite{Aharony:2008ug}.
It is believed that the $U(N)_k \times U(N+M)_{-k}$ ABJ theory gives the
low-energy description of the system of $N$ M2-branes and $M$ fractional
M2-branes probing $\bbC^4/\bbZ_k$, and its holographic dual geometry is
AdS$_4\times S^7/\bbZ_k$ in M-theory or AdS$_4\times \mathbb{CP}^3$ in
type IIA superstring \cite{Aharony:2008ug, Aharony:2008gk}.  It is also
conjectured that this theory has another dual description in terms of
the $\cN=6$ supersymmetric, parity-violating version of the Vasiliev
higher spin theory in four dimensions \cite{Chang:2012kt}.

Applying the localization method to the ABJ(M) theory on $S^3$ yields
\cite{Kapustin:2009kz} the so-called ABJ(M) matrix model, which has been
extensively studied recently and led to much insight into the
non-perturbative effects in the theory.
For instance, applying the standard large $N$ technique to the ABJ matrix
model reproduced the $N^{3/2}$ scaling of the free energy expected from
the gravity dual \cite{Drukker:2010nc, Marino:2011nm, Herzog:2010hf}.  For the ABJM
theory, the $1/N$ corrections were summed to all orders using
holomorphic anomaly equation at large 't Hooft coupling $\lambda=N/k$ in
the type IIA regime $k\gg 1$ and an Airy function behavior was found
\cite{Fuji:2011km}.  Furthermore, the powerful Fermi gas approach was
developed and reproduced the Airy function in the M-theory regime (large
$N$, finite $k$)\cite{Marino:2011eh}.  Based on the Fermi gas approach,
instanton corrections were examined  \cite{Hatsuda:2012dt,
Hatsuda:2013gj} and a cancellation mechanism between worldsheet and
membrane (D2) instantons was discovered.  More recently, the Fermi gas
approach was generalized to the ABJ theory in \cite{Matsumoto:2013nya} and
further studied in \cite{Honda:2014npa}.
In a different line of development, in \cite{Awata:2012jb}, the ABJ
matrix model was rewritten in a form generalizing the ``mirror
description'' for the ABJM theory \cite{Benini:2009qs, Kapustin:2010xq}.
This was achieved by starting with the $U(N_1)\times U(N_2)$ lens space
matrix model \cite{Marino:2002fk, Aganagic:2002wv} which is exactly
computable and analytically continuing it by sending $N_2\to -N_2$ to
reach the ABJ matrix model. As a result, partition function was written
as a ${\rm min}(N_1,N_2)$-dimensional contour integral and the residue
it picks up can be given an interpretation of perturbative or
non-perturbative contribution, according to its $k$ dependence.  Later,
this mirror expression was reproduced in a more direct way in
\cite{Honda:2013pea}.  This formulation is particularly suitable for
studying the duality with the higher spin theory \cite{Chang:2012kt,
HHOS}.

It was conjectured in \cite{Aharony:2008gk} that the ABJ theory has Seiberg
duality, which states that the following two theories are equivalent:
\begin{align}
U(N_1)_k\times U(N_2)_{-k}\quad = \quad U(2N_1-N_2+k)_k\times U(N_1)_{-k}\ ,
\label{SD_intro}
\end{align}
where we assumed $N_1\le N_2$.  This duality can be understood in the
brane realization of the ABJ theory \cite{Aharony:2008gk} as moving
5-branes past each other and creating/annihilating D3-branes between
them by the Hanany-Witten effect \cite{Hanany:1996ie}.  It is a special
case of the Giveon-Kutasov duality \cite{Giveon:2008zn} for more general
$\cN=2$ CS-matter theories,\footnote{An incomplete list of related
papers is \cite{Hanany:2008fj, Amariti:2009rb, Armoni:2009vv,
Evslin:2009pk, Jensen:2009xh, Herzog:2010hf, Benini:2011mf,
Amariti:2011uw, Closset:2012eq, Kim:2012uz, Xie:2013lya,
Dwivedi:2014awa, Amariti:2014ewa, Amariti:2014lla}. } which can be
regarded as the three-dimensional analog of the four-dimensional Seiberg
duality \cite{Giveon:1998sr}.
The equality of the partition function (up to a phase) between the dual
theories in \eqref{SD_intro} was proven in \cite{Kapustin:2010mh},
although one mathematical relation was assumed.  On the other hand, in
the ``mirror'' framework of \cite{Awata:2012jb}, the equality of the
dual partition function is more or less obvious \cite{Awata:2012jb,
Honda:2013pea} and, moreover, it was observed that the perturbative and
non-perturbative contributions get exchanged into each other under the
duality.

Wilson loops are the only observables in pure CS theory
\cite{Witten:1988hf} and, even in CS-matter theory, they are very
natural objects to consider.
It is known that there are two basic circular supersymmetric Wilson
loops for the ABJ(M) theory on $S^3$; the Wilson loop carrying a non-trivial
representation with respect to one of the two gauge groups preserves
$\frac{1}{6}$ of supersymmetry \cite{Drukker:2008zx, Chen:2008bp,
Rey:2008bh} while an appropriate combination of $\frac{1}{6}$-BPS
Wilson loops for two gauge groups preserves $\frac{1}{2}$ of
supersymmetry \cite{Drukker:2009hy}.
The bulk dual of the $1\over 2$-BPS Wilson loop is a fundamental string
while the bulk dual of the $1\over 6$-BPS Wilson loop is not completely
understood \cite{Drukker:2008zx, Chen:2008bp, Rey:2008bh}.\footnote{See
section \ref{SummaryDiscussions} for more discussion.}
By the localization method, these Wilson loops can be computed using the ABJ(M)
matrix model \cite{Kapustin:2009kz}, and the techniques developed for
partition function to study non-perturbative effects can be generalized
to Wilson loops, such as the large $N$ analysis \cite{Marino:2009jd,
Drukker:2010nc, Drukker:2011zy}, the Fermi gas approach
\cite{Klemm:2012ii}, and the cancellation between worldsheet and
membrane instantons \cite{Hatsuda:2013yua, Hatsuda:2013oxa}.

In this paper we study the $\frac{1}{6}$ and $\frac{1}{2}$-BPS Wilson
loops and their Seiberg duality in the ABJ theory, based on the approach
developed in \cite{Awata:2012jb} for partition function.  We mainly
focus on the representations with winding number $n$.
Starting with Wilson loops in lens space matrix model and analytically
continuing it, we obtain a new expression for the supersymmetric Wilson
loops in the ABJ theory in terms of $\mathrm{min}(N_1,N_2)$-dimensional
contour integrals.
As an application and as a way to check the consistency of our formula,
we study Seiberg duality on the ABJ Wilson loops.
The duality rule for Wilson loops is highly non-trivial because they are
non-local operators and not charged under the global symmetry.  How to
find Given-Kutasov duality relations for Wilson loops in general $\cN=2$
CS-matter theories, of which the ABJ theory is a special case, was proposed
in \cite{Kapustin:2013hpk} based on quantum algebraic relations.
We derive the duality rule of the supersymmetric Wilson loops using our
integral expression and confirm that it is consistent with the proposal
of \cite{Kapustin:2013hpk}.  We also discuss how the perturbative and
non-perturbative effects are mapped under  Seiberg duality.
Moreover, we provide a heuristic explanation of the duality rule from
the brane realization of the ABJ theory.  In an appropriate duality
frame, the Wilson loop is interpreted as the position of the branes and,
carefully following how the Hanany-Witten effect acts on it, we
reproduce the correct duality rule.

This paper is organized as follows. In section \ref{MainResults}, we
present the main results of the paper without proof: the integral
expression for the $\frac{1}{6}$-BPS and $\frac{1}{2}$-BPS Wilson loops
and the their transformation rules under Seiberg duality.  In section
\ref{derivation}, we explain how to derive the integral expression for
the ABJ Wilson loops by analytically continuing the lens space ones.  In
section \ref{SDsection}, we first give a heuristic, brane picture for
the Seiberg duality rule, and then present a rigorous proof using the
integral expression.  Section \ref{SummaryDiscussions} is devoted to a
summary and discussion.  Appendices contain further detail of
computations in the main text, as well as a discussion of Wilson loops
with general representation in Appendix \ref{WL_gen_rep} and an
alternative derivation of the Seiberg duality rule using the algebraic
approach of \cite{Kapustin:2013hpk} in Appendix \ref{KapustinWillett}.


\section{Main results}
\label{MainResults}

The Wilson loops we are concerned with are those preserving fractions of
supersymmetries in the ${\cal N}=6$ $U(N_1)_k\times U(N_2)_{-k}$ CSM
theory, also known as the ABJ(M) theory, saturating the
Bogomol'nyi-Prasad-Sommerfield (BPS) bound.  More specifically, we
consider two types of circular BPS Wilson loops on $S^3$, one preserving
a ${1\over 6}$-th and the other preserving a half of supersymmetries
\cite{Drukker:2008zx, Chen:2008bp, Rey:2008bh, Drukker:2009hy}. In the
main text of this paper, we restrict ourselves to the Wilson loops with
winding number $n$, where $n=1$ (or $n=-1$) corresponds to the
fundamental (or anti-fundamental) representation of $U(N_1)$ and/or
$U(N_2)$ gauge groups, and the Wilson loops in more general
representations will be discussed in Appendix \ref{WL_gen_rep}.

\medskip
The ABJ(M) theory consists of two 3$d$ ${\cal N}=2$ vector multiplets $(A_{\mu}^A, \sigma^A, \lambda^A, \bar{\lambda}^A, D^A)$ with $A=1,2$ which are the dimensional reduction of 4$d$ ${\cal N}=1$ vector multiplets and four bifundamental chiral multiplets, an $SU(4)_R$ vector, $(C_I, \psi_I, F_I)$ with $I=1,\cdots,4$ in the representation $(N_1, \bar{N}_2)$ and their conjugates.
The ${1\over 6}$-BPS Wilson loops of our interest are constructed as \cite{Berenstein:2008dc, Drukker:2008zx, Chen:2008bp, Rey:2008bh}
\be
{\cal W}^{\rm I}_{1\over 6}(N_1,N_2; R)_k:=\left\langle\Tr_R P\exp\int \left(iA^1_{\mu}\dot{x}^{\mu}
+{2\pi\over k}|\dot{x}|M^I_{\mbox{ }J}C_I\bar{C}^J\right)ds\right\rangle\ ,
\ee
where the path of the loop specified by the vector $x^{\mu}(s)$ is a
circle, and the matrix $M^I_{\mbox{ }J}$ is determined by the
supersymmetries preserved and one can choose it to be
$M=\diag(-1,-1,1,1)$.  These are the Wilson loops on the first gauge
group $U(N_1)$, but the ${\cal W}^{\rm II}_{1\over 6}(N_1,N_2; R)_k$,
those on the second gauge group $U(N_2)$, can be constructed similarly
by replacing $N_1\to N_2$, $A_{\mu}^1\to A_{\mu}^2$ and $C_I \to
\bar{C}^I$ ($\bar{C}^J \to C_J$).  The ${1\over 2}$-BPS Wilson loop is
constructed in \cite{Drukker:2009hy} and can be conveniently expressed
in terms of the supergroup $U(N_1|N_2)$ as
\be
{\cal W}_{1\over 2}(N_1,N_2;{\cal R})_k:=\left\langle\Tr_{\cal R}P\exp\left(i\int {\cal A}ds\right)\right\rangle\ ,
\ee
where ${\cal R}$ is a super-representation of $U(N_1|N_2)$ and ${\cal A}$ is the super-connection 
\be
{\cal A}=
\left(
\begin{array}{cc}
A^1_{\mu}\dot{x}^{\mu}
-i{2\pi\over k}|\dot{x}|N^I_{\mbox{ }J}C_I\bar{C}^J & \sqrt{2\pi\over k}|\dot{x}|\psi_I\bar{\eta}_I\\
 \sqrt{2\pi\over k}|\dot{x}|\eta_I\bar{\psi}^I & A^2_{\mu}\dot{x}^{\mu}
-i{2\pi\over k}|\dot{x}|N^I_{\mbox{ }J}C_I\bar{C}^J 
\end{array}
\right)
\ee
with the matrix $N^I_{\mbox{ }J}=\diag(-1,1,1,1)$, the (super-)circular path $(x^1,x^2)=(\cos s,\sin s)$, and  $\eta_I=(e^{is/2}, -ie^{-is/2})\delta^1_I$. 

\medskip
As mentioned, we are concerned with Wilson loops with winding number $n$ rather than in generic representations in this paper. Focusing on this class of Wilson loops, the application of the localization technique \cite{Pestun:2007rz} for the ${1\over 6}$-BPS Wilson loops with winding $n$ reduces to the finite dimensional integrals of the matrix model type \cite{Kapustin:2009kz}
\begin{align}
{\cal W}^{\rm I}_{{1\over 6}}(N_1,N_2;n)_k&
=\left\langle \sum_{j=1}^{N_1}e^{n \mu_j}\right\rangle
\label{1/6wilson0}
\end{align}
where the vev is with respect to the eigenvalue integrals
\begin{align}
\left\langle{\cal O}\right\rangle
 :=
 \cN_{\rm ABJ}
 \int\prod_{i=1}^{N_1}{d\mu_i\over 2\pi}\prod_{a=1}^{N_2}{d\nu_a\over 2\pi}
 {\Delta_{\rm sh}(\mu)^2\Delta_{\rm sh}(\nu)^2\over
 \Delta_{\rm ch}(\mu,\nu)^2}\,{\cal O}\,
 e^{-{1\over 2g_s}\left(\sum_{i=1}^{N_1}\mu_i^2-\sum_{a=1}^{N_2}\nu_a^2\right)}
 \label{ABJMM}
\end{align}
with the shorthand notations for the one-loop
determinant factors defined by 
\be
\Delta_{\rm sh}(\mu)=\prod_{1\le
i<j\le N_1}\left(2\sinh\left({\mu_i-\mu_j\over 2}\right)\right)\
,\quad\Delta_{\rm sh}(\nu)=\prod_{1\le a<b\le
N_2}\left(2\sinh\left({\nu_a-\nu_b\over 2}\right)\right) \ee and \be
\Delta_{\rm ch}(\mu,
\nu)=\prod_{i=1}^{N_1}\prod_{a=1}^{N_2}\left(2\cosh\left({\mu_i-\nu_a\over
2}\right)\right)\ .  
\ee
The coupling constant\footnote{Note that this is not the physical string
coupling constant.  The physical string coupling constant of the dual
type IIA string theory in AdS$_4\times \mathbb{CP}^3$ is $(g_s)_{\rm
physical}\sim N^{1/4}k^{-5/4}$.} $g_s$ is related to the CS level $k$ by
\begin{align}
 g_s:={2\pi i \over k},
\end{align}
while the factor ${\cal N}_{\rm ABJ}$ in front is the normalization factor \cite{Marino:2011nm}
\be
{\cal N}_{\rm ABJ}:={i^{-{\kappa\over 2}(N_1^2-N_2^2)}\over N_1!N_2!}\ ,\qquad
\kappa:=\sign k\ .
\ee

Meanwhile, the ${1\over 2}$-BPS Wilson loop
localizes to the supertrace \cite{Drukker:2009hy}
\begin{align}
{\cal W}_{1\over 2}(N_1,N_2;{\cal R})_k=\left\langle \str_{\cal R}^{}
\left(
\begin{array}{cc}
e^{\mu_i} & 0 \\
0 & -e^{\nu_a}
\end{array}
\right)\right\rangle
\end{align}
that yields, for the $n$ winding Wilson loop, a linear combination of ${1\over 6}$-BPS Wilson loops \cite{Klemm:2012ii}
\be
{\cal W}_{1\over 2}(N_1,N_2;n)_k={\cal W}^{\rm I}_{1\over 6}(N_1,N_2; n)_k
-(-1)^n{\cal W}^{\rm II}_{1\over 6}(N_1,N_2; n)_k\ .
\ee
Note that the integral \eqref{ABJMM} is a well-defined Fresnel integral
even for non-integral (but real) $k$ and thus gives a continuous
function $k$, although in the physical ABJ theory the CS level $k$ is
quantized to an integer.

We now present the results of our analysis of these matrix eigenvalue integrals.


\subsection{The Wilson loops in ABJ theory}
\label{WilsonLoopResults}

The ${1\over 6}$-BPS Wilson loops are only on the first $U(N_1)$ or the second $U(N_2)$ gauge group. Depending on whether $N_1\le N_2$ or $N_1\ge N_2$, their formula takes rather different forms. 

\paragraph{$\bullet $ The ${1\over 6}$-BPS $U(N_1)$ Wilson loop with $N_1\le N_2$ :}
In the case of $N_1\le N_2$, introducing the normalized Wilson loop $W^{\rm I}_{1\over 6}(N_1,N_2; n)_k$,\footnote{The normalization is essentially by the partition function.} we find the ${1\over 6}$-BPS Wilson loop with winding number $n$ on the first gauge group $U(N_1)$ to be
\begin{align}
W^{\rm I}_{1\over 6}(N_1,N_2; n)_k
 &:={{\cal W}^{\rm I}_{1\over 6}(N_1,N_2; n)_k\over {\cal W}^{\rm I}_{1\over 6}(N_1,N_2; 0)_k}
=q^{-{n^2\over 2}+n}{I(N_1,N_2;n)_k\over I(N_1,N_2;0)_k}\ ,
\label{1/6WilsonI1}
\end{align}
where
\begin{align}
 q:=e^{-g_s}=e^{-{2\pi i\over k}}
\end{align}
and
\begin{align}
I(N_1,N_2;n)_k&:={1\over N_1!}\sum_{l=1}^{N_1}\prod_{i=1}^{N_1}\left[{-1\over 2\pi i}\int_C{\pi ds_i\over\sin(\pi s_i)}\right]
q^{-ns_l+n(l-2)}\prod_{\substack{i=1 \\ i\ne l}}^{N_1}
{\left(q^{s_i-s_{l}-n}\right)_1\over \left(q^{s_i-s_{l}}\right)_1}
\label{IntegralI}\\
&\quad\times\prod_{i=1}^{N_1}\left[{(-1)^{M}\left(q^{s_i+1}\right)_M
\over\left(1+q^{n\delta_{il}}\right)\left(-q^{s_i+1+n\delta_{il}}\right)_{M}}\prod_{j=1}^{i-1}{\left(q^{s_i-s_j}\right)_{1}\over\left(-q^{s_i-s_j+n\delta_{il}}\right)_{1}}
\prod_{j=i+1}^{N_1}{\left(q^{s_j-s_i}\right)_{1}\over\left(-q^{s_j-s_i-n\delta_{il}}\right)_{1}}\right]
\nn
\end{align}
with $M:=|N_2-N_1|=N_2-N_1$ and the symbol $(a)_z:=(a;q)_z$ is a
shorthand notation for the $q$-Pochhammer symbol defined in Appendix
\ref{qanalogs}.  The choice of the integration contour $C$ will be
discussed in detail in Section \ref{IntRep}.  We note that the integral
expression $I(N_1,N_2;0)_k$ without winding agrees with
that of the
partition function in \cite{Awata:2012jb}.\footnote{The precise relation to the quantity $\Psi$ defined in
\cite{Awata:2012jb} is $I(N_1,N_2;0)_k=N_1 \Psi(N_1,N_2)_k$.}

\paragraph{$\bullet $ The ${1\over 6}$-BPS $U(N_1)$ Wilson loop with $N_1\ge N_2$ :}
In the case of $N_1\ge N_2$  the formula turns out to be slightly more involved and takes the form
\begin{align}
W^{\rm I}_{1\over 6}(N_1,N_2; n)_k:={{\cal W}^{\rm I}_{1\over 6}(N_1,N_2; n)_k\over {\cal W}^{\rm I}_{1\over 6}(N_2,N_1; 0)_k}
=q^{-{n^2\over 2}+n}{I^{(1)}(N_1,N_2;n)_k+I^{(2)}(N_1,N_2;n)_k\over I^{(2)}(N_1,N_2;0)_k}
\label{1/6WilsonI2}
\end{align}
that is only valid for $|n|\ge 1$ as will be elaborated in the comments below\footnote{The expression (\ref{1/6WilsonI1}), on the other hand, does not have this restriction.} and we defined
\begin{align}
I^{(1)}(N_1,N_2;n)_k: &={1\over N_2!}\sum_{c=0}^{n-1}
\prod_{a=1}^{N_2}\left[{-1\over 2\pi i}\int_{C_1\![c]}{\pi ds_a\over\sin(\pi s_a)}\right]
q^{n(2c-M)}{\left(q^{1-n}\right)_c\left(q^{1+n}\right)_{M-1-c}\over (q)_c(q)_{M-1-c}}
\prod_{a=1}^{N_2}{\left(q^{s_a+1}\right)_M\over 2\left(-q^{s_a+1}\right)_{M}}\nn\\
&
 \times \prod_{a=1}^{N_2}\left[{\left(-q^{s_a+1+c}\right)_1\left(q^{s_a+1+c-n}\right)_1
\over \left(q^{s_a+1+c}\right)_1\left(-q^{s_a+1+c-n}\right)_1}\prod_{b=1}^{a-1}{\left(q^{s_a-s_b}\right)_{1}
\over\left(-q^{s_a-s_b}\right)_{1}}
\prod_{b=a+1}^{N_2}{\left(q^{s_b-s_a}\right)_{1}\over\left(-q^{s_b-s_a}\right)_{1}}\right]\ ,
\label{IntegralII1}
\end{align}
and
\begin{align}
I^{(2)}(N_1,N_2;n)_k &:={1\over N_2!}\sum_{d=1}^{N_2}
\prod_{a=1}^{N_2}\left[{-1\over 2\pi i}\int_{C_2}{\pi ds_a\over\sin(\pi s_a)}\right]
q^{-ns_d+n(d-M-2)}
\prod_{\substack{a=1 \\ a\ne d}}^{N_2}{\left(q^{s_a-s_{d}-n}\right)_1\over \left(q^{s_a-s_{d}}\right)_1} \label{IntegralII2}\\
&\times\prod_{a=1}^{N_2}\left[{\left(q^{s_a+1+n\delta_{ad}}\right)_M
\over\left(1+q^{n\delta_{ad}}\right)\left(-q^{s_a+1}\right)_{M}}
\prod_{b=1}^{a-1}{\left(q^{s_a-s_b}\right)_{1}
\over\left(-q^{s_a-s_b+n\delta_{ad}}\right)_{1}}
\prod_{b=a+1}^{N_2}{\left(q^{s_b-s_a}\right)_{1}\over\left(-q^{s_b-s_a-n\delta_{ad}}\right)_{1}}\right]\nn
\end{align}
with $M:=|N_2-N_1|=N_1-N_2$. 
Note that $I^{(2)}(N_1,N_2;0)_k=I(N_2,N_1;0)_k$ and the normalization in (\ref{1/6WilsonI2}) differs from that in (\ref{1/6WilsonI1}) in that the ranks of the gauge groups $N_1$ and $N_2$ are exchanged.
The choice of the integration contours $C_1=\{C_1[c]\}$ ($c=0,\cdots,n-1$) and $C_2$ will be discussed in detail in Section \ref{IntRep}. We would, however, like to make a remark concerning the contour $C_1[c]$: There are subtleties in evaluating the integrals with the contour $C_1[c]$. In order to properly deal with them, we shall adopt the $\epsilon$-prescription shifting the parameter $M\to M+\epsilon$ with $\epsilon>0$ and the contour is placed between $s_a=-1-c$ and $-1-c-\epsilon$. Related comments will be made in Section \ref{ABJMlimit} below \eqref{limitresidue} and Appendix \ref{Cancellation}.

\paragraph{$\bullet $ The ${1\over 6}$-BPS $U(N_2)$ Wilson loops :}
It follows from the definition and symmetry that the ${1\over 6}$-BPS Wilson loop on the second gauge group $U(N_2)$ with winding number $n$ is related to that on the first gauge group $U(N_1)$ in a simple manner:
\begin{align}
W^{\rm II}_{1\over 6}(N_1,N_2; n)_k
=W^{\rm I}_{1\over 6}(N_2,N_1; n)_{-k}\ .\label{1/6WilsonII}
\end{align}

\paragraph{$\bullet $ The ${1\over 2}$-BPS Wilson loop :}
Meanwhile, the (normalized) ${1\over 2}$-BPS Wilson loop is given by a linear combination of two (normalized) ${1\over 6}$-BPS Wilson loops, one on the first and the other on the second gauge group,  and turns out to take a rather simple form
\begin{align}
W_{\half}(N_1,N_2; n)_k& :=W^{\rm I}_{1\over 6}(N_1,N_2; n)_{k}-(-1)^nW^{\rm II}_{1\over 6}(N_1,N_2; n)_{k}\nn\\
&=\left\{
\begin{array}{lcl}
 \displaystyle
(-1)^{n+1}q^{{n^2\over 2}-n}{I^{(1)}(N_2,N_1;n)_{-k}\over I^{(2)}(N_2,N_1;0)_{-k}} &\mbox{for}& N_1 \le N_2\\[2ex]
 \displaystyle
q^{-{n^2\over 2}+n}{I^{(1)}(N_1,N_2;n)_{k}\over I^{(2)}(N_1,N_2;0)_k} &\mbox{for}& N_1 \ge N_2\ ,
\label{1/2Wilson}
\end{array}
\right.
\end{align}
where the two terms $q^{-\half n^2+n}I^{(2)}(N_1,N_2;n)_k$ and $(-1)^nq^{\half n^2-n}I(N_2,N_1;n)_{-k}$ cancel out, as we will show later in Section \ref{1/2BPSproof}.

\paragraph{$\bullet$ Comments on the zero winding limit $n\to 0$ :}

There are a few subtleties to be addressed in the expressions
$I^{(1)}$ in (\ref{IntegralII1}) and $I^{(2)}$ in (\ref{IntegralII2}). They are linked to the
comments made below (\ref{S1}) and (\ref{S2}) concerning the ranges of
the summations and the discussions in Appendix \ref{Cancellation}. Here
we focus on the subtlety in the range of the sum $\sum_{c=0}^{n-1}$ in
(\ref{IntegralII1}). In its original form, the sum over $c$ is taken
from $0$ to $M-1$ as derived in (\ref{S1original}) in Appendix
\ref{analcontdetail}. However, when $n\ge 1$, we can rewrite the sum
(\ref{S1original}) $+$ (\ref{S2original}), after passing it to the
integral representation discussed in Section \ref{IntRep}, by the sum
(\ref{IntegralII1}) $+$ (\ref{IntegralII2}).  In particular, the upper
limit $M-1$ of the sum over $c$ can be replaced by $n-1$. For $n< M$
this relies on the fact that the factor $(q^{1-n})_c=0$ when $n\ge 1$
and $c\ge n$. Since the factor $(q^{1-n})_c$ does not vanish when $n<1$,
it implies that the $n=0$ limit of (\ref{IntegralII1}) that violates the
bound $n\ge 1$ would not coincide with ($M$ times) the integral
representation for the partition function in \cite{Awata:2012jb}.  In
fact, (\ref{IntegralII1}) in the $n=0$ limit simply vanishes, whereas
(\ref{IntegralII2}) reduces to ($N_2$ times) the integral representation
for the partition function.

\paragraph{$\bullet$ Comments on the bound on winding number $n$ :}
We observe that the integrals (\ref{IntegralI}) and (\ref{IntegralII2})
diverge when $|n|\ge {k\over 2}$.\footnote{We thank Nadav Drukker for
discussions on this point. This is also in accord with the singularities
at $k=1, 2$ observed in \cite{Klemm:2012ii}. See more comments on the ${1\over 2}$-BPS case.} As we will elaborate later, the contour $C$
for the integrals goes to imaginary infinity. For large imaginary $s_l$
and $s_c$, the integrands become asymptotically $\exp(2\pi
nis_l/k)/\sin(\pi s_l)$ and $\exp(2\pi nis_c/k)/\sin(\pi s_c)$,
respectively, that grow exponentially when $|n| > {k\over 2}$ and
approach a constant when $|n|={k\over 2}$.  This implies that ${1\over
6}$-BPS Wilson loops are only well-defined for $|n|< {k\over 2}$. 
In the ${1\over 2}$-BPS case, it might appear that the Wilson loop is well-defined for all
values of $n$, since the integral (\ref{IntegralII1}) converges for any
$n$. Note, however, first that since (\ref{IntegralII1}) is periodic under the shift
$n\to n+k$ owing to the properties, $q^k=1$ and $(q^{1-n})_{c\ge n}=0$,
the winding $n$ is restricted to the range $|n|\le {k\over 2}$. Moreover, as indicated in \eqref{IntegralII1ABJM}, a closer inspection shows that the ${1\over 2}$-BPS Wilson loop, in the ABJM limit, diverges at the winding $n={k\over 2}$ due to the factor $(1+q^n)^{-1}$ that comes from the pole at $s_a=-1-c$. This implies that in the case of $k=1, 2$, the only allowed winding is $n=0$ and thus the ${1\over 2}$-BPS Wilson loop does not exist.\footnote{We are indebted to Yasuyuki Hatsuda for discussions on this point, correcting our statements in the previous version of this paper.}

\paragraph{$\bullet$ Comments on the ABJM limit :}
In the ABJM limit the ranks of two gauge groups are equal, i.e., $N_1=N_2\Longleftrightarrow M=0$, and the two results (\ref{1/6WilsonI1}) and (\ref{1/6WilsonI2}) coincide as they should.
However, the way they coincide turns out to be very subtle. Naively, it may look that $I^{(1)}(N_1,N_2;n)_k$ vanishes and $I(N_1,N_2;n)_k$ and $I^{(2)}(N_1,N_2;n)_k$ coincide in the ABJM limit. However, a careful analysis of the ABJM limit reveals that $\lim_{N_1\to N_2}I^{(1)}(N_1,N_2;n)_k$ remains finite, but the sum $I^{(1)}(N_1,N_2;n)_k+I^{(2)}(N_1,N_2;n)_k$ coincides with $I(N_1,N_2;n)_k$ in the limit. As will be elaborated in Section \ref{ABJMlimit}, this is rooted in the difference of the integration contours $C$ in (\ref{1/6WilsonI1}) and $C_1, C_2$ in (\ref{1/6WilsonI2}). Since $C\ne C_2$, the integral $I^{(2)}(N_1,N_2;n)_k$ differs from $I(N_1,N_2;n)_k$ even in the ABJM limit, but this difference is canceled by $I^{(1)}(N_1,N_2;n)_k$.
Note that it is very important that $\lim_{N_1\to N_2}I^{(1)}(N_1,N_2;n)_k\ne 0$, since the ${1\over 2}$-BPS Wilson loop (\ref{1/2Wilson}) exists nonvanishing in the ABJM limit.


\subsection{Seiberg duality of the Wilson loops}

As we discussed in the introduction, the ABJ theory is conjectured to
possess Seiberg duality as given in \eqref{SD_intro}. In our previous
paper \cite{Awata:2012jb}, we have explicitly checked that the partition
functions of a dual pair are equal to each other up to a phase
factor. The precise form of the phase factor was first conjectured by
\cite{Kapustin:2010mh} and later derived in \cite{Willett:2011gp,
Benini:2011mf}. The difference of the phase factors in dual pairs is now
understood as an anomaly in large gauge transformations
\cite{Closset:2012vp}. We would like to emphasize that not only does our
formula for the partition function in \cite{Awata:2012jb} confirm the
proof in \cite{Kapustin:2010mh, Willett:2011gp} but it also allows us to
understand how Seiberg duality of the ABJ theory works in detail. In
particular, it was observed explicitly in \cite{Awata:2012jb} that the
perturbative and nonperturbative contributions to the partition function
are exchanged under the duality. In Section \ref{SDsection} we will see
the same property in the duality of Wilson loops.

\paragraph{$\bullet $ The duality map of ${1\over 6}$-BPS Wilson loops :}
Using our formulas, we find the maps between Wilson loops in the original and Seiberg dual theories:
\begin{align}
W^{\rm II}_{1\over 6}(N_1,N_2; n)_k&=-W^{\rm I}_{1\over 6}(\widetilde{N}_2,N_1; n)_k
-2(-1)^{n+1}W^{\rm II}_{1\over 6}(\widetilde{N}_2,N_1; n)_k\ ,\label{1/6duality}\\
W^{\rm I}_{1\over 6}(N_1,N_2; n)_k&=W^{\rm II}_{1\over 6}(\widetilde{N}_2,N_1; n)_k
\label{flavorduality}
\end{align}
for the ${1\over 6}$-th BPS Wilson loops, and the rank of the dual gauge group is denoted by $\widetilde{N}_2=2N_1-N_2+k$.

\paragraph{$\bullet $ The duality map of ${1\over 2}$-BPS Wilson loops :} For the $\half$-BPS Wilson loops, the duality map turns out to be very simple
\be
W_{\half}(N_1,N_2; n)_k = (-1)^n W_{\half}(\widetilde{N}_2,N_1; n)_k\ .
\label{1/2duality}
\ee

\medskip\noindent
Note that these three relations are not independent, but one of them can be derived from the other two. In the following sections, we will refer to (\ref{1/6duality}), (\ref{flavorduality}), and (\ref{1/2duality}) as the ${1\over 6}$-BPS Wilson loop duality, the flavor Wilson loop duality, and the $\half$-BPS Wilson loop duality, respectively.\footnote{Under the duality the $U(N_1)$ sector is inert and can be regarded as a \lq\lq flavor group.'' The Wilson loops in (\ref{flavorduality}) are only on the $U(N_1)$ \lq\lq flavor group,'' hence the name, the flavor Wilson loop duality.} 

\medskip
We will vindicate these maps by a heuristic yet very useful argument based on the brane configuration in Section \ref{branepicturesection} and an alternative proof that is an application of the proof by Kapustin and Willett \cite{Kapustin:2013hpk} in Appendix \ref{KapustinWillett}.

\section{The derivation of the results}
\label{derivation}
 
We follow the same strategy as that employed in the computation of the ABJ partition function \cite{Awata:2012jb}. The outline of the derivation is as follows:

\medskip\noindent
(1) We first compute Wilson loops in the $U(N_1)\times U(N_2)$ lens space matrix model, where the Wilson loops are defined in analogy to those in the ABJ theory. The matrix integrals are simply Gaussian and can be done exactly.

\medskip\noindent
(2) We then analytically continue the rank of one of the gauge groups from $N_2$ to $-N_2$. This maps the lens space matrix model to the ABJ matrix model \cite{Drukker:2010nc, Marino:2011nm}. As in the case of the partition function, the result so obtained is expressed in terms of formal series that is not well-defined in the regime of the strong coupling we are concerned with and only sensible as perturbative expansion with the generalized $\zeta$-function (polylogarithm) regularization assumed. 

\medskip\noindent
(3) Similar to the case of the partition function, we can render the formal series perfectly well-defined by means of the Sommerfeld-Watson transform and the resultant expression is given in terms of $\mbox{min}(N_1, N_2)$-dimensional integrals. This is a nonperturbative completion in the sense that what renders the formal series well-defined is an inclusion of nonperturbative contributions, as will be elaborated later. 

\subsection{The lens space Wilson loop}

We define the (un-normalized) Wilson loop on the first gauge group $U(N_1)$ with $n$ windings in the lens space matrix model by
\begin{align}
{\cal W}_{\rm lens}^{\rm I}(N_1,N_2;n)_k:=
\left\langle\sum_{j=1}^{N_1}e^{n \mu_j}\right\rangle\ ,
\label{WilsonLens}
\end{align}
where we have defined the expectation value of ${\cal O}$ by
\begin{align}
 \langle{\cal O}\rangle :=\cN_{\rm lens}
 \int\prod_{i=1}^{N_1}{d\mu_i\over 2\pi}\prod_{a=1}^{N_2}{d\nu_a\over 2\pi}
 \Delta_{\rm sh}(\mu)^2\Delta_{\rm sh}(\nu)^2\Delta_{\rm ch}(\mu,\nu)^2
 \,{\cal O}\,
 e^{-{1\over 2g_s}\left(\sum_{i=1}^{N_1}\mu_i^2+\sum_{a=1}^{N_2}\nu_a^2\right)}\ ,
 \label{vev}
\end{align}
where
the normalization constant $\cN_{\rm lens}$ is given by
\be
{\cal N}_{\rm lens}={i^{-{\kappa\over 2}(N_1^2+N_2^2)}\over N_1!N_2!}\ .
\ee
As will be shown in detail in
Appendix \ref{lensWL}, the eigenvalue integrals are simply Gaussian and
can be carried out exactly. Introducing the normalized Wilson loop, the
end result takes the form \be W_{\rm lens}^{\rm I}(N_1,N_2;n)_k:={{\cal
W}_{\rm lens}^{\rm I}(N_1,N_2;n)_k\over{\cal W}_{\rm lens}^{\rm
I}(N_1,N_2;0)_k}=q^{-{n^2\over 2}+n}{S(N_1,N_2;n)_k\over S(N_1,
N_2;0)_k}\ , \label{LensSpaceWilsonLoop} \ee where the function $S(N_1,
N_2;n)$ is given by
\begin{align}
S(N_1,N_2,n)_k&={1\over N_1!}\sum_{l=1}^{N_1}\sum_{1\le C_1,\cdots, C_{N_1}}q^{-2nC_l+n(N+l-1)}\prod_{\substack{k,l=1\\ k\neq l}}^{N_1}
{\left(q^{C_k-C_{l}-n}\right)_1\over \left(q^{C_k-C_{l}}\right)_1}\label{LensSpaceSfunction}\\
&\times\prod_{i=1}^{N_1}
 \left[{\left(-q^{1+n\delta_{il}}\right)_{C_i-1}\left(-q^{1-n\delta_{il}}\right)_{N-C_i}\over(q)_{C_i-1}(q)_{N-C_i}}
\prod_{j=1}^{i-1}{\left(q^{C_i-C_j}\right)_{1}\over\left(-q^{C_i-C_j+n\delta_{il}}\right)_{1}}
\prod_{j=i+1}^{N_1}{\left(q^{C_j-C_i}\right)_{1}\over\left(-q^{C_j-C_i-n\delta_{il}}\right)_{1}}\right]\ ,
\nn
\end{align}
where $N=N_1+N_2$. We note that this is actually a finite sum: There is no contribution from $C_i> N$, since the factor $1/(q)_{N-C_i}=\left(q^{1-(C_i-N)}\right)_{C_i-N}$ for $N-C_i<0$ vanishes. We have deliberately rewritten the result in the form of an infinite sum that is suitable for the analytic continuation in the next section.

\subsection{The analytic continuation}

The ${1\over 6}$-BPS Wilson loop in ABJ theory can be obtained from the lens space Wilson loop (\ref{LensSpaceWilsonLoop}) by means of the analytic continuation $N_2\to -N_2$. A little care is needed for the analytic continuation. Namely, the analytic continuation requires a regularization: We first replace $N_2$ by $-N_2+\epsilon$ and send $\epsilon\to 0$ in the end.
As mentioned in Section \ref{WilsonLoopResults}, we need to treat the cases $N_1\le N_2$ and $N_1\ge N_2$ separately. We leave most of the computational details in Appendix \ref{analcontdetail}.

\paragraph{$\bullet $ The ${1\over 6}$-BPS $U(N_1)$ Wilson loop with $N_1\le N_2$ :}

The function $S(N_1,N_2;n)_k$ in (\ref{LensSpaceSfunction}) is continued as
\begin{align}
S(N_1,-N_2+\epsilon,n)_k&
=\left(\epsilon\ln q\right)^{N_1}S^{\rm ABJ}(N_1,N_2,n)_k
+{\cal O}\left(\epsilon^{N_1+1}\right)\ ,
\label{N1lessN2analcont}
\end{align}
where we have defined 
\begin{align}
S^{\rm ABJ}(N_1,N_2,n)_k&:={1\over N_1!}\sum_{l=1}^{N_1}\sum_{1\le C_1,\cdots,C_{N_1}}q^{-nC_l+n(l-1)}\prod_{\substack{k,l=1 \\ k\ne l}}^{N_1}
{\left(q^{C_k-C_{l}-n}\right)_1\over \left(q^{C_k-C_{l}}\right)_1}\label{ABJSfunction1}\\
&\times\prod_{i=1}^{N_1}\left[{(-1)^{C_i+M}\left(q^{C_i}\right)_M
\over\left(1+q^{n\delta_{il}}\right)\left(-q^{C_i+n\delta_{il}}\right)_{M}}\prod_{j=1}^{i-1}{\left(q^{C_i-C_j}\right)_{1}\over\left(-q^{C_i-C_j+n\delta_{il}}\right)_{1}}
\prod_{j=i+1}^{N_1}{\left(q^{C_j-C_i}\right)_{1}\over\left(-q^{C_j-C_i-n\delta_{il}}\right)_{1}}\right]
\nn
\end{align}
with $M:=|N_2-N_1|=N_2-N_1$. 
Note that in contrast to the lens space case, there is no truncation of summations and this sum is really an infinite sum. In fact, this is not a convergent sum and becomes ill-defined for the value of $q=e^{-2\pi i/k}$ that is of our actual interest. Thus this expression is at best a formal series and we shall render it well-defined on the entire $q$-plane by means of a type of  Sommerfeld-Watson transform in the next section.

\medskip
Hence the analytic continuation yields an expression in terms of formal series
\be
\left[W^{\rm I}_{\rm ABJ}(N_1,N_2;n)_k\right]_{\rm naive}:=
\lim_{\epsilon\to 0}W_{\rm lens}^{\rm I}(N_1,-N_2+\epsilon;n)_k=q^{-{n^2\over 2}+n}{S^{\rm ABJ}(N_1,N_2;n)_k\over S^{\rm ABJ}(N_1, N_2;0)_k}
\label{formalABJ1/6WilsonLoop}
\ee
for $N_1\le N_2$. After implementing the generalized $\zeta$-function
(polylogarithm) regularization, this result agrees with the final result
(\ref{1/6WilsonI1}) only in perturbative expansion in the coupling
constant $g_s(=-\log q)$. Until Sommerfeld-Watson like transform is
performed, this result is not nonperturbatively complete.

\paragraph{$\bullet $ The ${1\over 6}$-BPS $U(N_1)$ Wilson loop with $N_1\ge N_2$ :} This case is slightly more involved than the previous case. The analytic continuation of $S(N_1,N_2;n)_k$ consists of two terms
\begin{align}
S(N_1,-N_2+\epsilon;n)_k&
=\left(\epsilon\ln q\right)^{N_2}\left(S^{\rm ABJ}_{(1)}(N_1,N_2;n)_k
+S^{\rm ABJ}_{(2)}(N_1,N_2;n)_k\right)
+{\cal O}\left(\epsilon^{N_2+1}\right)\ ,
\label{N1geN2Sn}
\end{align}
where we have defined
\begin{align}
S^{\rm ABJ}_{(1)}(N_1,N_2;n)_k&:={1\over N_2!}\sum_{c=0}^{n-1}\sum_{-c\le D_1,\cdots ,D_{N_2}}
q^{n(2c-M)}{\left(q^{1-n}\right)_c\left(q^{1+n}\right)_{M-1-c}\over (q)_c(q)_{M-1-c}}
\prod_{a=1}^{N_2}{\left(q^{D_a}\right)_M\over 2\left(-q^{D_a}\right)_M}\label{S1}\\
&\times \prod_{a=1}^{N_2}\left[{(-1)^{D_a}\left(-q^{D_a+c}\right)_1\left(q^{D_a+c-n}\right)_1
\over \left(q^{D_a+c}\right)_1\left(-q^{D_a+c-n}\right)_1}\prod_{b=1}^{a-1}{\left(q^{D_a-D_b}\right)_{1}
\over\left(-q^{D_a-D_b}\right)_{1}}
\prod_{b=a+1}^{N_2}{\left(q^{D_b-D_a}\right)_{1}\over\left(-q^{D_b-D_a}\right)_{1}}\right]\ ,\nn\\
S^{\rm ABJ}_{(2)}(N_1,N_2;n)_k&:={1\over N_2!}
\sum_{d=1}^{N_2}\sum_{\substack{1\le D_1,\cdots,D_{d-1},D_{d},\cdots, D_{N_2}\\-n+1\le D_d}}q^{-nD_d+n(d-M-1)}
\prod_{\substack{a=1 \\ a\ne d}}^{N_2}{\left(q^{D_a-D_{d}-n}\right)_1\over \left(q^{D_a-D_{d}}\right)_1} \label{S2}\\
&\times\prod_{a=1}^{N_2}\left[{(-1)^{D_a}\left(q^{D_a+n\delta_{ad}}\right)_M
\over\left(1+q^{n\delta_{ad}}\right)\left(-q^{D_a}\right)_{M}}\prod_{b=1}^{a-1}{\left(q^{D_a-D_b}\right)_{1}
\over\left(-q^{D_a-D_b+n\delta_{ad}}\right)_{1}}
\prod_{b=a+1}^{N_2}{\left(q^{D_b-D_a}\right)_{1}\over\left(-q^{D_b-D_a-n\delta_{ad}}\right)_{1}}\right]\nn
\end{align}
with $M:=|N_2-N_1|=N_1-N_2$ and we replaced the sum $\sum_{c=0}^{M-1}$ by $\sum_{c=0}^{n-1}$ provided that $n>0$ in the first line of (\ref{S1}). See the remark below (\ref{IntegralII2}) for the explanation. 
The caveat on convergence and well-definedness of the sum noted in the previous case applies to this case as well. Note that when $n=0$, (\ref{S1}) vanishes by definition.

\medskip
An important remark is in order: In the first line of (\ref{S1}), we replaced the sum $\sum_{c=0}^{M-1}$ by $\sum_{c=0}^{n-1}$ provided that $n>0$ and extended the range of $D_a$'s from $\sum_{1\le D_1,\cdots ,D_{N_2}}$ to $\sum_{-c\le D_1,\cdots ,D_{N_2}}$. In sync with this replacement and extension, we extended the range of $D_d$ from $\sum_{1\le D_d}$ to $\sum_{-n+1\le D_d}$ in the first line of (\ref{S2}). As shown in Appendix \ref{Cancellation}, the added contributions conspire to cancel out in the sum $S^{\rm ABJ}_{(1)}+S^{\rm ABJ}_{(2)}$, justifying the replacement and extensions we have made in (\ref{S1}) and (\ref{S2}).

\medskip
Similar to the previous case, the analytic continuation yields an expression in terms of formal series
\bea
\left[W^{\rm I}_{\rm ABJ}(N_1,N_2;n)_k\right]_{\rm naive}&:=&
\lim_{\epsilon\to 0}W_{\rm lens}^{\rm I}(N_1,-N_2+\epsilon;n)_k\nn\\
&=&q^{-{n^2\over 2}+n}{S^{\rm ABJ}_{(1)}(N_1,N_2;n)_k+S^{\rm ABJ}_{(2)}(N_1,N_2;n)_k\over S^{\rm ABJ}_{(2)}(N_1, N_2;0)_k}
\label{formalABJ1/6WilsonLoop}
\eea
for $N_1\ge N_2$. Note that $S^{\rm ABJ}_{(1)}(N_1,N_2;0)=0$ as inferred
from (\ref{S1}) and the denominator $S^{\rm ABJ}_{(2)}(N_1,N_2;0)=S^{\rm
ABJ}(N_2,N_1;0)$. Again this result is not complete as yet but agrees,
after the generalized $\zeta$-function regularization, with the final
result (\ref{1/6WilsonI2}) in perturbative expansion in the coupling constant $g_s(=-\log q)$.

\paragraph{$\bullet $ The ${1\over 6}$-BPS $U(N_2)$ Wilson loops :} As stated in the summary of the main results, it follows from the definition and symmetry that the Wilson loop on the second gauge group $U(N_2)$ is obtained from that on the first gauge group $U(N_1)$ as
\begin{align}
\left[W^{\rm II}_{\rm ABJ}(N_1,N_2; n)_k\right]_{\rm naive} =\left[W^{\rm I}_{\rm ABJ}(N_2,N_1; n)_{-k}\right]_{\rm naive}\ .
\end{align}

\paragraph{$\bullet $ The ${1\over 2}$-BPS Wilson loop :}
The ${1\over 2}$-BPS Wilson loop is given by a linear combination of two ${1\over 6}$-BPS Wilson loops, one on the first and the other on the second gauge group:
\begin{align}
\left[W^{\half}_{\rm ABJ}(N_1,N_2; n)_k\right]_{\rm naive}& :=W^{\rm I}_{\rm ABJ}(N_1,N_2; n)_{k}-(-1)^nW^{\rm II}_{\rm ABJ}(N_1,N_2; n)_{k}\nn\\
&=q^{-{n^2\over 2}+n}{S^{\rm ABJ}_{(1)}(N_1,N_2;n)_k
\over S^{\rm ABJ}_{(2)}(N_1,N_2;0)_k}
\label{formal1/2Wilson}
\end{align}
for $N_1\ge N_2$ where we used the fact $q^{-\half n^2+n}S^{\rm ABJ}_{(2)}(N_1,N_2;n)_k
=(-1)^nq^{\half n^2-n}S^{\rm ABJ}(N_2,N_1;n)_{-k}$ that we will show in Section \ref{1/2BPSproof}.

\medskip
In the next subsection we discuss a nonperturbative completion of the above naive results that were given in terms of the formal series $S^{\rm ABJ}$, $S^{\rm ABJ}_{(1)}$ and $S^{\rm ABJ}_{(2)}$. 

\subsection{The integral representation -- a nonperturbative completion}
\label{IntRep}

The analytic continuation in the previous section yielded tentative results for the ABJ Wilson loops that involve formal series (\ref{ABJSfunction1}), (\ref{S1}) and (\ref{S2}). These are non-convergent formal series, since the summands do not vanish as $C_i$'s and $D_a$'s run to infinity. If we are only interested in perturbative expansion in the coupling $g_s=-\log q$, implementing the generalized $\zeta$-function regularization
\be
\sum_{s=0}^{\infty}(-1)^ss^n=\left\{
\begin{array}{ll}
{\rm Li}_{-n}(-1)=(2^{n+1}-1)\zeta(-n)
=-{2^{n+1}-1\over n+1}B_{n+1} &\quad(\mbox{for}\quad n\ge 1)\ ,\\
1+{\rm Li}_0(-1) = \half=-B_1 &\quad(\mbox{for}\quad n=0)\ ,
\end{array}
\right.
\label{genZeta}
\ee
where ${\rm Li}_s(z)$ is the polylogarithm and $B_n$ are the Bernoulli numbers, the formal series can be rendered convergent as in the case of the partition function \cite{Awata:2012jb}. The $\zeta$-function regularized Wilson loops so obtained indeed reproduce the correct perturbative expansions in $g_s$.
However, when $q=e^{-g_s}$ is a root of unity with $g_s=2\pi i/k$ that is the value we are  actually interested in and beyond the perturbative regime, the sums (\ref{ABJSfunction1}), (\ref{S1}) and (\ref{S2}) diverge and require a nonperturbative completion. 

\medskip
Fortunately, as we have done so for the partition function \cite{Awata:2012jb}, these problems can be circumvented by introducing an integral representation similar to the Sommerfeld-Watson transform:\footnote{We thank Yoichi Kazama and Tamiaki Yoneya for pointing out a similarity of this prescription to the Sommerfeld-Watson transformation.}
\begin{align}
\sum_{1\le C_1,\cdots, C_{N_1}}\prod_{i=1}^{N_1}(-1)^{C_i+1}\left(\dots\right)
&\quad\longrightarrow\quad \prod_{i=1}^{N_1}\left[{-1\over 2\pi i}\int_C{\pi ds_i\over\sin(\pi s_i)}\right]\left(\dots\right)\ ,\label{integral1}\\
\sum_{1\le D_1,\cdots, D_{N_2}}\prod_{a=1}^{N_2}(-1)^{D_a+1}\left(\dots\right)
&\quad\longrightarrow\quad \prod_{a=1}^{N_2}\left[{-1\over 2\pi i}\int_C{\pi ds_a\over\sin(\pi s_a)}\right]\left(\dots\right)\ ,\label{integral2}
\end{align} 
where $C_i$ is replaced by $s_i+1$ and $D_a$ by $s_a +1$. As we will see shortly, \emph{this is a transformation that adds nonperturbative contributions missed in the formal series.}
Note, however, that this is a prescription that lacks a first principle derivation and needs to be justified. In the case of the partition function, this prescription has passed both perturbative and nonperturbative checks. In particular, the latter has confirmed the equivalence of Seiberg dual pairs up to the aforementioned phase factors  \cite{Awata:2012jb}. More recently, a direct proof of this prescription for the partition function was given by Honda by utilizing a generalization of Cauchy identity  \cite{Honda:2013pea}. In the Wilson loop case, although we are missing a similar derivation, the proof of Seiberg duality in Section \ref{SDsection} provides convincing evidence for this prescription.

\medskip
The contour $C$ of integration in (\ref{integral1}) and (\ref{integral2}) is chosen in order that (1) the perturbative expansion is correctly reproduced for small $g_s$ (corresponding to large $k$) and, (2) as we decrease $k$ continuously, the values of integrals remain continuous as functions of $k$. These requirements yield the contour $C$ parallel and left to the imaginary axis. To elaborate on it, we look into the pole structure of integrands: 

\paragraph{$\bullet $ The pole structure for $n=0$ :}
It is illustrative to first review the pole structure for the partition function, i.e., the $n=0$ case. In this case, without loss of generality, we can assume $N_1 \le N_2$.

\medskip It is very useful to classify the poles into two classes,
perturbative (P) and nonperturbative (NP) poles. We are interested in
the summand in (\ref{ABJSfunction1}) with $n=0$. By multiplying the
factor $\prod_{i=1}^{N_1}\left[-1/(2i\sin(\pi s_i))\right]$, this
becomes the integrand with the replacement $C_i \to s_i+1$. The P poles
come from the factor $\prod_{j=1}^{N_1}\left[(q^{s_j+1})_M/\sin(\pi
s_j)\right]$, whereas the NP poles are from the factors
$\prod_{j=1}^{N_1}1/(-q^{s_j+1})_M$ and $\prod_{j\ne
i}1/(-q^{s_j-s_i})_1$ for \emph{generic} (real non-integral) values of
$k$.

\begin{figure}[h!]
\centering \includegraphics[height=1.8in]{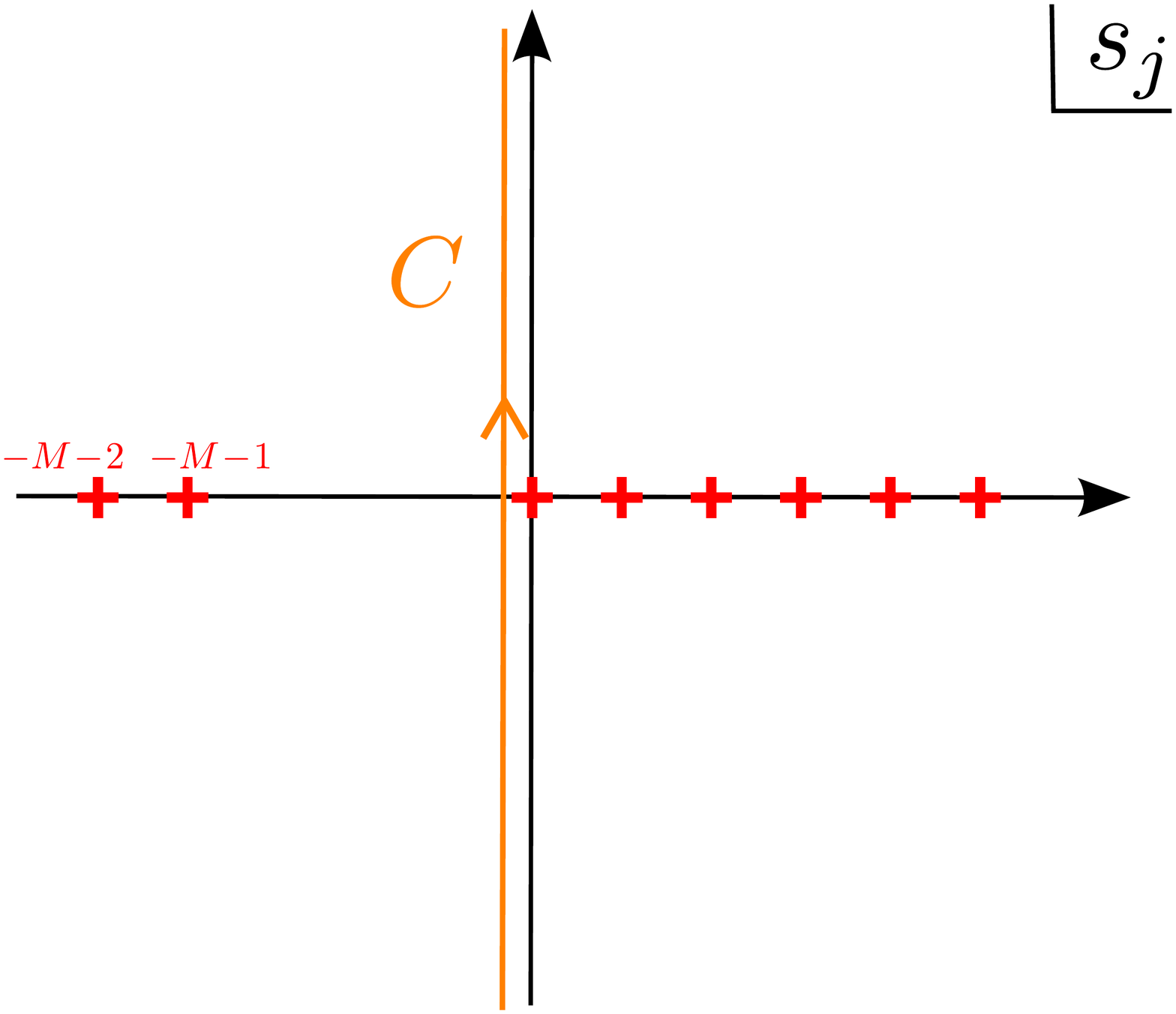} 
\hspace{1.5cm}
\includegraphics[height=1.8in]{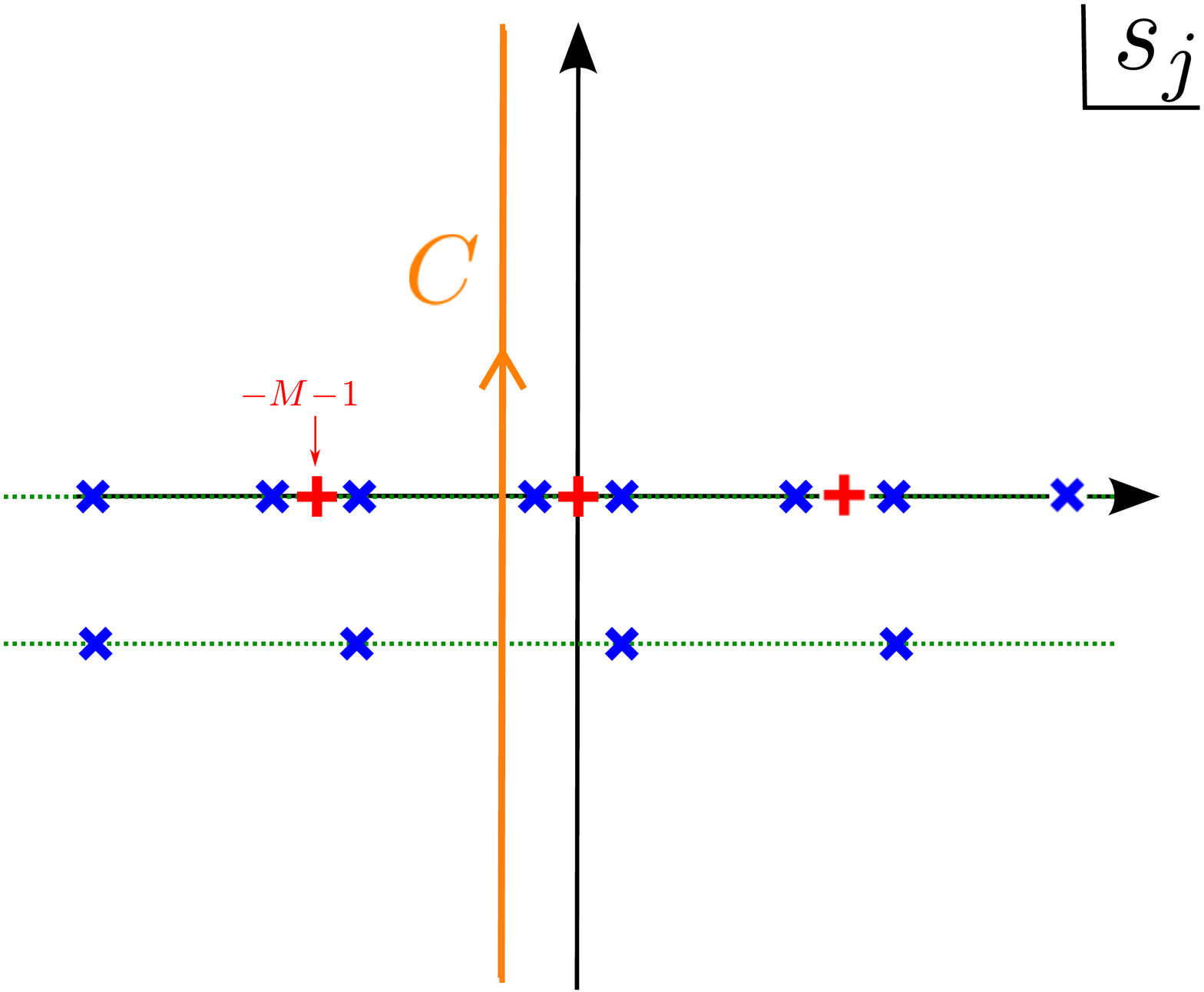} 
\caption{The
integration contour $C$ for the partition function: (A) The left figure corresponds to the large $k$ limit where only perturbative (P) poles indicated by red ``$+$'' are present. 
(B) The right figure is an example of the finite $k$ case ($k=3$ and $M=2$) where  nonperturbative (NP) poles indicated by blue ``$\times$'' are also present.  The green dotted line corresponds to $s_j=s_i-{k\over 2}$ mod $k$ with some $s_i$. Note that some of P poles and zeros of integrands coalesce and cancel out for integer $k$. }  \label{contourFig1}
\end{figure}

\medskip
Hence the P poles are at
\be
s_j \quad = \qquad \dots, -M-2, -M-1;\qquad 0, 1, 2, \dots
\label{Ppoles}
\ee
with the gap between $s_j = -M-1$ and $0$ as shown in the left figures of Fig.~\ref{contourFig1}. In \eqref{Ppoles}, we organized poles into groups separated by a semicolon to clarify this gap structure.
In the large $k$ limit, these are the only poles. The contour $C$ parallel to the imaginary axis can be placed anywhere in the gap. Indeed, enclosing the contour with an infinite semi-circle to the right in the complex $s_j$ plane, the residue integral reduces to the sum (\ref{ABJSfunction1}) with the generalized $\zeta$-function regularization (\ref{genZeta}) implemented automatically by the integral formula
\be
-{1\over 2\pi i}\int_C{\pi ds\over\sin(\pi s)}s^n =-{2^{n+1}-1\over n+1}B_{n+1}
\qquad (n\ge 0)\ ,
\ee
and the perturbative expansion in $g_s$ is correctly reproduced. 
It should now be clear why this class of poles are called P poles.

\medskip
The NP poles are at 
\begin{align}
s_j \quad &= \qquad {k\over 2}-M, {k\over 2}-(M-1), \dots, {k\over 2}-1\qquad{\rm mod}\quad k\\
s_j \quad &= \qquad s_i +{k\over 2}\qquad{\rm mod}\quad k
\end{align}
This class of poles are called NP poles because $k\propto 1/g_s$ and thus the residues are of order  $\exp(-1/g_s)$.
On the right panel of Fig.~\ref{contourFig1} shown is the case $k=3$ and $M=2$. Note that for \emph{integer} $k$ as opposed to \emph{generic} (non-integral) values of $k$, there are extra cancellations of the zeros and poles in the factor $\prod_{j=1}^{N_1}\left[(q^{s_j+1})_M/\sin(\pi s_j)\right]$ since the zeros
\be
s_j \quad=\qquad k - M, k-(M-1), \dots, k-1\qquad{\rm mod}\quad k
\ee  
can coincide with some of the P poles in (\ref{Ppoles}). More precisely, the gap between $s_j = -M-1$ and $0$ repeats itself periodically modulo $k$ and thus the P poles for an integer k appear at
\be
s_j \quad=\qquad 0, 1, \dots, k-M-1\qquad{\rm mod}\quad k\ .
\label{Ppoleskint}
\ee

\medskip
Now more important is the fact that for a given $M < k$ as we decrease $k$ continuously, the NP poles on the positive real axis move to the left. For a sufficiently large $k> 2M$, the NP pole closest to the origin is at $s_j={k\over 2}-M>0$. As $k$ is decreased from $k> 2M$ to $k<2M$, this pole crosses the imaginary axis to the left. As we decrease $k$ further, more NP poles cross the imaginary axis. 
For the partition function to be continuous in $k$, these poles should not cross the contour $C$ and therefore we need to shift the contour $C$ to the left so as to avoid the crossing of these NP poles that invade into the real negative region.
More precisely, the contour $C$ has to be placed between $s_j = -{k\over 2}-1\, (>-M-1)$ and ${k\over 2}-M$ when $(M\le)\, k < 2M$. 
This is a prescription that needs to be justified. In \cite{Awata:2012jb} it was checked that  Seiberg duality holds with this contour prescription, vindicating our integral representation as a nonperturbative completion. 

\paragraph{$\bullet $ The pole structure for $n> 0$ and $N_1\le N_2$ :}
It is straightforward to generalize the contour prescription to the case of Wilson loops. In this case, however, we need to discuss two cases $N_1\le N_2$ and $N_1> N_2$ separately. We start with the former that is simpler than the latter.
We are interested in the summand in (\ref{ABJSfunction1}). Again by multiplying the factor $\prod_{i=1}^{N_1}\left[-1/(2i\sin(\pi s_i))\right]$, this becomes the integrand with the replacement $C_i \to s_i+1$. For the P poles the relevant factor is the same as in the partition function, $\prod_{j=1}^{N_1}\left[(q^{s_j+1})_M/\sin(\pi s_j)\right]$. This yields the P poles for generic (non-integral) $k$ at
\be
s_j \quad = \qquad \dots, -M-2, -M-1;\qquad 0, 1, 2, \dots .
\label{Ppoles2}
\ee
A similar remark on the integer $k$ case applies to this case, and the gap between $s_j=-M-1$ and $0$ repeats itself periodically modulo $k$. This implies that the P poles for an integer $k$ appear at
\be
s_j \quad = \qquad 0, 1, \dots, k-M-1 \qquad{\rm mod}\quad k.
\label{Ppoles2kint}
\ee

\begin{figure}[h!]
\centering \includegraphics[height=1.8in]{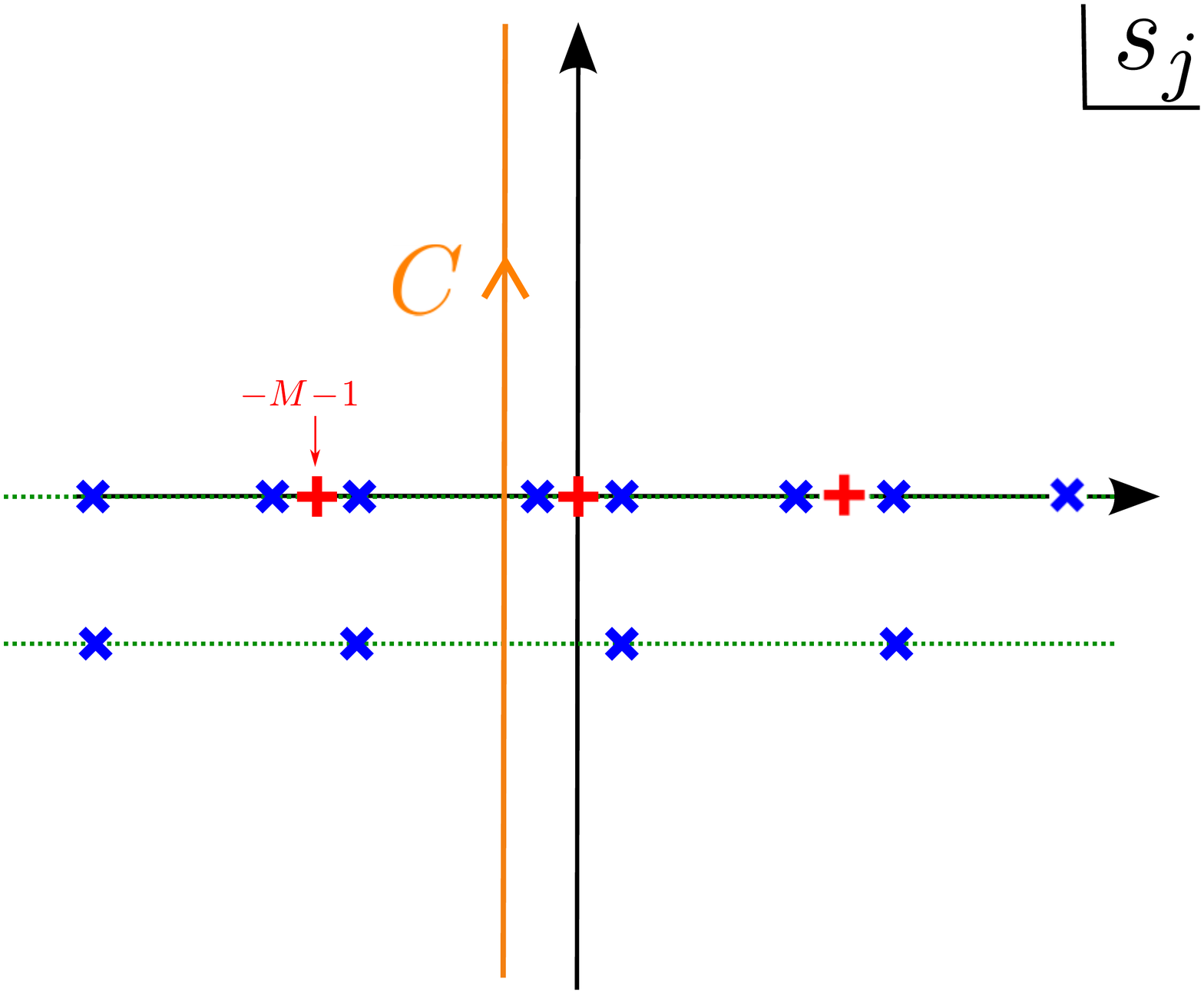} 
\hspace{1.5cm}
\includegraphics[height=1.8in]{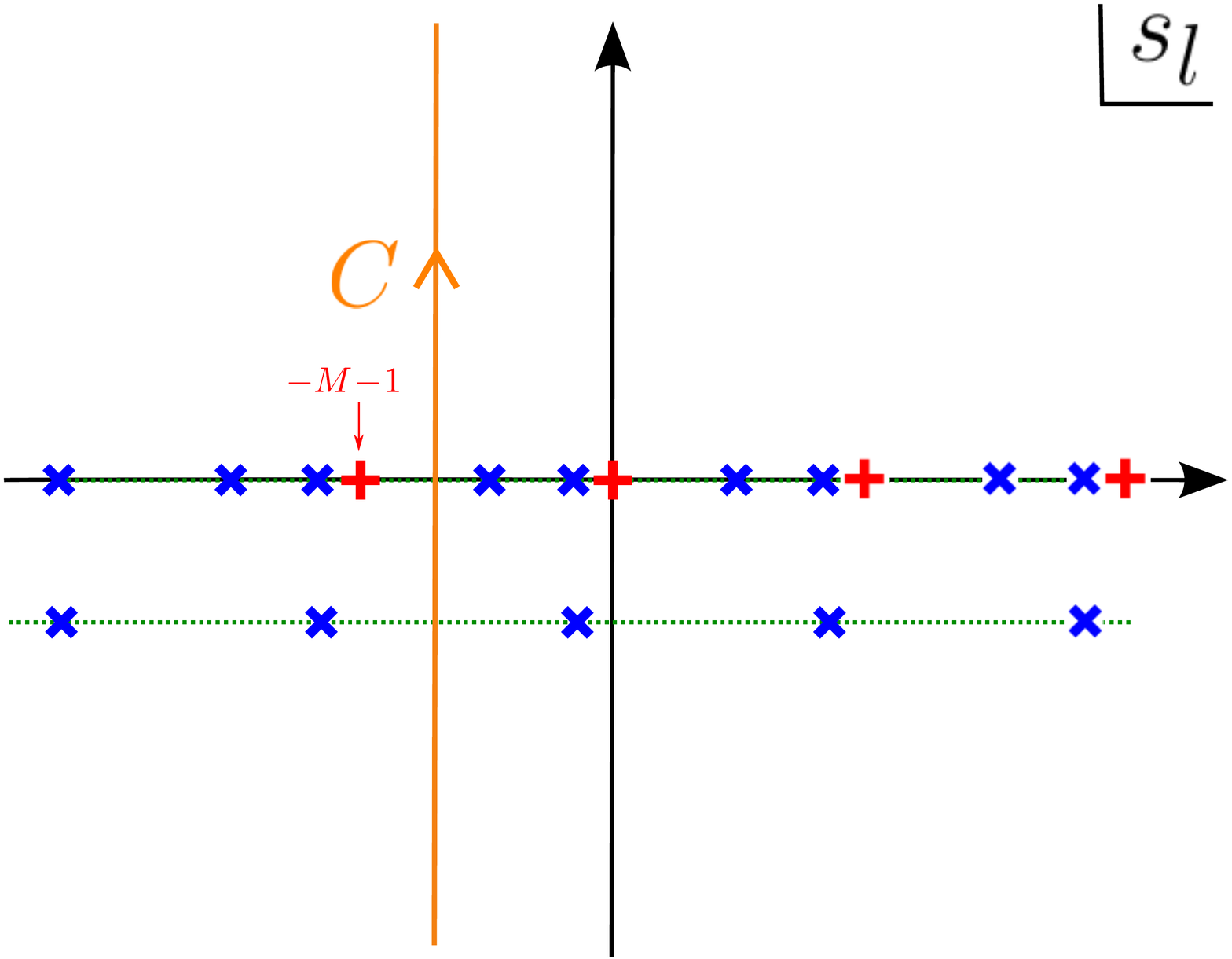} 
\caption{The
integration contour $C$ for Wilson loops with $N_1\le N_2$: Shown is the case, $k=3$, $M=2$ and $n=1$. (A) The left figure is for $s_j$ with $j\ne l$. The contour is placed between $s_j={\rm max}(-M-1,-{k\over 2}-1)$ and $s_j={\rm min}(0,{k\over 2}-M)$. (B) The right figure is for $s_j$ with $j=l$ and the contour is placed between $s_l={\rm max}(-M-1,-{k\over 2}-1-n)$ and $s_l={\rm min}(0,{k\over 2}-M-n)$.  The green dotted line corresponds to $s_j=s_i-{k\over 2}-n$ mod $k$ with some $s_i$. }  \label{contourFig2}
\end{figure}

\medskip
For the NP poles the relevant factors are $\prod_{j=1}^{N_1}1/(-q^{s_j+1+n\delta_{jl}})_M$ and $\prod_{j\ne i}1/(-q^{s_j-s_i+n\delta_{jl}})_1$ where $l$ runs from $1$ to $N_1$. The NP poles are thus at 
\begin{align}
 s_j \quad &= \qquad {k\over 2}-M-n\delta_{jl}, {k\over 2}-(M-1)-n\delta_{jl}, \dots, {k\over 2}-1-n\delta_{jl}\qquad{\rm mod}\quad k\label{NPpoles2kint}\\
 s_j \quad &= \qquad s_i +{k\over 2}-n\delta_{jl}\qquad{\rm mod}\quad k\ .
\end{align}
Note that the NP poles are simply shifted by $-n\delta_{jl}$ as compared to those for the partition function. Thus the pole structure differs from that of the partition function only for the integration variable $s_l$. 
As mentioned in the end of Section \ref{WilsonLoopResults}, the integral representation for the sum (\ref{ABJSfunction1}) is only well-defined for $|n|< {k\over 2}$. We thus restrict $n$ in this range. In Fig.~\ref{contourFig2} shown are both P and NP poles as well as the contour $C$ both for $j\ne l$ and $j=l$.
Similar to the partition function, as we decrease $k$, NP poles on the positive real axis move to the left. For $j= l$, in particular, when $k$ becomes smaller than $2(M+n)$ (for $n>0$), the NP pole closest to the P pole at the origin crosses the imaginary axis.
For Wilson loops to be continuous as a function of $k$, similar to the partition function, the contour $C$ has to be placed between $s_l ={\rm min}(-M-1,-{k\over 2}-1-n)$ and ${\rm max}(0,{k\over 2}-M-n)$ for $n>0$. Note that the bound $n<{k\over 2}$ ensures that $s_l=-M-1$ is left to $s_l={k\over 2}-M-n$.
For $j\ne l$ the contour is the same as that in the partition function.
The pole structure and the integration contour $C$ in this case are shown in Figure \ref{contourFig2}. 

\paragraph{$\bullet $ The pole structure for $n> 0$ and $N_1\ge N_2$ :}
In this case we are interested in the summands (\ref{S1}) and (\ref{S2}). Again by multiplying the factor $\prod_{a=1}^{N_2}\left[-1/(2i\sin(\pi s_a))\right]$, these summands become the integrands with the replacement $D_a \to s_a+1$.
We first discuss the pole structure of (\ref{S1}). This time the relevant factor for the P poles is different from the previous cases, 
$\prod_{a=1}^{N_2}\left[(q^{s_a+1})_M/\sin(\pi s_a)\times (q^{s_a+1+c-n})_1/(q^{s_a+1+c})_1\right]$ where $c$ runs from $0$ to $n-1$. This yields the P poles for generic $k$ at
\be
s_a \quad = \qquad \dots, -M-2, -M-1;\qquad -1-c; \qquad 0, \dots, -2-c+n;\qquad -c+n,\dots  
\label{Ppoles3}
\ee
Note that there is an additional pole at $s_a=-1-c$ in the gap between $s_a=-M-1$ and $0$ and a hole at $s_a = -1-c+n$. 
In the case of integer $k$, the gap, the additional point $s_a=-1-c$ and the hole $s_a=-1-c+n$ repeat themselves modulo $k$ and the P poles appear at
\begin{align}
 s_a \quad= \qquad-1-c;
 \qquad 0, 1,\dots, -2-c+n;\qquad -c+n,\dots,k-M-1\qquad{\rm mod}\quad k\ .
 \label{Ppoles3kint}
\end{align}
To be more precise, if $c$ is small enough and the hole at $s_a=-1-c+n$ falls into a gap, the hole is absent.

\begin{figure}[h!]
\centering \includegraphics[height=1.8in]{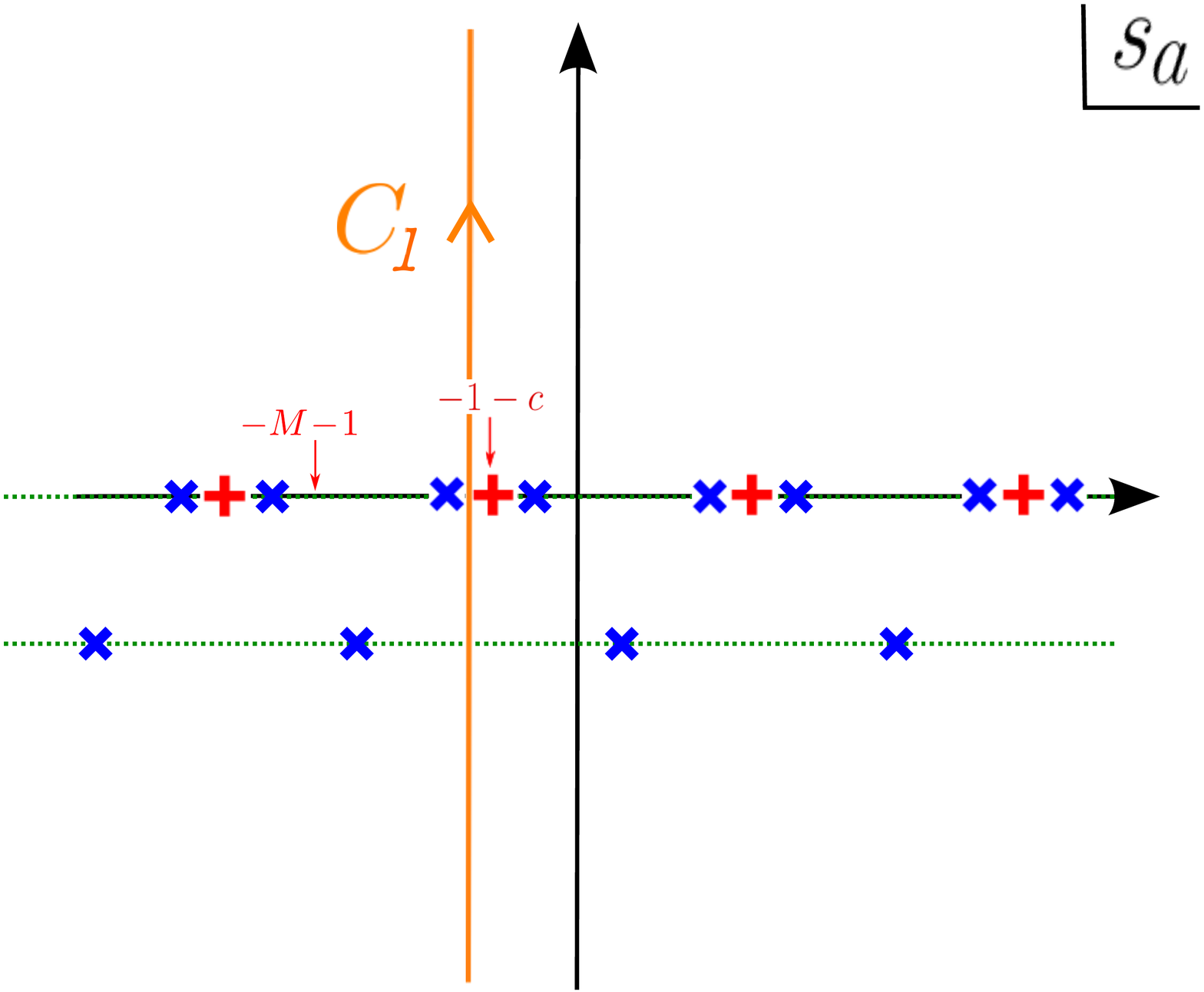} 
\caption{The integration contour $C_1[c]$ for Wilson loops with $N_1\ge N_2$:  The figure is for the integrals (\ref{IntegralII1}) associated with the sum (\ref{S1}). The green dotted line corresponds to $s_a=s_b+{k\over 2}$ mod $k$ with some $s_b$. Shown is the case $k=3$, $M=2$, $n=1$ and $c=0$.}  \label{contourFig3}
\end{figure}

\medskip
For the NP poles the relevant factors are $\prod_{a=1}^{N_2}1/(-q^{s_a+1})_M\times(-q^{s_a+1+c})_1/(-q^{s_a+1+c-n})_1$ and $\prod_{a\ne b}1/(-q^{s_a-s_b})_1$. The NP poles are thus at 
\begin{align}
 s_a \quad&=\qquad {k\over 2}-M, \dots, {k\over 2}-2-c;\qquad {k\over 2}-c,\dots, {k\over 2}-1; 
 \qquad {k\over 2}+(n-1)-c\qquad{\rm mod}\quad k\\
 s_a \quad&=\qquad s_b +{k\over 2}\qquad{\rm mod}\quad k\ .
\end{align}
The choice of integration contour $C_1$ is more involved than the previous cases: (1) In addition to the P poles at $s_a=0,1,\dots$ for a large $k$ (i.e., in the perturbative regime), we need to include, for a given $c$, the P pole at $s_a = -1-c$.\footnote{As we will show below, the contribution from this pole is canceled by a similar contribution in $I^{(2)}(N_1,N_2;n)_k$.} This means that we place the integration contour \emph{for a given} $c$ to the left of the P pole at $s_a=-1-c$. (2) As in the previous cases, we require the continuity with respect to $k$. The NP pole at $s_a={k\over 2}-M$ invades into the negative real axis, as $k$ is decreased, whereas the NP pole at $s_a=-{k\over 2}+(n-1)-c$ moves to the right. Hence the contour has to be placed to the left of  $s_a={\rm min}(-1-c, {k\over 2}-M)$ and the right of $s_a={\rm max}(-M-1, -{k\over 2}+(n-1)-c)$ for $c<M$ and to the left of $s_a=-1-c$ and the right of $s_a= -{k\over 2}+(n-1)-c$ for $c\ge M$.\footnote{Since $k\ge M$ and $n-1\ge c$, the NP pole at $s_a=-{k\over 2}+(n-1)-c$ is always left to the NP pole at $s_a={k\over 2}-M$. Similarly, the P pole at $s_a=-1-c$ is always right to the NP poles at $s_a=-{k\over 2}+(n-1)-c$ since $n<{k\over 2}$.}
The pole structure and the contour $C_1$ are illustrated in Figure \ref{contourFig3}. Note that the $C_1$ depends on $c$ and may thus be denoted as $C_1=\{C_1[c]\}$ ($c=0,\dots, n-1)$.

Next we turn to the pole structure of (\ref{S2}). For the P pole the relevant factor is 
$\prod_{a=1}^{N_2}\left[(q^{s_a+1+n\delta_{ad}})_M/\sin(\pi s_a)\right]$ where $d$ runs from $1$ to $N_2$. This yields for generic $k$ the P poles at 
\be
s_a \quad=\qquad \dots, -M-2-n\delta_{ad}, -M-1-n\delta_{ad};
 \qquad -n\delta_{ad}, 1-n\delta_{ad}, \dots
\ .
\ee
Note that the P poles are shifted by $-n\delta_{ad}$ as compared to those in the $N_1<N_2$ case. For an integer $k$ the gap between $s_a=-M-1-n\delta_{ad}$ and $-n\delta_{ad}$ repeats itself periodically modulo $k$ and the P poles appear at
\be
s_a \quad =\qquad -n\delta_{ad}, 1-n\delta_{ad}, \dots, 
k-M-1-n\delta_{ad}
\qquad{\rm mod}\quad k\ .
\ee
It could happen that if the winding $n$ is sufficiently large, $k-M-1-n$ becomes negative.

\begin{figure}[h!]
\centering 
\includegraphics[height=1.8in]{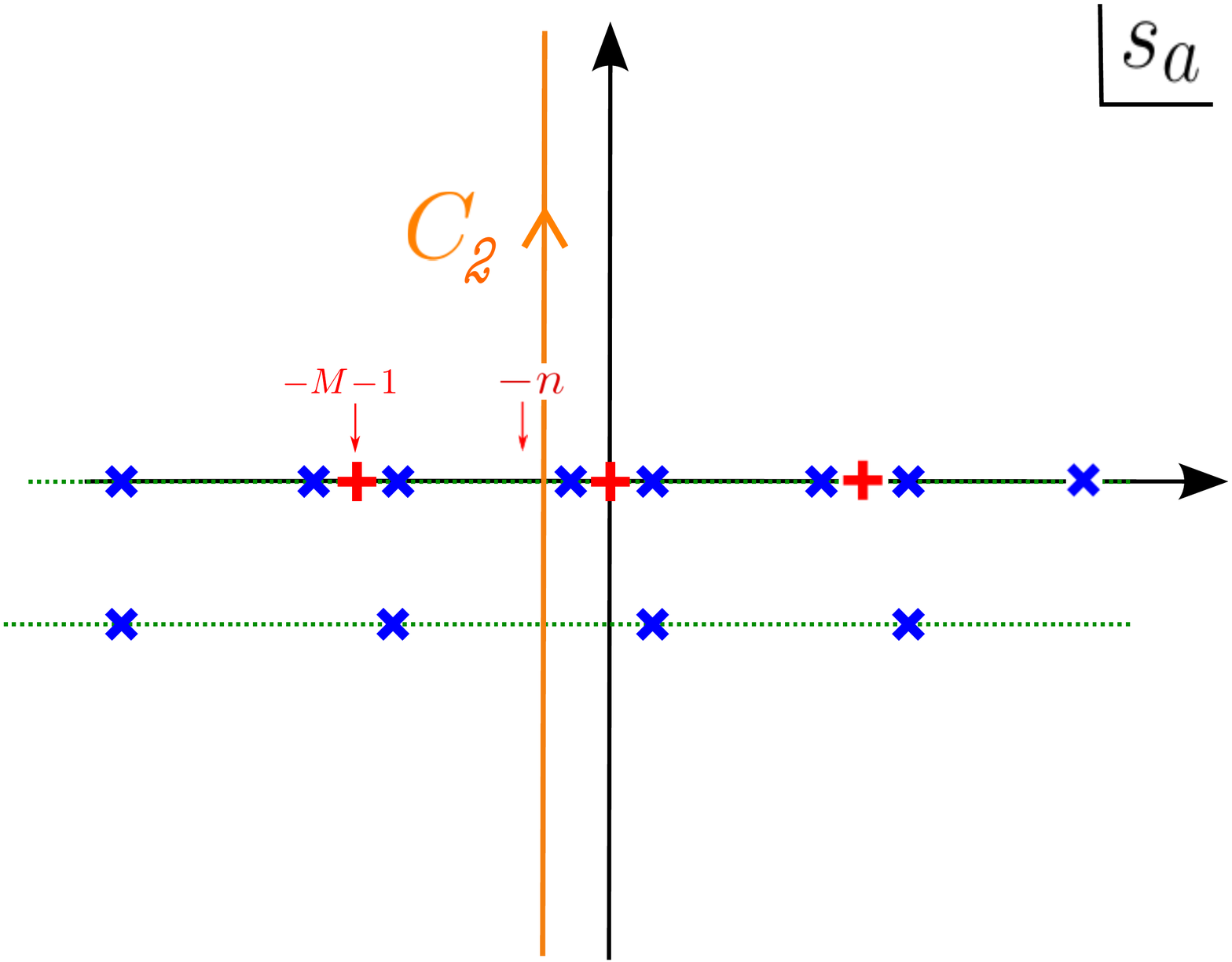} 
\hspace{1.5cm}
\includegraphics[height=1.8in]{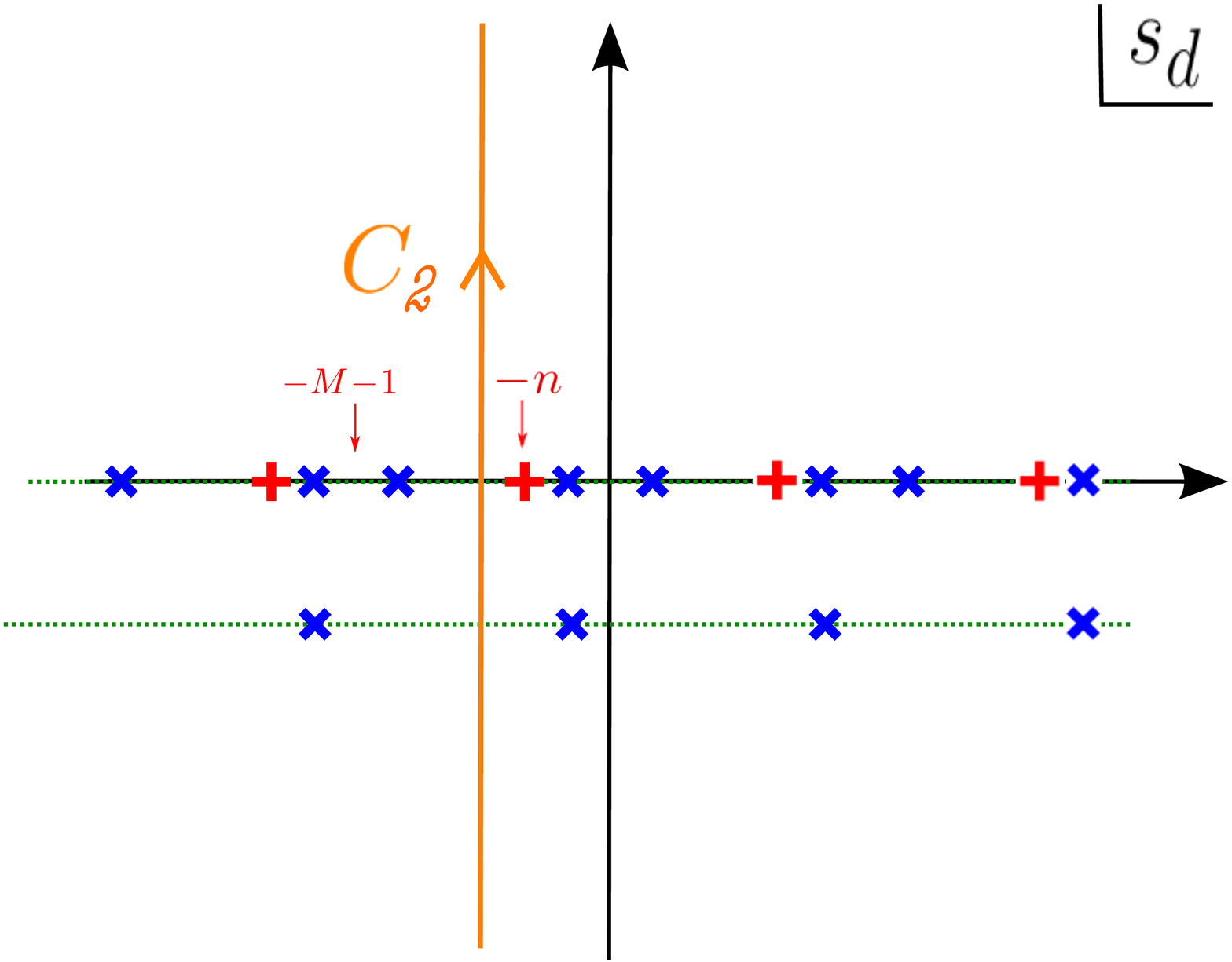} 
\caption{The
integration contour $C_2$ for Wilson loops with $N_1\ge N_2$:  (A) The left figure is the pole structure and the contour for $a\ne d$. (B) The right figure is for $a= d$ and the contour is shifted by $-n$ as compared to that in (A).  The green dotted line corresponds to $s_a=s_b-{k\over 2}-n\delta_{ad}$ mod $k$ with some $s_b$. The contours are the same. Shown is the case $k=3$, $M=2$ and $n=1$.}  \label{contourFig4}
\end{figure}

\medskip
For the NP poles the relevant factors are $\prod_{a=1}^{N_2}1/(-q^{s_a+1})_M$ and $\prod_{a\ne b}1/(-q^{s_a-s_b+n\delta_{ad}})_1$. The NP poles are thus at 
\begin{align}
 s_a \quad &= \qquad {k\over 2}-M, {k\over 2}-(M-1), \dots, {k\over 2}-1\qquad{\rm mod}\quad k\\
 s_a \quad &= \qquad s_b +{k\over 2}-n\delta_{ad}\qquad{\rm mod}\quad k\ .
\end{align}
The choice of integration contour $C_2$ is similar to the $N_1\le N_2$ case except that the contour for the variable $s_d$ is to the left of $s_d={\rm min}(-n, {k\over 2}-M)$ and the right of $s_d ={\rm max}(-M-1, -{k\over 2}-1)$ and picks up, in particular, the residues from the P poles at $s_d=-1,\dots, -n$. 
The pole structure and the contour $C_2$ are illustrated in Figure \ref{contourFig4}.

Again this is a prescription that lacks a first principle derivation. In the case of $N_1\le N_2$, similar to the partition function, for a large $k$, enclosing the contour with an infinite semi-circle to the right in the complex $s_j$ plane, the residue integral reduces to the sum (\ref{ABJSfunction1}), with the generalized $\zeta$-function regularization (\ref{genZeta}) implemented automatically, and the perturbative expansion in $g_s$ is correctly reproduced. In the case of $N_1\ge N_2$, however, the way this prescription works is more subtle even for a large $k$. Each of the two integral expressions (\ref{IntegralII1}) and (\ref{IntegralII2}) picks up \emph{extra} perturbative contributions from the poles at $s=-1, -2,\dots, -n$.
For ${1\over 6}$-BPS Wilson loops these extra contributions cancel out in the sum of (\ref{IntegralII1}) and (\ref{IntegralII2}), as shown in Appendix \ref{Cancellation}, thereby reproducing the correct perturbative expansion in $g_s$.\footnote{The perturbative equivalence of ${1\over 6}$-BPS Wilson loops of the lens space and ABJ matrix models, via the analytic continuation, has been established and checked by direct perturbative calculations.} 
As the nonperturbative test, we show in the next section that  Seiberg duality holds with our prescription, where it becomes clear that the inclusion of the P poles at $s=-1, -2,\dots, -n$ is necessary.


\subsection{Remarks on the ABJM limit}
\label{ABJMlimit}

As noted in Section \ref{WilsonLoopResults}, there are subtleties in taking the ABJM limit $M\to 0$, in particular, in the formula (\ref{IntegralII1}) for the case $N_1\ge N_2$.
There are two points to be addressed; (1) the agreement of the two formulae (\ref{1/6WilsonI1}) and (\ref{1/6WilsonI2}), and (2) the ${1\over 2}$-BPS Wilson loop (\ref{1/2Wilson}) in the ABJM limit.

\medskip
To address the first point, notice that the integrands of (\ref{IntegralI}) and (\ref{IntegralII2}) in the limit $M\to 0$ become identical. However, as remarked before, the contours $C$ and $C_2$ are different. Now since the factors $\prod_{i=1}^{N_1}[1/(-q^{s_i+1+n\delta_{il}})_M]$ in (\ref{IntegralI}) and $\prod_{a=1}^{N_2}[1/(-q^{s_a+1})_M]$ in (\ref{IntegralII2}) are absent, there are no NP poles on the real axis. Hence the difference due to the contours $C$ and $C_2$ only comes from the residues at the P poles $s_a=-1,-2,\dots, -n$ in (\ref{IntegralII2}). Therefore, in order for (\ref{1/6WilsonI1}) and (\ref{1/6WilsonI2}) to agree, these residues have to be canceled by (\ref{IntegralII1}). To see it, we carefully take the $M\to 0$ limit of (\ref{IntegralII1}):
\begin{align}
I^{(1)}(N_1,N_2;n)_k&={1\over N_2!}\sum_{c=0}^{n-1}
\prod_{a=1}^{N_2}\left[{-1\over 2\pi i}\int_{C_1\![c]}{\pi ds_a\over\sin(\pi s_a)}\right]
q^{n(2c-M)}{\left(q^{1-n}\right)_c\left(q^{1+n}\right)_{M-1-c}\over (q)_c(q)_{M-1-c}}
\prod_{a=1}^{N_2}{\left(q^{s_a+1}\right)_M\over 2\left(-q^{s_a+1}\right)_{M}}\nn\\
&\times \prod_{a=1}^{N_2}\left[{\left(-q^{s_a+1+c}\right)_1\left(q^{s_a+1+c-n}\right)_1
\over \left(q^{s_a+1+c}\right)_1\left(-q^{s_a+1+c-n}\right)_1}\prod_{b=1}^{a-1}{\left(q^{s_a-s_b}\right)_{1}
\over\left(-q^{s_a-s_b}\right)_{1}}
\prod_{b=a+1}^{N_2}{\left(q^{s_b-s_a}\right)_{1}\over\left(-q^{s_b-s_a}\right)_{1}}\right]\ ,
\label{IntegralII1v2}
\end{align}
In particular, we focus on the factors near the P pole for a selected variable $s_d$ at $s_d=-1-c$ as $M\to 0_+$,
\bea
\lim_{s_d\to -1-c}{(q^{1-n})_c(q^{1+n})_{M-1-c}\over (q)_c(q)_{M-1-c}}{(q^{s_d+1})_M\over (q^{s_d+1+c})_1}
&=&{q^{-nc}(q^{n-c})_M\over (q^n)_1(q^{-c})_c(q)_{M-1-c}}\lim_{s_d\to -1-c}{(q^{s_d+1})_M\over  (q^{s_d+1+c})_1}\nn\\
&=&{q^{-nc}(q^{n-c})_M\over (q^n)_1}
\stackrel{M\to 0_+}{\longrightarrow}{q^{-nc}\over 1-q^n}
\label{limitresidue}
\eea
where we used $(q^{1-n})_c(q^{1+n})_{M-1-c}=(-1)^cq^{-nc+\half c(c+1)}(q^{n-c})_M/(q^n)_1$ and $\lim_{\epsilon\to 0}(q^{\epsilon-c})_M/(q^{\epsilon})_1=(q^{-c})_c(q)_{M-1-c}$. 
Note that in the $M\to 0_+$ limit these factors vanish away from the P pole $s_d=-1-c$, i.e., if the limit $s_d\to -1-c$ is not taken in the first line. Therefore, in the ABJM limit the only contribution comes from the residues at the P poles $s_d=-1-c$ where $c=0,\dots,n-1$. 

\medskip
An important remark is in order: When $c< M$, the pole at $s_d=-1-c$ is clearly a simple pole, since the factor $(q^{s_d+1+c})_1$ that appear in the first line of \eqref{limitresidue} is canceled by the same factor in the $q$-Pochhammer symbol $(q^{s_d+1})_M$. However, when $c\ge M$, it is subtle, because there may not seem to be no apparent cancellation of these factors. Nevertheless, we treat the pole at $s_d=-1-c$ as a simple pole. As it turns out, the proper way to deal with this subtlety is to adopt an $\epsilon$-prescription for the parameter $M$. Namely, $M$ is always kept off an integral value by the shift $M\to M+\epsilon$ with $\epsilon>0$. The factor $(q^{s_d+1})_M$ is always assumed to be $(q^{s_d+1})_{M+\epsilon}$ with a non-integral index in our calculations and is defined by (\ref{maux20Sep12}) for a non-integral $M+\epsilon$. With this prescription, the factor $(q^{s_d+1+c})_1$ is always canceled by the same factor in $(q^{s_d+1})_{M+\epsilon}$, making the pole at $s_d=-1-c$ a simple pole.
However, there is one more subtlety in this prescription to be clarified. Namely, there appears a pole at $s_d=-1-c-\epsilon$ even with this prescription when $c\ge M$. If this pole were included within the contour, the contribution \eqref{limitresidue} would have been canceled. In other words, it would have been the same as treating the pole at $s_d=-1-c$ as a double pole. Thus our $\epsilon$-prescription involves a particular choice of the contour $C_1[c]$, i.e., to place it between $s_d=-1-c$ and $-1-c-\epsilon$ so as to avoid the latter pole.
This is a very subtle point and so much of a detail but is absolutely necessary for getting sensible results.

\medskip
Without loss of generality, we can choose the index $d$ to be $1$, since the expression is invariant under permutations of $s_a$'s. This yields
\begin{align}
I^{(1)}(N,N;n)_k&={1\over 2^{N-1}(N-1)!}\sum_{c=0}^{n-1}
{(-q^n)^{c}\over 1+q^{n}}
\prod_{a=2}^{N}\left[{-1\over 2\pi i}\int_{C_1\![c]}{\pi ds_a\over\sin(\pi s_a)}\right]
{\left(q^{s_a+1+c}\right)_1\left(q^{s_a+1+c-n}\right)_1
\over \left(-q^{s_a+1+c}\right)_1\left(-q^{s_a+1+c-n}\right)_1}\nn\\
&\quad\times 
\prod_{b=2}^{a-1}
{\left(q^{s_a-s_b}\right)_{1}\over\left(-q^{s_a-s_b}\right)_{1}}
\prod_{b=a+1}^{N}
{\left(q^{s_b-s_a}\right)_{1}\over\left(-q^{s_b-s_a}\right)_{1}}\ .
\label{IntegralII1ABJM}
\end{align}
As shown in Appendix \ref{Cancellation}, this is exactly canceled out by the sum of residues in $I^{(2)}(N,N;n)_k$ at the P poles $s_1=-1,\dots,-n$.
Hence the formula (\ref{1/6WilsonI2}) in the ABJM limit reduces to $I^{(2)}(N,N;n)_k$ with the contour $C_2$ being replaced by the contour $C$. This proves the agreement of  (\ref{1/6WilsonI1}) and (\ref{1/6WilsonI2}) in the ABJM limit.
We also note that the formula (\ref{IntegralII1ABJM}), when multiplied by $q^{-\half n^2+n}$,  yields the ${1\over 2}$-BPS Wilson loop (\ref{1/2Wilson}) in the ABJM limit (up to a normalization). We now discuss more on the ${1\over 2}$-BPS Wilson loop in the next subsection.

\subsection{The ${1\over 2}$-BPS Wilson loop}
\label{1/2BPSproof}

The ${1\over 2}$-BPS Wilson loop is given by (\ref{1/2Wilson}) that follows from the equality 
\be
q^{-\half n^2+n}I^{(2)}(N_1,N_2;n)_k=(-1)^nq^{\half n^2-n}I(N_2,N_1;n)_{-k}\ ,
\label{II2relation}
\ee
where $N_1\ge N_2$.
In order to show this identity, we recall that 
\begin{align}
I^{(2)}(N_1,N_2;n)_k &={1\over N_2!}\sum_{d=1}^{N_2}
\prod_{a=1}^{N_2}\left[{-1\over 2\pi i}\int_{C_2}{\pi ds_a\over\sin(\pi s_a)}\right]
q^{-ns_d+n(d-M-2)}
\prod_{\substack{a=1 \\ a\ne d}}^{N_2}{\left(q^{s_a-s_{d}-n}\right)_1\over \left(q^{s_a-s_{d}}\right)_1} \label{IntegralII2v2}\\
&\quad\times\prod_{a=1}^{N_2}\left[{\left(q^{s_a+1+n\delta_{ad}}\right)_M
\over\left(1+q^{n\delta_{ad}}\right)\left(-q^{s_a+1}\right)_{M}}
\prod_{b=1}^{a-1}{\left(q^{s_a-s_b}\right)_{1}
\over\left(-q^{s_a-s_b+n\delta_{ad}}\right)_{1}}
\prod_{b=a+1}^{N_2}{\left(q^{s_b-s_a}\right)_{1}\over\left(-q^{s_b-s_a-n\delta_{ad}}\right)_{1}}\right]\nn
\end{align}
and
\begin{align}
&I(N_2,N_1;n)_{-k} ={1\over N_2!}\sum_{d=1}^{N_2}\prod_{a=1}^{N_2}\left[{-1\over 2\pi i}\int_C{\pi ds_a\over\sin(\pi s_a)}\right]
q^{ns_d-n(d-2)}\prod_{\substack{a=1 \\ a\ne d}}^{N_2}
{\left(q^{-s_a+s_d+n}\right)_1\over \left(q^{-s_a+s_d}\right)_1}
\label{IntegralIconjugate}\\
&\quad
 \times\prod_{a=1}^{N_2}\left[{(-1)^{M}\left(q^{-s_a-1};q^{-1}\right)_M
\over\left(1+q^{-n\delta_{ad}}\right)\left(-q^{-s_a-1-n\delta_{ad}};q^{-1}\right)_{M}}
\prod_{b=1}^{a-1}{\left(q^{-s_a+s_b}\right)_{1}\over\left(-q^{-s_a+s_b-n\delta_{ad}}\right)_{1}}
\prod_{b=a+1}^{N_2}{\left(q^{-s_b+s_a}\right)_{1}\over\left(-q^{-s_b+s_a+n\delta_{ad}}\right)_{1}}\right]\ ,\nn
\nn
\end{align}
where $M:=|N_2-N_1|=N_1-N_2$. In fact, it is straightforward to check that these two are related, precisely as in (\ref{II2relation}), by the change of variables, 
\be
t_a= -s_a-1-M-n\delta_{ad}\ ,
\label{ttos}
\ee
for a given $d$, where $s_a$'s are the variables in the latter (\ref{IntegralIconjugate}) and $t_a$'s be identified with those in the former (\ref{IntegralII2v2}). 
The contour $C$ is placed in the intervals, $-{k\over 2}-1 < s_a < {\rm min}(0,{k\over 2}-M)$ for $a\ne d$ and ${\rm max}(-M-1, -{k\over 2}-1-n) < s_d < {\rm min}(0,{k\over 2}-M-n)$. By the above change of variables, this becomes precisely the contour $C_2$, where ${\rm max}(-M-1, -{k\over 2}-1) < t_a < {k\over 2}-M$ for $a\ne d$ and ${\rm max}(-M-1-n, -{k\over 2}-1) < t_d <{\rm min}(-n, {k\over 2}-M )$.\footnote{The $-1$ in (\ref{ttos}) compensates the orientation flip of the contour.} 

\medskip
This proves that the ${1\over 2}$-BPS Wilson loop is given by $q^{-{n^2\over 2}+n}I^{(1)}(N_1,N_2;n)_k$ up to the normalization.


\section{Seiberg duality -- derivations and a proof}
\label{SDsection}

There is a duality between two ABJ theories \cite{Aharony:2008gk}. Schematically, when $N_2 > N_1$, the following ABJ theories are equivalent:
\be
U(N_1)_k\times U(N_2)_{-k}\quad = \quad U(2N_1-N_2+k)_k\times U(N_1)_{-k}\ .
\label{SD2}
\ee
The partition functions of the two theories agree up to a phase \cite{Kapustin:2010mh, Willett:2011gp, Closset:2012vp, Awata:2012jb}. It was further understood in \cite{Awata:2012jb} how the perturbative and nonperturbative contributions to the partition function are exchanged under the duality map. 

\medskip
The Wilson loops, in contrast, are not invariant under the duality. The mapping rule for $\half$-BPS Wilson loops in general representations in ${\cal N}=2$ CSM theories with a simple gauge group has been studied by Kapustin and Willett \cite{Kapustin:2013hpk}. These Wilson loops correspond to ${1\over 6}$-BPS Wilson loops in the ABJ theory. Our results are consistent with their rule and slightly generalize it to the case where the flavor group is gauged. Similar to the case of the partition function \cite{Awata:2012jb}, our formulae for the Wilson loops allow us to understand an important aspect of the duality, namely, how the perturbative and nonperturbative contributions are exchanged under the duality map. 

\medskip
In this section we provide a proof of the duality map by analyzing our expressions (\ref{1/6WilsonI1}), (\ref{1/6WilsonI2}) and (\ref{1/2Wilson}) for the Wilson loops. 
To this end, let us recall our results for the duality map of the Wilson loops, (\ref{1/6duality}), (\ref{flavorduality}), and (\ref{1/2duality}).
\paragraph{$\bullet $ The ${1\over 6}$-BPS Wilson loop duality :}
\begin{align}
W^{\rm II}_{1\over 6}(N_1,N_2; n)_k=-W^{\rm I}_{1\over 6}(\widetilde{N}_2,N_1; n)_k
-2(-1)^{n+1}W^{\rm II}_{1\over 6}(\widetilde{N}_2,N_1; n)_k\ ,\label{1/6duality2}
\end{align}
\paragraph{$\bullet $ The flavor Wilson loop duality :}
\begin{align}
W^{\rm I}_{1\over 6}(N_1,N_2; n)_k=W^{\rm II}_{1\over 6}(\widetilde{N}_2,N_1; n)_k\ ,
\label{flavorduality2}
\end{align}
\paragraph{$\bullet $ The ${1\over 2}$-BPS Wilson loop duality :} 
\be
W_{\half}(N_1,N_2; n)_k = (-1)^nW_{\half}(\widetilde{N}_2,N_1; n)_k\ ,
\label{1/2duality2}
\ee
where we denoted the rank of the dual gauge group by $\widetilde{N}_2=2N_1-N_2+k$.
These three relations are not independent, but one of them can be derived from the other two by using the relation between the ${1\over 2}$- and ${1\over 6}$-BPS Wilson loops
\be
W_{\half}(N_1,N_2; n)_k=W^{\rm I}_{1\over 6}(N_1,N_2; n)_{k}-(-1)^nW^{\rm II}_{1\over 6}(N_1,N_2; n)_{k}\ .\label{1/21/6relation}
\ee

\medskip
Before going into a rigorous technical proof, we provide a heuristic yet very useful and intuitive way to understand how the Wilson loops would be mapped. 

\subsection{The brane picture -- a heuristic derivation%
\protect\footnote{We thank Kazutoshi Ohta for explaining to us the brane
picture in the work \cite{Kitao:1998mf, Ohta:2012ev} and for very
useful discussions.}  } \label{branepicturesection}

In Refs.\ \cite{Aharony:2008ug, Aharony:2008gk}, the brane realization
of the ABJ(M) theory was proposed.  The brane content of this
configuration is given by the following:
\begin{align}
\begin{array}{lcl}
 \text{D3-brane}  &:&0126\\
 \text{NS5-brane}  &:&012345\\
 \text{$(1,k)$ 5-brane}&:&012
  \left[{3\atop 7}\right]_\theta
  \left[{4\atop 8}\right]_\theta
  \left[{5\atop 9}\right]_\theta
\end{array}\label{mbxq12Apr13}
\end{align}
Here, $x^6$ is periodically identified, and $\left[{3\atop
7}\right]_\theta$ means the direction on the 3-7 plane with an angle
$\theta$ with the $x^3$ axis, where $\tan\theta=k$.  For $U(N_1)_k\times
U(N_2)_{-k}$ theory with $N_2-N_1=M>0$, there are $N_1$ D3-branes
between an NS5-brane and a $(1,k)$ 5-brane, and $N_2=N_1+M$ D3-branes
between the $(1,k)$ and the NS5.  This system realizes 3D supersymmetric
field theory that lives in the 012 directions and flows in the IR to the
ABJ SCFT\@.  Seiberg duality corresponds to moving the NS5 and the
$(1,k)$ branes past each other and, during the process, $k$ D3-branes
are created by the Hanany-Witten effect \cite{Hanany:1996ie} while $M$
D3-branes are annihilated, in the end leaving D3-branes realizing the
dual theory \eqref{SD2}.

\begin{figure}[htbp]
\begin{quote}
\begin{center}
 \includegraphics[height=4cm]{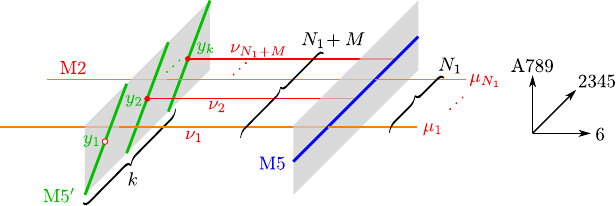} \caption{M-theory
 representation of ABJ theory with Wilson loops (presented is the case
 with $k=3,N_1=2,N_2=4,M=2$).  The left and right ends of the
 configuration are identified.  The position of the M2-branes in the
 $x^2$ direction corresponds to the Wilson loop.  Among $k$ places in
 which fractional M2-branes can end, the occupied ones are denoted by
 $\bullet$ and the unoccupied ones by $\circ$.  \label{fig:Mlift1}}
\end{center}
\end{quote}
\end{figure}

For our purpose, it is convenient to consider the following M-theory
lift of the configuration \eqref{mbxq12Apr13}, in which Wilson loops are
geometrically realized \cite{Kitao:1998mf, Ohta:2012ev}.  Assume that we
have non-trivial Wilson loop along e.g.\ $x^2$, namely $\int dx^2
A_2\neq 0$.\footnote{The relation between the Wilson loop for ABJ theory
on flat space and that for ABJ theory on $S^3$ is not clear.  This is
one of the reasons why the argument presented here is heuristic.}  If we
T-dualize the configuration \eqref{mbxq12Apr13} along $x^2$ and further
lift it to 11 dimensions, we obtain
\begin{align}
\begin{split}
 \begin{array}{lcl}
 \rm M2   &:& 016\\
 \rm M5   &:& 012345\\
 \rm M5'  &:& 01
  \left[{2\atop\rm A}\right]_\theta 
  \left[{3\atop    7}\right]_\theta
  \left[{4\atop    8}\right]_\theta
  \left[{5\atop    9}\right]_\theta
 \end{array}
\end{split}
\end{align}
where ``A'' denotes the 11th direction.  Note that the $(1,k)$ 5-brane
has lifted to an M5-brane (denoted by M5$'$) that is tilted in four
2-planes with the same angle $\theta$.
In Figure \ref{fig:Mlift1}, we
schematically described this configuration.
Because M5$'$-branes are tilted in the $x^2$-$x^{\rm A}$ plane, there
are only $k$ places in which ``fractional'' M2-branes can stretch
between M5$'$ and M5.  Note that only one fractional M2-brane can exist
in one place because of the s-rule \cite{Kitao:1998mf, Ohta:2012ev}.
$M=N_2-N_1$ fractional M2-branes are distributed among these $k$ places.
On the other hand, $N_1$ ``entire'' M2-branes are going around the $x^6$
direction and they do not have to sit in these places but can be
anywhere.

Because of the Wilson loop, different M2-branes are located at different
positions along the $x^2$ direction.  Let the $x^2$ coordinate of the
$N_1$ M2-branes between M5 and M5$'$ be $\mu_j$, $j=1,\dots,N_1$, and
that of the $N_2=N_1+M$ M2-branes (both fractional and entire) between
M5$'$ and M5 be $\nu_a$, $a=1,\dots,N_2$ (see Figure
\ref{fig:Mlift1}).\footnote{\label{ftnt:munu} Note that there is no
direct relation between the $\mu,\nu$ here and the ones that appear in
the ABJ matrix model \eqref{1/6wilson0}.  In the computation of Wilson
loops using localization \cite{Kapustin:2009kz}, at saddle points, gauge
fields (which correspond to $\mu,\nu$ here) vanish and Wilson loops get
contribution only from auxiliary fields (which correspond to $\mu,\nu$
in the ABJ matrix model). } Furthermore, let the $x^2$ coordinate of the
$k$ places in which fractional M2-branes can end be $y_\alpha$,
$\alpha=1,\dots,k$.  If the radius of the $x^2$ direction is $2\pi$, we
have $y_\alpha={2\pi \alpha\over k}+{\rm const}$ and, for $n\in\bbZ$,
\begin{align}
 \sum_{\alpha=1}^k e^{iny_\alpha}
 \propto \sum_{\alpha=1}^k e^{i 2\pi n\alpha/k}
 =\begin{cases}
   0 & (n\neq 0 \mod{k})\\
   k &(n=0 \mod{k})
  \end{cases}\label{eqo13Apr13}
\end{align}

\begin{figure}[htbp]
\begin{quote}
\begin{center}
\includegraphics[height=4cm]{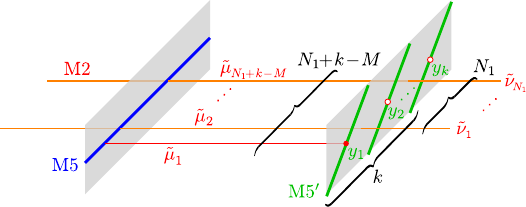}
\caption{The Seiberg dual configuration of the configuration in Figure \ref{fig:Mlift1}.
\label{fig:Mlift2}}
\end{center}
\end{quote}
\end{figure}

As mentioned above, Seiberg duality corresponds to moving M5 and M5$'$
past each other.  In this process, $M$ fractional M2-branes get
annihilated, and $k$ fractional M2-branes are created, leaving
$\Nt_2=N_1-M+k=2N_1-N_2+k$.  The resulting configuration is shown in Figure
\ref{fig:Mlift2}.
Let the $x^2$ coordinate of the $\Nt_2$ M2-branes (both fractional and
entire) between M5 and M5$'$ be $\mut_1,\dots,\mut_{\Nt_2}$, and that of the
$N$ M2-branes between M5$'$ and M5 be
$\nut_1\dots,\nut_{N_1}$ (see Figure \ref{fig:Mlift2}).

In the original configuration in Figure \ref{fig:Mlift1}, among the $k$
spots $\{1,\dots, k\}$ at which fractional M2-branes can end, let the
occupied ones be $O_1,\dots,O_M$ and the unoccupied ones be
$U_1,\dots,U_{k-M}$.  Of course, $\{O_1,\dots,O_M\}+
\{U_1,\dots,U_{k-M}\}=\{1,\dots,k\}$. Clearly,
\begin{align}
  \{\nu_1,\dots,\nu_{N_2}\}
 &=\{\mu_1,\dots,\mu_{N_1}\}+ \{y_{O_1}^{},\dots,y_{O_M}^{}\}.
\label{dsm28Apr14}
\end{align}
In the dual theory, the position of the
entire M2-branes are unchanged, while the occupied and unoccupied spots
for fractional M2-branes are interchanged.  Therefore,
\begin{align}
 \{\mut_1,\dots,\mut_{\Nt_2}\}
 &=\{\mu_1,\dots,\mu_{N_1}\}+ \{y_{U_1}^{},\dots,y_{U_{k-M}}^{}\}\notag\\
 &=\{\mu_1,\dots,\mu_{N_1}\}+ 
 \{y_{1},\dots,y_{k}\}-\{y_{O_1}^{},\dots,y_{O_{M}}^{}\},
 \label{fhes28Apr14}
\\
 \{\nut_1,\dots,\nut_{N_1}\}
 &=\{\mu_1,\dots,\mu_{N_1}\}.
\label{fhyt28Apr14}
\end{align}

\subsubsection{Fundamental representation}

Now we want to use this picture to give a very heuristic explanation of
Seiberg duality \eqref{SD2}.  Consider the original configuration in
Figure \ref{fig:Mlift1}.  The Wilson loop in the
fundamental representation simply measures the position of the
M2-branes.  Therefore, naively, we have\footnote{Note that this is very
rough and heuristic; thus ``$\sim$''.  In reality, $\nu_a$ is a variable to be
integrated over and is not localized at $y_\alpha^{}$. Even if there is
a sense in which they are localized at $y_\alpha^{}$, we should sum over
all possible ways to distribute $M$ fractional M2-branes over $k$
positions.}
\begin{align}
W^{\rm I }_{1\over 6}(N_1,N_2; n)
 &\sim  \sum_{j=1}^{N_1}e^{in\mu_j},\label{figa28Apr14}\\
W^{\rm II}_{1\over 6}(N_1,N_2; n)
 & \stackrel{?}{\sim} \sum_{a=1}^{N_2} e^{in\nu_a}
 =\sum_{j=1}^{N_1} e^{in\mu_j}+\sum_{\alpha=1}^M e^{iny_{O_\alpha}^{}},\label{cgj28Apr14}
 \end{align}
where in the second equation we used \eqref{dsm28Apr14}.
Actually, it turns out that, in order to reproduce the explicit results
obtained in the current paper, we must set $\nu_a\to \nu_a+\pi$ by hand  so that
\eqref{cgj28Apr14} is replaced by
\begin{align}
W^{\rm II}_{1\over 6}(N_1,N_2; n)
 & \sim \sum_{a=1}^{N_2} e^{in(\nu_a+\pi)}
 =(-1)^n\left(\sum_{j=1}^{N_1} e^{in\mu_j}+\sum_{\alpha=1}^M e^{iny_{O_\alpha}^{}}\right).\label{dhh28Apr14}
\end{align}
In the dual theory, using \eqref{fhes28Apr14} and \eqref{fhyt28Apr14}, we obtain
\begin{align}
  W^{\rm I }_{1\over 6}(\Nt_2,N_1; n)
 &\sim (-1)^n\sum_{j=1}^{\Nt_2} e^{in\mut_j}
= (-1)^n\left(\sum_{j=1}^{N_1} e^{in\mu_j}+
 \sum_{\alpha=1}^k e^{iny_\alpha}-\sum_{\alpha=1}^{M} e^{iny_{O_\alpha}^{}}\right)\notag\\
 &= (-1)^n\left(\sum_{j=1}^{N_1} e^{in\mu_j}-\sum_{\alpha=1}^{M} e^{iny_{O_\alpha}^{}}\right),
 \label{dor28Apr14}\\
 W^{\rm II}_{1\over 6}(\Nt_2,N_1; n)
 &
 \sim \sum_{a=1}^{N_1} e^{in\nut_a}
 =  \sum_{j=1}^{N_1} e^{in\mu_j},\label{dtv28Apr14}
\end{align}
where we introduced another ad hoc rule $\mut_j\to \mut_j+\pi $ just as
we did in \eqref{dhh28Apr14}. Also, in the second line of
\eqref{dor28Apr14}, we used \eqref{eqo13Apr13}, assuming that $n\neq 0$ mod $k$.  Therefore, comparing
\eqref{figa28Apr14}, \eqref{dhh28Apr14} and \eqref{dor28Apr14},
\eqref{dtv28Apr14}, we ``derived'' the following duality relations:
\begin{align}
 W^{\rm I}_{1\over 6}(N_1,N_2; n)
 &=W^{\rm II}_{1\over 6}(\Nt_2,N_1; n),\label{ftxs13May14}
 \\
 W^{\rm II}_{1\over 6}(N_1,N_2; n) +
 W^{\rm I}_{1\over 6}(\Nt_2,N_1; n)
 &=2(-1)^n W^{\rm II}_{1\over 6}(\Nt_2,N_1; n).\label{fubl13May14}
\end{align}
This means that the 1/2-BPS Wilson loop defined by
\begin{align}
 W_{1\over 2}(N_1,N_2;n)
 &:=
 W^{\rm I}_{1\over 6}(N_1,N_2; n)
 -(-1)^n W^{\rm II}_{1\over 6}(N_1,N_2; n)\label{fzg28Apr14}
\end{align}
is expected to have the following simple transformation rule:
\begin{align}
 W_{1\over 2}(N_1,N_2;n)
 =
 (-1)^n
 W_{1\over 2}(\Nt_2,N_1;n).\label{ghnp28Apr14}
\end{align}
Note that the above arguments are based on the identity
\eqref{eqo13Apr13} and are valid only for $n\ne 0$ mod $k$.  We do not
expect to get correct equations by setting $n=0$ in the above duality
relations, as we commented in subsection
\ref{WilsonLoopResults}.\footnote{Actually, equations
\eqref{ftxs13May14} and \eqref{ghnp28Apr14} still give correct equations
if we set $n= 0$, but \eqref{fubl13May14} does not.}

Although the ad hoc rule $\nu_a\to \nu_a + \pi$, $\mut_j\to \mut_j+\pi$
was crucial to reproduce the correct transformation rule for Wilson
loops, its physical meaning is unclear.  It is somewhat reminiscent of
the fact that, in the ABJ matrix model at large $N_1,N_2$
\cite{Drukker:2010nc}, the eigenvalue distribution for $U(N_2)$ is
offset relative to the $U(N_1)$ eigenvalue distribution on the complex
eigenvalue plane, but further investigations are left for future
research.  Since the arguments given in this subsection are meant to be
only heuristic, we simply accept the rule as a working assumption and
proceed.
In passing, we note that, with the above ad-hoc rule, the 1/2-BPS Wilson loop
\eqref{fzg28Apr14} can be understood simply as supertrace as follows:
\begin{align}
 W_{1\over 2}(N_1,N_2;n)
 &\sim
 \sum_j e^{in\mu_j} - \sum_a e^{in\nu_a}
 =\tr X^n-\tr Y^n=\str Z^n,\label{vim28Apr14}
\end{align}
where $X=\diag(e^{i\mu_j})$,  $Y=\diag(e^{i\nu_a})$, and $Z=\begin{pmatrix}X&\\&Y\end{pmatrix}$.

\subsubsection{More general representations}

The above heuristic method to guess the Seiberg duality relation can be
generalized to more general representations.\footnote{For a $U(N)$
representation with a Young diagram $\lambda$, the Wilson loop is
$S_\lambda(e^{i\mu_1},\dots,e^{i\mu_N})$, where
$S_\lambda(x_1,\dots,x_N)$ is the Schur polynomial
\cite{Drukker:2009hy}.  For example,
$S_{\Yboxdim1ex{\yng(1)}}=\sum_{i} x_i$,
$S_{\Yboxdim1ex{\yng(2)}}=\sum_{i\le j}x_ix_j$,
$S_{\Yboxdim1ex{\yng(1,1)}}=\sum_{i< j}x_ix_j$.}
For example, in the original $U(N_1)_k\times U(N_2)_{-k}$ theory,
consider the following 1/6-BPS Wilson loops:
\begin{align}
\begin{split}
  W_{\tiny\yng(2)~\bullet\,}
 &
 \sim \sum_{i\le j}^{N_1} e^{i(\mu_i+\mu_j)}
 =
 \half \left(\sum_{i=1}^{N_1} e^{i\mu_i}\right)^2
 +\half \sum_{i=1}^{N_1} e^{2i\mu_i}
 =: 
 \half (e^{i\mu})^2 + \half e^{2i\mu}
 \\
 W_{\tiny\yng(1,1)~\bullet\,}
 &
 \sim \sum_{i< j}^{N_1} e^{i(\mu_i+\mu_j)}
 =
 \half \left(\sum_{i=1}^{N_1} e^{i\mu_i}\right)^2
 -\half \sum_{i=1}^{N_1} e^{2i\mu_i}
 =: 
 \half (e^{i\mu})^2 - \half e^{2i\mu}
 \\
 W_{\tiny\bullet~\yng(2)}
 &
 \sim \sum_{a\le b}^{N_2} e^{i(\nu_a+\nu_b)}
 =
 \half \left(\sum_{a=1}^{N_2} e^{i\nu_a}\right)^2
 +\half \sum_{a=1}^{N_2} e^{2i\nu_a}
 \\
 &
 =
 \half \left(\sum_{i=1}^{N_1} e^{i\mu_i}+\sum_{\alpha=1}^M e^{iy_{O_\alpha}^{}}\right)^2
 +\half \left(\sum_{i=1}^{N_1} e^{2i\mu_i}+\sum_{\alpha=1}^M e^{2iy_{O_\alpha}^{}}\right)\\
 &=: 
 \half (e^{i\mu})^2
 +\half(e^{iy_O^{}})^2+e^{i\mu}e^{iy_O^{}}
 +\half e^{2i\mu}
 +\half e^{2iy_O^{}},
 \\
 W_{\tiny\bullet~\yng(1,1)}
 &
 \sim \sum_{a<b}^{N_2} e^{i(\nu_a+\nu_b)}
 =: 
 \half (e^{i\mu})^2
 +\half(e^{iy_O^{}})^2+e^{i\mu}e^{iy_O^{}}
 -\half e^{2i\mu}
 -\half e^{2iy_O^{}},
 \\
 W_{\tiny\yng(1)\,\yng(1)}
 &
 \sim -\sum_{i=1}^{N_1} e^{i\mu_i} \sum_{a=1}^{N_2}  e^{i\nu_a}
 =: 
 -(e^{i\mu})^2
 -e^{i\mu}e^{iy_O^{}},
\end{split}
\end{align}
where $W_{R\Rt}$ denotes the Wilson loop in the representations $R$ and
$\Rt$ for $U(N_1)$ and $U(N_2)$, respectively, and  ``$\bullet$'' means the
trivial representation.  Also, in the last expression of each line, we
used a schematic notation, whose meaning is defined by the immediately
preceding expression.  In the dual $U(\Nt_2)_k\times U(N_1)_{-k}$
theory, we have
\begin{align}
\begin{split}
  \Wt_{\tiny\yng(2)~\bullet}
 &
 \sim
 \half (e^{i\mu})^2
 +\half(e^{iy_O^{}})^2-e^{i\mu}e^{iy_O^{}}
 +\half e^{2i\mu}
 -\half e^{2iy_O^{}},
 \\
 \Wt_{\tiny~\yng(1,1)~\bullet}
 &
 \sim
 \half (e^{i\mu})^2
 +\half(e^{iy_O^{}})^2-e^{i\mu}e^{iy_O^{}}
 -\half e^{2i\mu}
 +\half e^{2iy_O^{}},
 \\
 \Wt_{\tiny\bullet~\yng(2)}
 &
 \sim 
 \half (e^{i\mu})^2
 +\half e^{2i\mu},\qquad
 \Wt_{\tiny\bullet~\yng(1,1)}
 \sim 
 \half (e^{i\mu})^2
 -\half e^{2i\mu},
 \\
 \Wt_{\tiny\yng(1)\,\yng(1)}
 &
 \sim
 -(e^{i\mu})^2
 +e^{i\mu}e^{iy_O^{}}.
\end{split}
\end{align}
The duality relation between $W$ and $\Wt$ is readily found to be
\begin{align}
 W=S\Wt,\qquad
 W=\left(\begin{array}{l}
    W_{\tiny\yng(2)~\bullet}\\
    W_{\tiny\yng(1,1)~\bullet}\\
    W_{\tiny~\bullet~\yng(2)}\\
    W_{\tiny~\bullet~\yng(1,1)}\\
    W_{\tiny\yng(1)\,\yng(1)}\\
   \end{array}\right),\qquad
S
=\left(
\begin{array}{ccccc}
 0 & 0 & 1 & 0 & 0 \\
 0 & 0 & 0 & 1 & 0 \\
 0 & 1 & 3 & 1 & 2 \\
 1 & 0 & 1 & 3 & 2 \\
 0 & 0 & -2 & -2 & -1 \\
\end{array}
\right).
\label{itjo29Apr13}
\end{align}

The combinations that have simple transformation rule are
\begin{align}
 W_{\tiny\yng(2)}^{1/2}
 &:= W_{\tiny\yng(2)~\bullet~} + W_{\tiny \bullet~\yng(1,1)} + W_{\tiny\yng(1)\,\yng(1)},\qquad
 W_{\tiny\yng(1,1)}^{1/2}
 := W_{\tiny\yng(1,1)~\bullet~} + W_{\tiny \bullet~\yng(2)} + W_{\tiny\yng(1)\,\yng(1)},\label{fyve28Apr14}
\end{align}
which transform as
\begin{align}
  W_{\tiny\yng(2)}^{1/2}=\Wt_{\tiny\yng(2)}^{1/2},\qquad
  W_{\tiny\yng(1,1)}^{1/2}=\Wt_{\tiny\yng(1,1)}^{1/2}.
\end{align}
These are precisely the 1/2-BPS Wilson
loops derived in \cite{Drukker:2009hy}.

Just as in \eqref{vim28Apr14}, we can write these in terms of supertrace
as follows:
\begin{align}
\begin{split}
  W_{\tiny\yng(2)}^{1/2}
 &=\half(\str Z)^2+\half\str (Z^2)
 =\str_{\tiny\yng(2)}^{}Z,\\
 W_{\tiny\yng(1,1)}^{1/2}
 &=\half(\str Z)^2-\half\str (Z^2)
 =\str_{\tiny\yng(1,1)}^{}Z.
\end{split}
\end{align}
Note that the right hand side is nothing but supertrace in the
respective representations.
More generally, the 1/2-BPS Wilson loop for general representation $R$ is
given by\footnote{Note that the 1/2-BPS Wilson loops derived in
\cite{Drukker:2009hy} are
\begin{align}
 W_R^{1/2}
 =\str_R^{}\begin{pmatrix}
	    e^{\mu_{\rm MM}}&\\ &-e^{\nu_{\rm MM}}
	   \end{pmatrix}
\end{align}
where $\mu_{\rm MM}^{},\nu_{\rm MM}^{}$ are the $\mu,\nu$ that appear in matrix
model (see footnote \ref{ftnt:munu}). Our ad hoc rule is to replace this with
\begin{align}
 W_R^{1/2}=
 \str_R^{}\begin{pmatrix}e^{i\mu}&\\ &e^{i\nu}\end{pmatrix}
 =\str_R^{} Z
\end{align}
where $\mu,\nu$ are the positions of the M2-branes.
}
\begin{align}
 W^{1/2}_R \sim \str_R^{} Z=P_R(\str (Z^n)),
\end{align}
where $P_R$ is a polynomial in $\str(Z^n)$ ($n=1,\dots,|R|$) obtained
from the Schur polynomial.  Each term in the polynomial contains a
product of $|R|$ $Z$'s.  The results from the previous section, more
specifically \eqref{ghnp28Apr14} and \eqref{vim28Apr14}, say that
$\str(Z^n)\to (-1)^n \str(Z^n)$ under duality.  Therefore, the
transformation law for the general 1/2-BPS Wilson loop is
\begin{align}
 W^{1/2}_R \to (-1)^{|R|}\, W^{1/2}_R.
\end{align}

\bigskip
None of the above is a derivation of Seiberg duality for 1/6-BPS
Wilson loops but is merely a motivation for it.  However, the fact that
it predicts a simple duality law for 1/2-BPS Wilson loops is
evidence that the 1/6-BPS duality relation is also correct.

\subsection{A rigorous derivation and proof}
\label{dualitygen}

Since the mapping (\ref{flavorduality2}) for the flavor Wilson loop and
(\ref{1/2duality2}) for the ${1\over 2}$-BPS Wilson loop are simple as
compared to the mapping (\ref{1/6duality2}) for ${1\over 6}$-BPS Wilson
loops, it is the best strategy to prove (\ref{flavorduality2}) and
(\ref{1/2duality2}) and then infer (\ref{1/6duality2}) from them.  In
due course, we will also see manifestly the exchange of perturbative and
nonperturbative contributions under the duality.

\paragraph{$\bullet $ The flavor Wilson loop duality :}
We first prove the flavor Wilson loop duality:
\begin{align}
W^{\rm I}_{1\over 6}(N_1,N_2; n)_k=W^{\rm II}_{1\over 6}(\widetilde{N}_2,N_1; n)_k
=W^{\rm I}_{1\over 6}(N_1,\widetilde{N}_2; n)_{-k}
\label{flavorduality3}
\end{align}
which amounts to the equality
\be
{q^{-{n^2\over 2}+n}I(N_1,N_2;n)_k\over I(N_1,N_2;0)_k}={q^{{n^2\over 2}-n}I(N_1,\widetilde{N}_2;n)_{-k}\over I(N_1,\widetilde{N}_2;0)_{-k}}\ .
\label{flavortobeproved}
\ee
The basic idea for the proof is to show that (1) the integrands in the numerators are identical, i.e., they share exactly the same zeros and poles and the same asymptotics up to the normalization, and (2) the contours are equivalent.  
The explicit forms of $I(N_1,N_2;n)_k$ and $I(N_1,\widetilde{N}_2;n)_{-k}$ are given by
\begin{align}
I(N_1,N_2;n)_k&={1\over N_1!}\sum_{l=1}^{N_1}\prod_{i=1}^{N_1}\left[{-1\over 2\pi i}\int_C{\pi ds_i\over\sin(\pi s_i)}\right]
q^{-ns_l+n(l-2)}\prod_{\substack{i=1 \\ i\ne l}}^{N_1}
{\left(q^{s_i-s_{l}-n}\right)_1\over \left(q^{s_i-s_{l}}\right)_1}
\label{IntegralIdual}\\
&\quad\times\prod_{i=1}^{N_1}\left[{(-1)^{M}\left(q^{s_i+1}\right)_M
\over\left(1+q^{n\delta_{il}}\right)\left(-q^{s_i+1+n\delta_{il}}\right)_{M}}\prod_{j=1}^{i-1}{\left(q^{s_i-s_j}\right)_{1}\over\left(-q^{s_i-s_j+n\delta_{il}}\right)_{1}}
\prod_{j=i+1}^{N_1}{\left(q^{s_j-s_i}\right)_{1}\over\left(-q^{s_j-s_i-n\delta_{il}}\right)_{1}}\right]
\nn
\end{align}
\begin{align}
I(N_1,\widetilde{N}_2;n)_{-k}&={1\over N_1!}\sum_{l=1}^{N_1}\prod_{i=1}^{N_1}\left[{-1\over 2\pi i}\int_{\widetilde{C}}{\pi ds_i\over\sin(\pi s_i)}\right]
q^{ns_l-n(l+M-3)}\prod_{\substack{i=1 \\ i\ne l}}^{N_1}
{\left(q^{-s_i+s_l+n}\right)_1\over \left(q^{-s_i+s_l}\right)_1}
\label{IntegralIconjugatedual}\\
&\quad\times\prod_{i=1}^{N_1}\left[{\left(q^{s_i+1}\right)_{k-M}
\over\left(1+q^{n\delta_{il}}\right)\left(-q^{s_i+1+n\delta_{il}}\right)_{k-M}}
\prod_{j=1}^{i-1}{\left(q^{-s_i+s_j}\right)_{1}\over\left(-q^{-s_i+s_j-n\delta_{il}}\right)_{1}}
\prod_{j=i+1}^{N_1}{\left(q^{-s_j+s_i}\right)_{1}\over\left(-q^{-s_j+s_i+n\delta_{il}}\right)_{1}}\right] ,\nn
\nn
\end{align}
where $M:=|N_2-N_1|=N_2-N_1$.
At first glance the zeros and poles of (\ref{IntegralIdual}) and (\ref{IntegralIconjugatedual}) differ in the $M$-dependence, since $M$ is replaced by $k-M$ in the latter. However, introducing the dual variables in the latter
\be
\widetilde{s}_i=-s_i+{k\over 2}-1-n\delta_{il}\ ,
\label{dualmap}
\ee
we can easily see that they actually agree. As simple as it may look, we stress that this is a very important map and can be regarded as the duality transformation, as we now justify.
In terms of the original variables $s_i$, the poles of the integrand in (\ref{IntegralIconjugatedual}) appear at
\bea
{\bf P:}\quad s_i &=& 0, 1,\dots, M-1\quad{\rm mod}\quad k\ ,\\
{\bf NP:}\quad
s_i &=& -{k\over 2}+M-n\delta_{il}, \dots, {k\over 2}-1-n\delta_{il}\quad{\rm mod}\quad k\ ,\\
{\bf (NP):}\quad
s_i &=& s_j +{k\over 2}-n\delta_{il}\quad{\rm mod}\quad k\ .
\eea
In terms of the dual variables $\widetilde{s}_i$, these poles are mapped to 
\bea
{\bf NP:}\quad
\widetilde{s}_i &=& {k\over 2}-M-n\delta_{il}, \dots, {k\over 2}-2-n\delta_{il}, {k\over 2}-1-n\delta_{il}\quad{\rm mod}\quad k\ ,\\
{\bf P:}\quad
\widetilde{s}_i &=& 0, 1,\dots, k-M-1 \quad{\rm mod}\quad k\ ,\\
{\bf (NP):}\quad
\widetilde{s}_i &=& \widetilde{s}_j +{k\over 2}-n\delta_{il}\quad{\rm mod}\quad k\ .
\eea
These are precisely the same as the poles of the integrand in (\ref{IntegralIdual}) and thus the two integrands share exactly the same poles. 
We emphasize that the P and NP poles are exchanged  by the duality transformation (\ref{dualmap}). 
In terms of the original variables, the contour $\widetilde{C}$ is placed in the intervals, ${\rm max}(M-k-1,-{k\over 2}-1)<s_i<{\rm min}(0,-{k\over 2}+M)$ for $i\ne l$ and ${\rm max}(-M-1,-{k\over 2}-1-n)<s_l<{\rm min}(0,-{k\over 2}+M-n)$. In terms of the dual variables, this becomes the intervals, 
${\rm max}(-{k\over 2}-1,-M-1)<\widetilde{s}_i<{\rm min}({k\over 2}-M,0)$ for $i\ne l$ and ${\rm max}(-{k\over 2}-1,-M-1-n)<s_l<{\rm min}({k\over 2}-M,-n)$. Hence, the contours $C$ and $\widetilde{C}$ are equivalent.\footnote{The $-1$ in (\ref{dualmap}) compensates the orientation flip of the contour.} 

\medskip
Meanwhile, the zeros appear only from the factors $(q^{s_i-s_j})_1$ and $(q^{s_i-s_l-n})_1$ and do not depend explicitly on $M$. It is easy to check that the zeros are invariant under the duality transformation (\ref{dualmap}).  Indeed, the factors that depend on the differences $s_i-s_j$ in (\ref{IntegralIdual}) and $\widetilde{s}_i-\widetilde{s}_j$ in (\ref{IntegralIconjugatedual}) only differ from each other by the factor $q^{2n(l-1)}$ multiplying the latter.

\medskip
It remains to find the asymptotics of the integrands. The asymptotics to be compared with are those at $s_i\to i\infty$ in (\ref{IntegralIdual}) and $\widetilde{s}_i\to i\infty$ in (\ref{IntegralIconjugatedual}), and for the factors that depend on the differences of the variables we only need to care about the factor $q^{2n(l-1)}$. Collecting various factors together and taking into account the orientations of the contours, it is straightforward to find that the latter asymptotics is $i^{k}q^{-n^2+2n}$ times the former. Since the factor $i^k$ is canceled by the same factor coming from the normalization, this completes the proof of the equality (\ref{flavortobeproved}).

\paragraph{$\bullet $ The ${1\over 2}$-BPS Wilson loop duality :}
We next prove the duality for the ${1\over 2}$-BPS Wilson loop:
\be
W_{\half}(N_1,N_2; n)_k = (-1)^nW_{\half}(\widetilde{N}_2,N_1; n)_k
\label{1/2duality3}
\ee
which from (\ref{1/2Wilson}) amounts to the equality
\be
{q^{{n^2\over 2}-n}I^{(1)}(N_2,N_1;n)_{-k}\over I^{(2)}(N_2,N_1;0)_{-k}}
=-{q^{-{n^2\over 2}+n}I^{(1)}(\widetilde{N}_2,N_1;n)_k\over I^{(2)}(\widetilde{N}_2,N_1;0)_k}\ .
\label{1/2BPStobeproved}
\ee
The explicit forms of $I^{(1)}(N_2,N_1;n)_{-k}$ and $I^{(1)}(\widetilde{N}_2,N_1;n)_{k}$ are given by
\begin{align}
&I^{(1)}(N_2,N_1;n)_{-k}={1\over N_1!}\sum_{c=0}^{n-1}
\prod_{i=1}^{N_1}\left[{-1\over 2\pi i}\int_{C_1\![c]}{\pi ds_i\over\sin(\pi s_i)}\right]
q^{n}{(q^{1-n})_c(q^{1+n})_{M-1-c}\over (q)_c(q)_{M-1-c}}
\prod_{i=1}^{N_1}{\left(q^{-s_i-1};q^{-1}\right)_M\over 2\left(-q^{-s_i-1};q^{-1}\right)_{M}}\nn\\
&\qquad\times \prod_{i=1}^{N_1}\left[{\left(-q^{-s_i-1-c}\right)_1\left(q^{-s_i-1-c-n}\right)_1
\over \left(q^{-s_i-1-c}\right)_1\left(-q^{-s_i-1-c-n}\right)_1}\prod_{j=1}^{i-1}{\left(q^{-s_i+s_j}\right)_{1}
\over\left(-q^{-s_i+s_j}\right)_{1}}
\prod_{j=i+1}^{N_1}{\left(q^{-s_j+s_i}\right)_{1}\over\left(-q^{-s_j+s_i}\right)_{1}}\right]\ ,
\label{IntegralII1v3}
\end{align}
\begin{align}
I^{(1)}(\widetilde{N}_2,N_1;n)_{k}&={1\over N_1!}\sum_{c=0}^{n-1}
\prod_{i=1}^{N_1}\left[{-1\over 2\pi i}\int_{\widetilde{C}_1\![c]}{\pi ds_i\over\sin(\pi s_i)}\right]
q^{n(2c+M)}{(q^{1-n})_c(q^{1+n})_{k-M-1-c}\over (q)_c(q)_{k-M-1-c}}
\prod_{i=1}^{N_1}{\left(q^{s_i+1}\right)_{k-M}\over 2\left(-q^{s_i+1}\right)_{k-M}}\nn\\
&\times \quad\prod_{i=1}^{N_1}\left[{\left(-q^{s_i+1+c}\right)_1\left(q^{s_i+1+c-n}\right)_1
\over \left(q^{s_i+1+c}\right)_1\left(-q^{s_i+1+c-n}\right)_1}\prod_{j=1}^{i-1}{\left(q^{s_i-s_j}\right)_{1}
\over\left(-q^{s_i-s_j}\right)_{1}}
\prod_{j=i+1}^{N_1}{\left(q^{s_j-s_i}\right)_{1}\over\left(-q^{s_j-s_i}\right)_{1}}\right]\ .
\label{IntegralII1v4}
\end{align}
Similar to the flavor Wilson loop duality, introducing the duality transformation in (\ref{IntegralII1v3})
\be
\widetilde{s}_i=-s_i-{k\over 2}-1\ ,
\label{dualmap2}
\ee
it becomes evident that the two integrands share the same zeros and poles.
In terms of the original variables $s_i$, the poles of the integrand in (\ref{IntegralII1v3}) appear modulo $k$ at
\bea
{\bf P:}\quad s_i &=& -1-c;\quad 0,\dots, -2-c+n;\quad -c+n,\dots, k-M-1\ ,\\
{\bf NP:}\quad
s_i &=& {k\over 2}-M, \dots, {k\over 2}-2-c;\quad {k\over 2}-c,\dots,{k\over 2}-1;\quad {k\over 2}+n-1-c\ ,\\
{\bf (NP):}\quad
s_i &=& s_j +{k\over 2}\ .
\eea
In terms of the dual variables $\widetilde{s}_i$, these poles are mapped, after shifting by $+k$, to 
\bea
{\bf NP:}\quad
\widetilde{s}_i &=& -{k\over 2}+M, \dots, {k\over 2}-2-\widetilde{c};\quad {k\over 2}-\widetilde{c},\dots,{k\over 2}-1;\quad {k\over 2}+n-1-\widetilde{c}\ ,\\
{\bf P:}\quad
\widetilde{s}_i &=& -1-\widetilde{c};\quad 0,\dots, -2-\widetilde{c}+n;\quad -\widetilde{c}+n,\dots, M-1\ ,\\
{\bf (NP):}\quad
\widetilde{s}_i &=& \widetilde{s}_j +{k\over 2}\ ,
\eea
where $\widetilde{c}=(n-1)-c$.
Indeed, these are exactly the poles of the integrand in (\ref{IntegralII1v4}). 
We again stress that the P and NP poles are exchanged by the duality transformation (\ref{dualmap2}).
The contour $C_1$ is placed in the interval, ${\rm max}(-M-1, -{k\over 2}+n-1-c)< s_i<{\rm min}(-1-c, {k\over 2}-M)$, that is mapped to 
${\rm max}(-{k\over 2}+n-1-\widetilde{c}, -(k-M)-1)< \widetilde{s}_i< {\rm min}({k\over 2}-(k-M), -1-\widetilde{c})$.\footnote{\label{footnoteC1}More precisely speaking, these ranges are for $c<M$ as discussed in Section \ref{IntRep}, but for $c\ge M$ the intervals are not conditional, $-{k\over 2}+n-1-c< s_i<-1-c$ and $-{k\over 2}+n-1-\widetilde{c}< \widetilde{s}_i< -1-\widetilde{c}$.} Hence the contour in terms of the variables $\widetilde{s}_i$ is equivalent to $\widetilde{C}_1$.

to the left of $s_a=-1-c$ and the right of $s_a= -{k\over 2}+(n-1)-c$ for $c\ge M$

\medskip
As for the zeros, similar to the flavor Wilson loop case, they appear only from the factors that depend on the differences $\widetilde{s}_i-\widetilde{s}_j$ in (\ref{IntegralII1v3}) and $s_i-s_j$ in (\ref{IntegralII1v4}). In fact, these factors are exactly the same in (\ref{IntegralII1v3}) and (\ref{IntegralII1v4}).

\medskip
In order to examine the asymptotics, we first note that\footnote{In the simplest case $n=1$, only $c=0$ gives a nonvanishing result in (\ref{sindepfactor}). The left hand side yields $q^{M}{(q^2)_{k-M-1}\over (q)_{k-M-1}}=-{1-q^M\over 1-q}$ that equals the right hand side, $-{(q^2)_{M-1}\over (q)_{M-1}}$, with a minus sign. More generally, we find
\begin{align}
\lim_{\epsilon\to 0}{(q^{\epsilon})_1(q^{n-\widetilde{c}})_{k-M}
\over(q^{\epsilon-\widetilde{c}})_{k-M}}
&=\lim_{\epsilon\to 0}{(q^{\epsilon})_1(q^{c+1})_{k-M}
\over(q^{\epsilon-n+c+1})_{k-M}}
=\lim_{\epsilon\to 0}q^{-Mn}{(q^{\epsilon})_1(1-q^{M-c})\cdots(1-q^{M-c+n-1})\over(1-q^{-c})\cdots(q^{-\epsilon})_1\cdots(1-q^{-c+n-1})}\nn\\
&=-q^{-nM}\lim_{\epsilon\to 0}{(q^{\epsilon})_1(q^{n-c})_M\over (q^{\epsilon-c})_M}\ .
\end{align}} 
\bea
q^{n(2\widetilde{c}+M)}{(q^{1-n})_{\widetilde{c}}(q^{1+n})_{k-M-1-\widetilde{c}}\over (q)_{\widetilde{c}}(q)_{k-M-1-\widetilde{c}}}
&=& q^{n(n-1)+n(M-c)}\lim_{\epsilon\to 0}{(q^{\epsilon})_1(q^{n-\widetilde{c}})_{k-M}
\over(q^n)_1(q^{\epsilon-\widetilde{c}})_{k-M}}\nn\\
&=&-q^{n(n-1)}{(q^{1-n})_{c}(q^{1+n})_{M-1-c}\over (q)_{c}(q)_{M-1-c}}\ .
\label{sindepfactor}
\eea
The asymptotics at $\widetilde{s}_i\to i\infty$ in (\ref{IntegralII1v3}) and $s_i\to i\infty$ in (\ref{IntegralII1v4}) then differ only by the factor $-i^{k}q^{-n^2+2n}$ and the factor $i^{k}$ is canceled by the same factor from the normalization. This completes the proof of (\ref{1/2BPStobeproved}).

\paragraph{$\bullet $ The ${1\over 6}$-BPS Wilson loop duality :}
Having proved the two of the duality maps (\ref{flavorduality2}) and (\ref{1/2duality2}), the definition of the ${1\over 2}$-BPS Wilson loop (\ref{1/21/6relation}) implies the duality map (\ref{1/6duality2}) for the ${1\over 6}$-BPS Wilson loops: An easiest way to see it is to add $(-1)^{n+1}W^{\rm I}_{1\over 6}(N_1,N_2;n)_k$ to the both sides of (\ref{1/6duality2}),
\bea
W_{\half}(N_1,N_2;n)&=&-\left(W^{\rm I}_{1\over 6}(\widetilde{N}_2,N_1; n)_k-(-1)^nW^{\rm I}_{1\over 6}(N_1,N_2;n)_k\right)\\
&&-2(-1)^{n+1}\left(W^{\rm II}_{1\over 6}(\widetilde{N}_2,N_1; n)_k-W^{\rm I}_{1\over 6}(N_1,N_2;n)_k\right)\ .\nn
\eea
Using the flavor Wilson loop duality (\ref{flavorduality2}) and the relation (\ref{1/21/6relation}), this yields the ${1\over 2}$-BPS Wilson loop duality, 
\be
W_{\half}(N_1,N_2;n)=(-1)^nW_{\half}(\widetilde{N}_2,N_1;n)\ .
\ee 
This thereby proves the ${1\over 6}$-BPS Wilson loop duality (\ref{1/6duality2}).


\subsection{The simplest example -- the $U(1)_k\times U(N)_{-k}$ theory}
\label{U1UNexample}

It is illustrative to work out the simplest case, the duality between the $U(1)_k \times U(N)_{-k}$ and $U(2+k-N)_k \times U(1)_{-k}$ ABJ theories, since the integral is one-dimensional and the integration can be explicitly carried out. 
Apart from its simplicity, the $U(1)_k \times U(N)_{-k}$ may also be relevant to the study of Vasiliev's higher spin theory with $U(1)$ symmetry in the 't Hooft limit, $N, k\to \infty$ with $N/k$ fixed\cite{Chang:2012kt}. Here we perform the integrals analytically and explicitly check  Seiberg duality for some small values of $N$, $k$ and $n$. 
In this connection, in the simplest case of $N=1$, a check against the direct integral  is provided for arbitrary $k$ and $n$ in Appendix \ref{U1U1check}.

\subsubsection{The flavor Wilson loop}

We first consider the duality of flavor Wilson loops that are $\frac{1}{6}$-BPS Wilson loops  on the ``flavor group'' $U(1)$. The  $\frac{1}{6}$-BPS flavor Wilson loops are given by
\begin{align}
W^{\mathrm{I}}_{\frac{1}{6}} (1,N;n)_k &= q^{-\frac{1}{2}n^2+n} \frac{I(1,N;n)_k}{I(1,N;0)_k}\ , \label{W1N}  \\
 W^{\mathrm{II}}_{\frac{1}{6}} (\widetilde{N},1;n)_{k} &=W^{\mathrm{I} }_{\frac{1}{6}} (1,\widetilde{N};n )_{-k}
 = q^{\frac{n^2}{2} -n} \frac{I(1,\widetilde{N};n )_{-k}}{I(1,\widetilde{N};0)_{-k}} \label{WN1}\ ,
\end{align}
where the dual gauge group $\widetilde{N}=2+k-N$ and the integral expression takes the form
\begin{align}
I(1,N;n)_k &=\frac{(-1)^{N-1} q^{-n}}{1+q^n}\frac{-1}{2 \pi i} \int_C \frac{\pi ds}{\sin(\pi s)} q^{-ns} \frac{  (q^{s+1})_{N-1} }{(-q^{s+1+n})_{N-1}}\ .
\end{align}
The function $I(1,N;0)_k$ appearing in the denominators is essentially the partition function and has the following property under Seiberg duality \cite{Kapustin:2010mh, Awata:2012jb}:
\begin{align}
 \label{eq:zratio}
I(1,N;0)_k = i^{-k}I(1,\widetilde{N};0)_{-k}\ . 
\end{align}
Meanwhile, the flavor Wilson loop duality is given by \eqref{flavorduality}
\be
W^{\mathrm{I}}_{\frac{1}{6}} (1,N;n)_k = W^{\mathrm{II}}_{\frac{1}{6}}(\widetilde{N},1;n)_{k}
\ee
that implies
\begin{align}
I(1,N;n)_k=i^{-k} q^{n^2 -2n} I(1,\widetilde{N};n)_{-k}\ .
\label{flavortobechecked}
\end{align}
This generalizes the relation \eqref{eq:zratio} for the partition function, and we are going to check this relation explicitly for some small values of $N$, $k$ and $n$.

\begin{figure}[H]
\begin{center}
  \includegraphics[width=10cm]{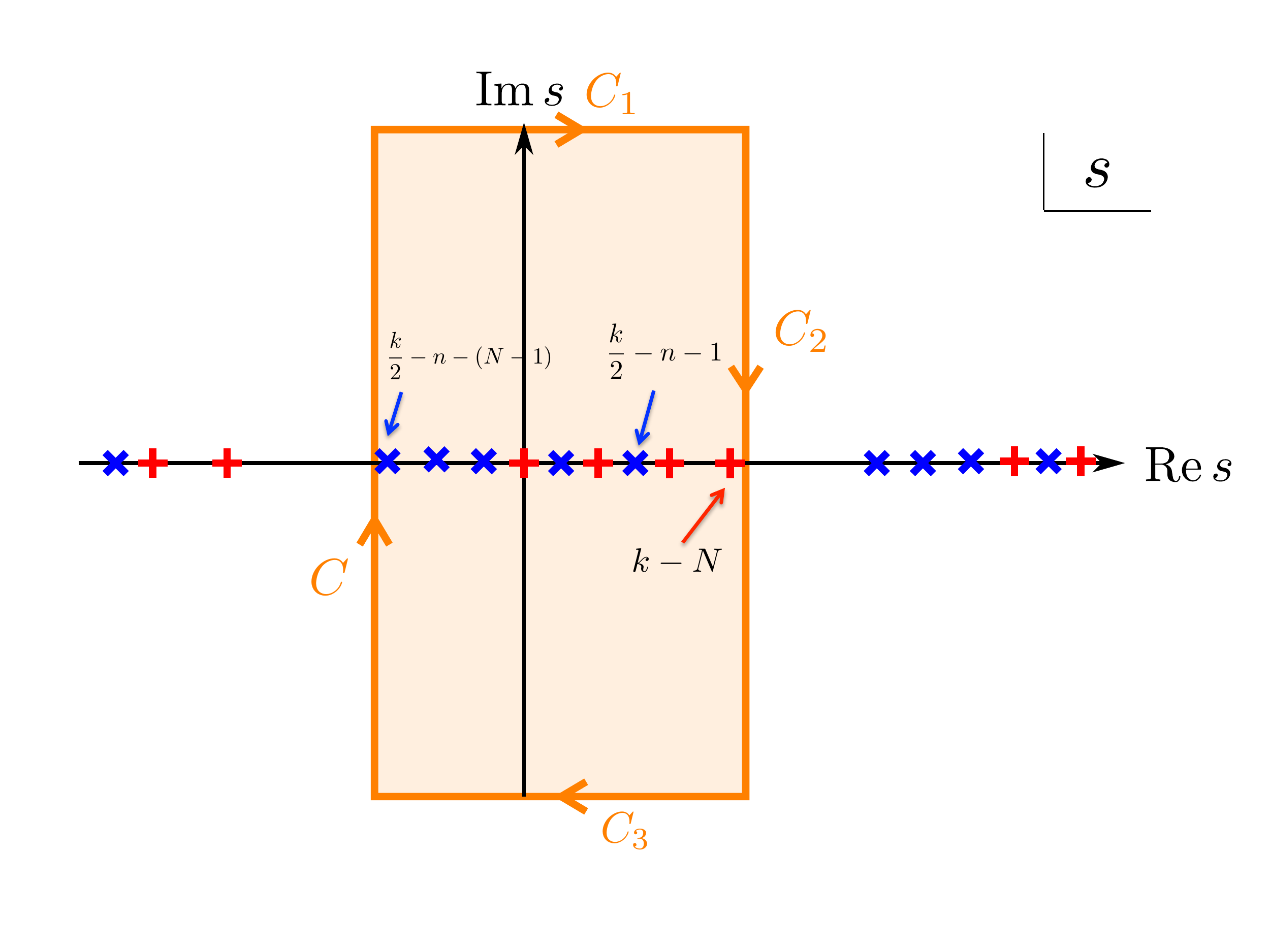} 
  \caption{The integration contour: Shown is a deformed contour where $C_2$ is shifted to the left from the vertical line running from $s=\frac{3k}{2}-n-(N-1)+i \infty -\epsilon$ to $\frac{3k}{2}-n-(N-1)-i \infty-\epsilon$, exploiting the absence of poles on the strip between $s=k-N$ and $\frac{3k}{2}-n-(N-1)$. }
  \label{contour1}
  \end{center}
\end{figure}

We first calculate the odd $k$ case, since it is simpler than the even $k$ case. The integrand flips the sign under the shift of integration variable, $s \rightarrow s+k$, in this case. Thanks to this property, the integrals can be evaluated by considering the closed contour depicted in Fig.\ref{contour1}: Let us denote the original contour by $C$ and the closed contour by  $C'$ that consists of $C+C_1+C_2+C_3$ where $C_2$ is parallel to $C$ and shifted by $k$. It is easy to see that there are no contributions from the contours $C_1$ and $C_3$, and the integral along $C_2$ is precisely the same as that along the original contour $C$ because the integrand only flips the sign under the shift $s\rightarrow s+k$.  It is then clear that the integral along the closed contour $C'$ is twice that along $C$:  
\begin{align}
\int_C (\cdots) =\frac{1}{2} \int_{C'} (\cdots)\ . 
\end{align}
Thus the integrals can be evaluated by residue calculations.
%
%
%
%
\noindent
It is straightforward to carry out the calculation and we find that
\begin{align}
I(1,N;n)_{{\rm odd}\,k} 
&= \frac{(-1)^{N-1} q^{-n} }{2(1+q^n)}\Biggl[ \sum_{s=0}^{k-N} (-1)^s  q^{-ns } \frac{(q^{s+1} )_{N-1}  }{(-q^{s+1+n})_{N-1} } \\ 
&\hspace{3cm}+\frac{ik}{2} \sum_{c=1}^{N-1} (-1)^{\frac{k+1}{2} +c} q^{n(n+c)} \prod_{b=1,b\neq c} ^{N-1} \frac{(-q^{b-c-n})_1 }{ (q^{b-c})_1}  \Biggr]\ ,\nn
\end{align}
where the first term is the contribution from the P poles at $s=0,1,\cdots,k-N$, while the second term is from the NP poles at $s=\frac{k}{2} -n-b,\, (b=1,\cdots,N-1)$.

The even $k$ case requires more considerations. In contrast to the odd $k$ case, the integrand is periodic under the shift $s \rightarrow s+k$. In addition, some of P and NP poles merge into double poles, since ${k\over 2}$ is an integer. We can, however, apply a similar trick as that used in \cite{Okuyama:2011su, Awata:2012jb}. For an illustration of this trick, let us consider a generic integral of the form $\int_C ds \, f(s)$ with $f(s)$ being periodic under the shift $s\to s+k$. The trick is instead to consider the integrand $g(s) =f(s)(s+a)$ with $a$ being an arbitrary constant. This integrand shifts as $g(s+k)=g(s)+kf(s)$ when $s$ is shifted by $k$. 
Thus the integral $\int_{C'}dsg(s)$ along the closed contour $C'=C+C_1+C_2+C_3$ yields $k\int_{C}dsf(s)$ provided that there are no contributions from $C_1$ and $C_3$.
This way the even $k$ case can also be evaluated by residue calculations. Note that the result does not depend on the choice of an arbitrary constant $a$.

In the following we only show the case when $\frac{k}{2}-(N-1)-n>0$. The other case, $\frac{k}{2}-(N-1)-n < 0$, however, can be easily derived in a similar manner.
The poles encircled by the contour appear at
\begin{align}
s=\begin{cases} 
\,\,\, 0,\cdots,\frac{k}{2}-N-n  \qquad(\mbox{simple})\\
\,\,\, \frac{k}{2}-(N-1)-n ,\cdots,\frac{k}{2}-1-n \qquad (\mbox{double})\\
\,\,\, \frac{k}{2}-n,\cdots,k-N\qquad (\mbox{simple})
\end{cases}\ .
\end{align}
The residue evaluation then yields
\begin{align}
I(1,N;n)_{{\rm even}~k}  
&=-\frac{(-1)^{N-1} q^{-n}}{1+q^n} \frac{1}{2k} \Biggl[ \sum_{\substack{s=0,\cdots,\frac{k}{2}-N-n \\ \frac{k}{2}-n,\cdots,k-N} } (-1)^s q^{-ns} \frac{(q^{s+1})_{N-1}}{(-q^{s+1+n})_{N-1} }(s+a)  \label{evenI} \\
&  +\sum_{b=1,\cdots,N-1} \lim_{s\rightarrow \frac{k}{2} -b-n} \frac{d}{ds} \left\{  \left(s-\left(\frac{k}{2}-b-n\right)\right)^2 \frac{\pi q^{-ns}}{ \sin \pi s} \frac{(q^{s+1})_{N-1}}{(-q^{s+1+n})_{N-1}}  (s+a) \right\}\Biggr]\ ,\nn
\end{align}
where the first line is the contribution from simple poles and the second line from double poles. 
These simple poles are a subgroup of P poles in \eqref{Ppoles2kint}, while the double poles are composed of P and NP poles, i.e., those P poles in \eqref{Ppoles2kint} coalescing NP poles in \eqref{NPpoles2kint}.

On the dual side, the integral $I(1,\widetilde{N};n)_{-{\rm odd}~k}$ can be calculated similarly as
\begin{align}
I(1,\widetilde{N};n)_{-{\rm odd}\,k} &=\frac{ q^{n(\widetilde{N}+1)}}{2(1+q^n)} \Biggl[ \sum_{s=0}^{k-\widetilde{N}} (-1)^s q^{ns} \frac{(q^{s+1})_{\widetilde{N}-1}}{(-q^{s+1+n})_{\widetilde{N}-1}}\\
&\hspace{3cm}-\frac{ik}{2} \sum_{b=1}^{\widetilde{N}-1} (-1)^{\frac{k-1}{2}+b} q^{-n^2 -bn} \prod_{a=1,a \neq b}^{\widetilde{N}-1} \frac{(-q^{a-b-n})_1}{(q^{a-b})_1}  \Biggr]\ .\nn
\end{align}
In the even $k$ case we only show result for the case when $\frac{k}{2}-(\widetilde{N}-1)-n>0$:
\begin{align}
I(1,\widetilde{N};n)_{-{\rm even}\,k} &=\frac{q^{n(\widetilde{N}+1)}}{2k(1+q^n)}  \Biggl[ \sum_{\substack{s=0,\cdots,\frac{k}{2}-\widetilde{N}-n \\ \frac{k}{2}-n ,\cdots,k-\widetilde{N}  } } (-1)^s q^{ns} \frac{(q^{s+1})_{\widetilde{N}-1}}{(-q^{\widetilde{N}+1+n})_{\widetilde{N}-1}}(s+a)   \\
& +\sum_{s=\frac{k}{2}-\widetilde{N}-n+1,\cdots,\frac{k}{2}-n-1  }  \lim_{s'\rightarrow s} \frac{d}{ds'} \left\{  \frac{\pi (s'-s)^2}{\sin(\pi s')} q^{ns'} \frac{  (q^{s'+1})_{\widetilde{N}-1} }{(-q^{s'+1+n})_{\widetilde{N}-1}} (s'+a)\right\}  \Biggr]\ .\nn
\end{align}
We have now collected necessary data to explicitly check the flavor Wilson loop duality. 
%
\subsubsection*{$\bullet$ Numerical checks}
%
Let us check \eqref{flavortobechecked} explicitly for the duality $U(1)_5 \times U(2)_{-5} =U(5)_5 \times U(1)_{-5}$ with winding $n=2$. The results are
\begin{align}
I(1,2;2)_5 &=\underbrace{( 1.46 + 0.47 i)+(0.90 - 0.29 i )+(0.45 - 0.62 i)+(-0.58 i)}_{\rm P} \nonumber \\
&\qquad+ \underbrace{(-0.73 + 1.01 i) }_{\rm NP}\ ,\label{fW12}\\
i^{-5} I(1,5;2)_{-5} &= \underbrace{(-0.73 + 1.01 i)}_{\rm P} \nonumber \\
&\qquad  +\underbrace{(1.46 + 0.47 i)+(0.90 - 0.29 i) +(0.45 - 0.62 i) +( -0.58 i) }_{\rm NP}\ .
\label{fW51}
\end{align}
Both in the original and dual theories, the first line is the perturbative contribution, while the second line is the nonperturbative contribution. We can explicitly see that the perturbative and nonperturbative contributions are exchanged under  Seiberg duality.

Next we consider the even $k$ case, $U(1)_{4} \times U(2)_{-4} =U(4)_4 \times U(1)_{-4} $ with winding $n=1$. For the original theory we have
\begin{align}
I(1,2;1)_{4} 
&=\frac{i-1}{16} \left[ \underbrace{\frac{(4-4 i)+(3+i) a \pi }{2 \pi }}_{\rm P+NP}+  \underbrace{ (-1-i) (1+a)+\left(\frac{i}{2}-\frac{1}{2}\right) (2+a)  }_{\rm P} \right] \label{I(12)1} \\
&=\frac{i-1}{16}\left[-2+\frac{2-2 i}{\pi }  \right], \label{I(12)}
\end{align}
where the first term in the first line comes from the double pole at $s=0$ and the rest are from the simple P poles at $s=1,2$.
Note that the $a$-dependence is canceled out, as it should. For the dual theory we have
\begin{align}
\hspace{-.55cm}
iI(1,4,1)_{-4} 
&= \frac{i-1}{16}\left[ \underbrace{\frac{(4-4 i)+(3+i)(-a') \pi }{2 \pi } }_{\rm P+NP}+  \underbrace{(-1 - i) (1 - a') +\left(\frac{i}{2}-\frac{1}{2}\right) (2-a') }_{\rm NP} \right] \label{I(14)1}\\
&= \frac{i-1}{16} \left[ -2 +\frac{2-2 i}{\pi }  \right]\ ,\label{I(14)}
\end{align}
where the first term in the first line comes from the double pole at $s=0$ and the rest are from the simple NP poles at $s=-1,-2$. The $a'$-dependence is canceled out similar to the previous case. However, we observe that the pole-by-pole maps agree if we identify $a=-a'$ even though the constants $a$ and $a'$ do not seem to carry any physical meaning. As indicated, the P and NP poles are exchanged in the original \eqref{I(12)1} and dual \eqref{I(14)1} Wilson loops.
These examples provide evidence for the flavor Wilson loop duality \eqref{flavorduality}.

\subsubsection{The $\frac{1}{2}$-BPS Wilson loop}

We next turn to the $\frac{1}{2}$-BPS Wilson loops with winding $n$  \eqref{1/2Wilson} 
\begin{align}
W_{\frac{1}{2}} (1,N;n)_k &=(-1)^{n+1} q^{\frac{n^2}{2} -n} \frac{I^{(1)} (N,1;n)_{-k}}{I^{(2)} (N,1;0)_{-k}}\ , \\
W_{\frac{1}{2}} (\widetilde{N},1;n)_k &=q^{-\frac{n^2}{2} +n } \frac{I^{(1)}(\widetilde{N},1;n )_k }{I^{(2)}(\widetilde{N},1;0)_{k}}
\end{align}
where the integral expression takes the form
\begin{align}
I^{(1)}(N,1;n)_{k}=&\sum_{c=0}^{n-1}
q^{n(2c-N+1)}{(q^{1-n})_c(q^{1+n})_{N-2-c}\over (q)_c(q)_{N-2-c}}
{-1\over 2\pi i}\int_{C_1\![c]}{\pi ds\over\sin(\pi s)}
{\left(q^{s+1}\right)_{N-1}\over 2\left(-q^{s+1}\right)_{N-1}}\nn\\
&\qquad\times{\left(-q^{s+1+c}\right)_1\left(q^{s+1+c+n}\right)_1
\over \left(q^{s+1+c}\right)_1\left(-q^{s+1+c+n}\right)_1}\ .
\label{I1U1}
\end{align}
The normalization factor has the relations $I^{(2)} (N,1;0)_{-k}=(-1)^{N-1}   I^{(2)}(N,1;0)_k = I(1,N;0)_k $ and obeys the duality relation
\begin{align}
I^{(2)} (N,1;0)_{-k}=i^{-k}I^{(2)}(\widetilde{N},1;0)_{k}\ .\label{Irelation}
\end{align}
Meanwhile, the ${1\over 2}$-BPS Wilson loop duality is given by \eqref{1/2duality2}
\begin{align}
W_{\frac{1}{2}} (1,N;n)_k &=(-1)^nW_{\frac{1}{2}} (\widetilde{N},1;n)_k
\end{align}
that, together with \eqref{Irelation}, implies
\be
I^{(1)}(N,1,n)_{-k}=-i^{-k}q^{-n^2+2n}I^{(1)}(\widetilde{N},1;n)_k\ .
\label{1/2tobechecked}
\ee
This is the relation we are going to check explicitly for some small value of $N$, $k$ and $n$.

Similar to the previous case, the integrand for odd $k$ is anti-periodic under the shift $s\to s+k$, whereas it is periodic for even $k$. In the latter case, some of P and NP poles merge into double poles. We can thus apply the same technique as that used in the previous case to this case.

In the odd $k$ case, the poles encircled by the contour appear at 
\begin{align}
{\bf P:}&\qquad s = -1-c;\quad 0,1,\cdots,-2-c+n;\quad -c+n,\cdots,k-N~~ \mbox{mod} ~k \\
{\bf NP:} &\qquad s = \frac{k}{2} -b+n \delta_{b,1+c}~~\mbox{mod}~k \qquad(b=1,\cdots,N-1)
\end{align}
for $I^{(1)} (1,N;n)_{-k}$ and a given $c$ and 
\begin{align}
{\bf P:}&\qquad s = -n+c;\quad 0,\cdots,c-1;\quad c+2,\cdots,N-2~~ \mbox{mod} ~k \\
{\bf NP:} &\qquad s = \frac{k}{2} -b-n \delta_{b,n-c}~~\mbox{mod}~k \qquad(b=1,\cdots,k-N+1)
\end{align}
for $I^{(1)} (\widetilde{N},1;n)_k$ and a given $\tilde{c}=(n-1)-c$.
As discussed, the change of the integration variable, $\tilde{s}=-s+\frac{k}{2} -1$, precisely exchanges P and NP poles in the two theories.  It is straightforward to carry out the residue integrals and we find 
\begin{align}
&I^{(1)}(N,1;n)_{\pm {\rm odd}\,k} \nonumber \\
&= \frac{1}{2}\sum_{c=0}^{n-1} \left[ \left(\sum_{s=0}^{-2-c+n} +\sum_{s=-c+n}^{k-N}  \right) (-1)^s q^{n (2c-N+1) }\frac{(q^{1-n})_c(q^{1+n})_{N-2-c}}{(q)_c(q)_{N-2-c}}   \right.  \\
&\times \frac{(q^{s+1})_{N-1}}{2(-q^{s+1})_{N-1}} \frac{(-q^{s+1+c})_1(q^{s+1+c-n})_1}{(q^{s+1+c})_1(-q^{s+1+c-n})_1} +(-1)^{c+1}q^{n(c-N+1)}\frac{(q^{n-c})_{N-1} (q^{-n})_1 }{(-q^{-c})_{N-1} (q^n)_1 (-q^{-n})_1}  \nonumber \\
&\left.\mp ik \sum_{b=1}^{N-1}(-1)^{\frac{k-1}{2} +b-n\delta_{b,1+c}}\frac{q^{n (2c-N+1) }(q^{1-n})_c(q^{1+n})_{N-2-c}}{(q)_c(q)_{N-2-c}}   \frac{\prod_{a=1}^{N-1} (-q^{a-b+n(\delta_{b,1+c} -\delta_{a,1+c} )+1 })_1 }{4 \prod_{a=1}^{N-1} (q^{a-b+n(\delta_{b,1+c} -\delta_{a,1+c})+1 })_1} \right]\nn
\end{align}
where $q=\mathrm{e}^{\pm\frac{2\pi i}{k}}$ for $\mp k$ with an abuse of notation. The first two lines are the contributions from P poles and the last line is those from NP poles.

%

In the even $k$ case we only show the case when $\frac{k}{2}>N-1+n$. The other case can be calculated in a similar manner. The poles encircled by the contour appear at
\begin{align}
s=\begin{cases} 
\,\,\, -1-c;\quad 0,\cdots,\frac{k}{2}-N  \qquad(\mbox{simple})\\
\,\,\, \frac{k}{2}-c+n ,\cdots,k-N  \qquad(\mbox{simple})\\
\,\,\, \frac{k}{2}-(N-1),\cdots,-2-c-n;\quad -c+n,\cdots, \frac{k}{2}-1-c+n  \qquad(\mbox{double})
\end{cases}\ .
\end{align}
Similar to the previous case \eqref{evenI}, we find that
\begin{align}
&I^{(1)}(N,1;n)_{\mathrm{even}\,k} \nn\\
&=\frac{-1}{2k} \sum_{c=0}^{n-1} \Biggl[ \sum_{\substack{s=0,\cdots, \frac{k}{2}-N \\ \frac{k}{2} -c+n,\cdots,k-N;-1-c }}(-1)^s q^{n (2c -N+1)}
\frac{(q^{1-n})_c (q^{1+n})_{N-2-c}}{(q)_c (q)_{N-2-c}} \frac{(q^{s+1})_{N-1}}{2(-q^{s+1})_{N-1}}
\nn\\ 
&\times \frac{(-q^{s+1+c})_1(q^{s+1+c-n})_1}{(q^{s+1+c})_1(-q^{s+1+c-n})_1}(s+a)
+\sum_{\substack{s=\frac{k}{2} -(N-1) ,\cdots,-2-c+n \\
-c+n ,\cdots, \frac{k}{2} -1-c+n }} \lim_{s'\rightarrow s} \frac{d}{ds'}\Biggl\{  (s'-s)^2 \frac{\pi}{\sin \pi s}q^{n (2c -N+1)}  \nn\\
& \times \frac{(q^{1-n})_c (q^{1+n})_{N-2-c}}{(q)_c (q)_{N-2-c}} \frac{(q^{s+1})_{N-1}}{2(-q^{s+1})_{N-1}} \frac{(-q^{s+1+c})_1(q^{s+1+c-n})_1}{(q^{s+1+c})_1(-q^{s+1+c-n})_1}(s+a) \Biggr\} \Biggr]
\end{align}
where $a$ is an arbitrary constant.

\subsubsection*{$\bullet$ A numerical check}
%
As an explicit check of \eqref{1/2tobechecked}, we consider the example, the duality $U(1)_5 \times U(2)_{-5} =U(5)_5 \times U(1)_{-5}$ with winding $n=2$. For the original theory we have 
\begin{align}
&I^{(1)}(2,1;2)_{-5} \nonumber \\
&\qquad = \frac{1}{4} \left( \sum_{s=0,2,3,4}(-1)q^{2} \frac{(q^{1-s})_1}{(-q^{1-s})_1} -5i q^{2} +\frac{2q(1-q)}{(1+q)(1+q^2)} \right) \\
&\qquad = -\frac{1}{4} \Biggl[ \underbrace{ (1.80 - 2.48 i) + (-0.42 + 0.58 i)+(0.42 - 0.58 i)+(-1.80 + 2.48 i) }_{\rm P} \nonumber \\
&\qquad  \qquad \qquad\qquad \qquad \qquad \qquad \qquad \qquad +\underbrace{(2.93 - 4.04 i)}_{\rm NP}  +\underbrace{(-2.35 i)}_{\rm P} \Biggr] \label{W21}
\end{align}
where $q=\mathrm{e}^{-\frac{2\pi i}{5}}$. In \eqref{W21} the first line is the contributions from P poles for $c=0$ at $s=-1,0,2,3$, the first term in the second line from NP poles for $c=0$ at $s=\frac{k}{2}+1$, and the last term from the P pole  for $c=1$ at $s=-2$.
For the dual theory we have
\begin{align}
&iI^{(1)} (5,1;2)_{5} \nonumber \\
&=\frac{i}{4} \left(\frac{-(q^3)_2^2}{(-q^3)_2 (-q)_2} +5i q^{2} \frac{(-q^3)_3}{(q)_3}+\sum_{b=0,1,3,4} \frac{5i (q^3)_2  }{2(q)_2}  \frac{(-1)^b(q^{2-b})_1}{(-q^{2-b})_1} \frac{\prod_{a=0}^5 (-q^{a-b})_1  }{\prod_{a=1,a\neq b}^{5} (q^{a-b})_1 }   \right) \\
&=-\frac{1}{4} \Biggl[ \underbrace{(2.93 -4.04i )}_{\rm P}   +\underbrace{(-2.35i )}_{\rm NP} \label{W51} \\
&\qquad \qquad \qquad \qquad  +\underbrace{(  1.80-2.48i) +( - 0.42+0.58i ) +( + 0.42 -0.58i) +(- 1.80+2.48i ) }_{\rm NP}
\Biggr]\ . \nn
\end{align}
The first term is the contributions from P poles for $c=1$ at $s=-2$, the second term from the NP pole for $c=0$ at $s=\frac{k}{2}+1$, and the second line from NP poles for $c=1$ at $s=\frac{k}{2}-b,\,\, (b=0,1,3,4)$. It is clear that in \eqref{W21} and \eqref{W51} the P and NP poles are interchanged under the duality as expected. Thus this provides evidence for the $\frac{1}{2}$-BPS Wilson loop duality \eqref{1/2duality2}.
By deduction, this result, combined with the flavor Wilson loop duality between \eqref{fW12} and \eqref{fW51}, also constitutes evidence for the ${1\over 6}$-BPS Wilson loop duality \eqref{1/6duality2}.

\section{Summary and discussions}
\label{SummaryDiscussions}

In the current paper, we discussed the Wilson loops of the
ABJ theory and studied their properties,
generalizing the techniques developed in \cite{Awata:2012jb} for
partition function.
In more detail, the objects of our interest were the circular
$1\over 6$- and $1\over 2$-BPS Wilson loops with winding number $n$ in the
$U(N_1)_k\times U(N_2)_{-k}$ ABJ theory on $S^3$.  By the localization
technique, the Wilson loop can be represented as an ordinary integral
with $N_1$ variables $\mu_i$ and $N_2$ variables $\nu_a$, corresponding
to the eigenvalues of $U(N_1)$ and $U(N_2)$ adjoint matrices.  Rather
than directly evaluating this ABJ matrix integral, we followed
\cite{Awata:2012jb} and started instead with the Wilson loop in the lens
space matrix integral, which is related to the ABJ one by the analytic
continuation $N_2\to -N_2$.  Because the lens space Wilson loop can be
computed exactly, by continuing it back to the ABJ theory by setting
$N_2\to -N_2$, we arrived at an infinite sum expression for the ABJ
Wilson loop.  Actually, this infinite sum is only formal and does not
converge.  However, by means of a Sommerfeld-Watson transformation, we
turned it into a convergent integral of ${\rm min}(N_1,N_2)$ variables
and successfully obtained the ``mirror'' description of the ABJ Wilson
loop, generalizing that for partition function.  The final expressions
are given in \eqref{1/6WilsonI1} for the $1\over 6$-BPS Wilson loop with
$N_1\ge N_2$, in \eqref{1/6WilsonI2} for the $1\over 6$-BPS Wilson loop
with $N_1\le N_2$, and in \eqref{1/2Wilson} for the $1\over 2$-BPS
Wilson loop.

The ABJ theory is conjectured to possess a Seiberg-like duality given in
\eqref{SD_intro}.  Based on the integral expressions for the ABJ Wilson
loops, we showed that the Wilson loops have non-trivial transformation
rule, given by \eqref{1/6duality}--\eqref{1/2duality}.  This result is
consistent with the result by Kapustin and Willett
\cite{Kapustin:2013hpk} and slightly generalize it to the case where the
flavor group is gauged.
We also presented a heuristic explanation of the Seiberg-like duality based on
the brane construction of the ABJ theory, followed by a rigorous proof
based on the integral representation of Wilson loops.  The brane picture
is heuristic but quite powerful and can be used to predict the duality
rule for Wilson loops with general representations.  We also presented
another derivation of the duality in Appendix \ref{KapustinWillett}.

Our method to start from the lens space theory and analytically continue
it to ABJ theory involves subtleties associated with a Sommerfeld-Watson
transformation to rewrite a divergent sum in terms of a well-defined contour
integral.  In particular, this rewriting has possible ambiguities in the
choice of integration contours including the $\epsilon$-prescription for
the parameter $M$. It is necessary to keep $M$ slightly away from an
integral value by the shift $M\to M+\epsilon$ with $\epsilon>0$ in the
course of calculations. In sync with this shift the contour $C_1[c]$ has
to be placed between $s=-1-c$ and $-1-c-\epsilon$ so as to avoid the
pole at $s=-1-c-\epsilon$.  Although the choice we made is
well-motivated by the continuity in $k$ and necessary to obtain sensible
results and Seiberg duality provides strong evidence in support for it,
a direct derivation is certainly desirable.  The approaches taken in
\cite{Honda:2013pea, Matsumoto:2013nya} presumably provide promising
directions for that purpose.

Actually, however, this weakness of our approach can be turned around
and regarded as its strength.  The infinite sum we encounter in the
intermediate stage can be understood as giving a perturbative expansion
of a gauge theory quantify which, by itself, is incomplete and
divergent.  Rewriting it in terms of a finite contour integral can be
thought of as supplementing it with non-perturbative corrections to make
it well-defined and complete; more precisely, summing over P poles
corresponds to summing up perturbative expansion and including NP poles
corresponds to adding non-perturbative corrections.  We emphasize that
it is very rare that we can carry out this non-perturbative completion
in non-trivial field theories and ABJ Wilson loops provide explicit and
highly non-trivial examples for it.

In \cite{Hatsuda:2012dt}, the partition function of ABJM theory was
evaluated using the Fermi gas approach in detail and a cancellation
mechanism was found between non-perturbative contributions.  Namely, for
certain values of $k$, the contribution from worldsheet instantons
diverges but, when that happens, the contribution from membrane
instantons also diverges and they cancel each other to produce a finite
result.  This was generalized to 1/2-BPS Wilson loops in ABJM theory in
\cite{Hatsuda:2013yua} and to ABJ theory in \cite{Matsumoto:2013nya,
Honda:2014npa}.
This phenomenon is reminiscent of what is happening in our formulation,
in which partition function and Wilson loops are expressed as contour
integrals.  The integrals can be evaluated by summing over the residue
of P and NP poles, which are generically simple poles. As we change $k$
continuously, at some integral values of $k$, two such simple poles can
collide and become a double pole.  For this to happen, the residue of
each simple pole must diverge but their sum must remain
finite.\footnote{For example, consider ${1\over 2\epsilon}({1\over
x+\epsilon}-{1\over x-\epsilon})={1\over x^2-\epsilon^2}$ where
$\epsilon\to 0$.}  It is reasonable to conjecture that this cancellation
of residues is closely related to the cancellation mechanism of
\cite{Hatsuda:2012dt}. We leave this fascinating possibility for future
study.

We observed that the $1\over 6$-BPS Wilson loop diverges for $n\ge
{k\over 2}$. Actually, because the lens space Wilson loop
\eqref{LensSpaceSfunction} is invariant under $n\to n+k$, we can define
the analytically continued $1\over 6$-BPS ABJ Wilson loops to have this
periodicity in $n$ as $n\cong n+k$.  Then the $1\over 6$-BPS ABJ Wilson
loop is divergent only for $n={k\over 2}$ mod $k$, which can be checked
in explicit expressions such as \eqref{W11}.  On the other hand, the
$1\over 2$-BPS Wilson loop is finite for all $n$.
In the type IIA bulk dual, in $AdS_4\times \mathbb{CP}^3$, the $1\over
2$-BPS Wilson loop corresponds to a fundamental string extending along $AdS_2$
inside $AdS_4$ and sitting at a point inside $\mathbb{CP}^3$
\cite{Drukker:2008zx, Chen:2008bp, Rey:2008bh}.   There is no problem
having $n$ such fundamental strings, which must correspond to the
$1\over 2$-BPS Wilson loop with arbitrary winding $n$.
On the other hand, for generic $n$, the $1\over 6$-BPS Wilson loop has
been argued to correspond to smearing the above fundamental string over
$\mathbb{CP}^1\subset \mathbb{CP}^3$ \cite{Drukker:2008zx}, which seems
a bit unnatural for an object as fundamental as a Wilson loop.  However,
particularly for $n={k\over 2}$ mod $k$, there is a $1\over 6$-BPS
configuration in which a D2-brane is along $S^1\subset \mathbb{CP}^3$
and carries fundamental string charge dissolved in worldvolume flux
\cite{Drukker:2008zx}.
So, it is tempting to conjecture that, for $n\neq {k\over 2}$ mod $k$,
there is some different configuration dual to the $1\over 6$-BPS Wilson
loop which becomes the D2-brane configuration at $n={k\over 2}$.  The
divergence is presumably related to this phase transition.  It would be
interesting to actually find such a brane configuration.

In \cite{Chang:2012kt}, it was conjectured that the $U(N_1)_k\times
U(N_2)_{-k}$ ABJ theory in the fixed $N_1$, large $N_2,k$ limit is dual
to the $\cN=6$ supersymmetric, parity-violating version of the Vasiliev
higher spin theory (where we assumed $N_1\ll N_2$).  Being based on
$N_1$-dimensional integral, our formulation is particularly suited for
studying this limit. So, it is very interesting to use our results to
evaluate Wilson loops in the higher spin limit and compare them with
predictions from the Vasiliev side.  It is also interesting to see if
our approach can be applied to more general CSM theories with less
supersymmetry, such as the necklace quiver \cite{Honda:2014ica}.


\section*{Acknowledgments}

We would like to thank Nadav Drukker, Masazumi Honda, Sanefumi Moriyama, Kazutoshi Ohta
and Kazumi Okuyama for useful discussions.
MS thank the IPhT, CEA-Saclay for hospitality where part of
this work was done.
%
SH would like to thank YITP, Nagoya University and KIAS for their
hospitality where part of this work was done.  The work of MS was
supported in part by Grant-in-Aid for Young Scientists (B) 24740159 from
the Japan Society for the Promotion of Science (JSPS)\@.

\appendix

\section{The q-analogs}
\label{qanalogs}
The results in the main text are given in terms of $q$-Pochhammer symbols. In this appendix we provide the definitions and some useful formulae and properties of related quantities.
Roughly, a $q$-analog is a generalization of a quantity to include a new
parameter $q$, such that it reduces to the original version in the $q\to
1$ limit.  In this appendix, we will summarize definitions of various
$q$-analogs and their properties relevant for the main text.

\paragraph{$\bullet$ $\boldsymbol{q}$-number:}  
For $z\in\bbC$, the $q$-number of $z$ is defined by
\begin{align}
 [z]_q:={1-q^z\over 1-q},\qquad 
\end{align}

\paragraph{$\bullet$ $\boldsymbol{q}$-Pochhammer symbol:}

For $a\in\bbC$, $n\in\bbZ_{\ge 0}$, the $q$-Pochhammer symbol $(a;q)$ is defined by
\begin{align}
 (a;q)_n&:= \prod_{k=0}^{n-1}(1-aq^k)
 =(1-a)(1-aq)\cdots(1-aq^{n-1})
 ={(a;q)_\infty\over (aq^n;q)_\infty}.
\end{align}
For $z\in\bbC$, $(a;q)_z$ is defined by the last expression:
\begin{align}
 (a;q)_z&:=
 {(a;q)_\infty\over (aq^z;q)_\infty}
 =\prod_{k=0}^\infty{1-aq^k\over 1-aq^{z+k}}
 .\label{maux20Sep12}
\end{align}
This in particular means
\begin{align}
 (a;q)_{-z}&={1\over (aq^{-z};q)_z}.
\end{align}
For $n\in \bbZ_{\ge 0}$,
\begin{align}
 (a;q)_{-n}&={1\over (aq^{-n};q)_n}={1\over \prod_{k=1}^n(1-a/q^k)}.
\end{align}

Note that the $q\to 1$ limit of the $q$-Pochhammer symbol is not the usual
Pochhammer symbol but only up to factors of $(1-q)$:
\begin{align}
  \lim_{q\to 1}{(q^a;q)_n\over (1-q)^n}&=
 a(a+1)\dots(a+n-1).
\end{align}

We often omit the base $q$ and simply write $(a;q)_\nu$ as
$(a)_\nu$.\footnote{We will not use the symbol $(a)_\nu$ to denote the
usual Pochhammer symbol.}

Some useful relations involving $q$-Pochhammer symbols are
\begin{align}
 (a)_\nu&={(a)_z\over (aq^\nu)_{z-\nu}}=(a)_z (aq^z)_{\nu-z},\label{hcey4Sep12}\\
 (q)_\nu&=(1-q)^{\nu}\Gamma_q(\nu+1),\label{kjzv20Sep12}\\
 (q^\mu)_\nu&={(q)_{\mu+\nu-1}\over (q)_{\mu-1}} =(1-q)^\nu{\Gamma_q(\mu+\nu)\over \Gamma_q(\mu)},\\
 (aq^\mu)_\nu&
 =(aq^\mu)_{z-\mu}(aq^z)_{\mu+\nu-z}
 ={(aq^\mu)_z\over (aq^{\mu+\nu})_{z-\nu}}
 ={(aq^z)_{\mu+\nu-z}\over (aq^z)_{\mu-z}},
 \label{jwdu20Sep12}
\end{align}
where $\mu,\nu,z\in\bbC$ and $\Gamma_q(z)$ is the $q$-Gamma function
defined below. For $n\in\bbZ$, we have the following formulae which
``reverse'' the order of the product in the $q$-Pochhammer symbol:
\begin{align}
 (aq^z)_n&=(-a)^n q^{zn+\half n(n-1)}(a^{-1}q^{1-n-z})_n,\label{jzmp20Sep12}\\
 (\pm q^{-n})_n&=(\mp 1)^n q^{-\half n(n+1)}(\pm q)_n.\label{hcit4Sep12}
\end{align}
If $\nu=n+\epsilon$ with $|\epsilon|\ll 1$, the correction to this is of
order $\cO(\epsilon)$:
\begin{align}
 (aq^z)_{n+\epsilon}&=(-a)^nq^{zn+{1\over 2}n(n-1)}(a^{-1}q^{1-n-z})_n(1+\cO(\epsilon)),\qquad a\neq 1.
\end{align}
Here we assumed that $a\neq 1$ and $a-1\gg \cO(\epsilon)$,

\paragraph{$\bullet$ $\boldsymbol{q}$-factorials:} For $n\in\bbZ_{\ge 0}$, the $q$-factorial is given by
\begin{align}
 [n]_q!:= [1]_q[2]_q\cdots[n]_q={(q)_n\over (1-q)^n},\qquad [0]_q!=1,
\qquad
 [n+1]_q!=[n]_q[n-1]_q!~.
\end{align}
\paragraph{$\bullet$ $\boldsymbol{q}$-Gamma function:} For $z\in\bbC$, the $q$-Gamma function $\Gamma_q(z)$ is defined by
\begin{align}
 \Gamma_q(z+1)&:= (1-q)^{-z}\prod_{k=1}^\infty {1-q^{k}\over
 1-q^{z+k}}.
\end{align}
The $q$-Gamma function satisfies the following relations:
\begin{align}
 \Gamma_q(z)&
 =(1-q)^{1-z}{(q)_\infty\over (q^z)_\infty}
 =(1-q)^{1-z}(q)_{z-1},\\
 \Gamma_q(z+1)&=[z]_q\Gamma_q(z),\\
 \Gamma_q(1)&=\Gamma_q(2)=1,\qquad
 \Gamma_q(n)=[n-1]_q! \quad (n\ge 1).
\end{align}
The behavior of $\Gamma_{q}(z)$ near non-positive integers is
\begin{align}
 \Gamma_q(-n+\epsilon)
 &={(-1)^{n+1}(1-q)q^{\half n(n+1)}\over \Gamma_q(n+1)\,\log q}
 {1\over \epsilon}+\cdots,\qquad
  \Gamma_q(n+1)=[n]_q!~,\quad 
\end{align}
where $n\in\bbZ_{\ge 0}$, and $\epsilon\to 0$.   As $q\to 1$, this reduces to the formula for the
ordinary $\Gamma(z)$,
\begin{align}
 \Gamma(-n+\epsilon)&={(-1)^{n}\over \Gamma(n+1)}{1\over \epsilon}+\cdots,\qquad
 \Gamma(n+1)=n!~.
\end{align}

\paragraph{$\bullet$ $\boldsymbol{q}$-Barnes $\boldsymbol{G}$ function:}
For $z\in\bbC$, the $q$-Barnes $G$ function is defined by \cite{Nishizawa}
\begin{align}
 G_2(z+1;q)&:=(1-q)^{-{1\over 2}z(z-1)}
 \prod_{k=1}^\infty 
 \biggl[\biggl({1-q^{z+k}\over 1-q^k}\biggr)^k(1-q^k)^z
 \biggr].
\end{align}
Some of its properties are
\begin{gather}
 G_2(1;q)=1,\qquad G_2(z+1;q)=\Gamma_q(z)G_2(z),\\
 G_2(n;q)
 =\prod_{k=1}^{n-1}\Gamma_q(k)
 =\prod_{k=1}^{n-2} [k]_q!
 =(1-q)^{-\half(n-1)(n-2)}\prod_{j=1}^{n-2}(q)_j
 =\prod_{k=1}^{n-2}[k]_q^{n-k-1},\\
 \prod_{1\le A<B\le n}(q^A-q^B)=
 q^{{1\over 6}n(n^2-1)}(1-q)^{\half n(n-1)}G_2(n+1;q).
 \label{ehwt6Nov12}
\end{gather}
The behavior of $G_2(z;q)$ near non-positive integers is
\begin{align}
 G_2(-n+\epsilon;q)
 ={(-1)^{\half (n+1)(n+2)}G_2(n+2;q)\,(\log q)^{n+1}
 \over q^{{1\over 6}n(n+1)(n+2)} (1-q)^{n+1}}\epsilon^{n+1}+\cdots,
 \label{ih4Nov12}
\end{align}
where $n\in \bbZ_{\ge 0}$, and $\epsilon\to 0$. As $q\to 1$, this
reduces to the formula for the ordinary $G_2(z)$,
\begin{align}
 G_2(-n+\epsilon)
 &=(-1)^{\half n(n+1)}G_2(n+2)\epsilon^{n+1}+\cdots.
 \label{il4Nov12}
\end{align}
We note that the Vandermonde determinant can be expressed by the $q$-Barnes $G$-function 
\begin{align}
\Delta(N):=\prod_{1\le A< B\le N}\left(q^A-q^B\right)=q^{{1\over 6}N(N^2-1)}(1-q)^{\half N(N-1)}G_2(N+1;q)
\label{qBarnesVande}
\end{align}
that follows from
\be
{\Delta(N+1)\over\Delta(N)}=q^{\half N(N+1)}\prod_{l=1}^N(1-q^l)=q^{\half N(N+1)}(1-q)^N\Gamma_q(N+1)\ .
\ee

\section{The computational details}
\label{Details}

Some details of the calculations in the main text are given in this
appendix. In particular, we provide relevant details in the calculation
of the Wilson loops in the lens space matrix model and those of their
analytic continuation to the Wilson loops in the ABJ(M) matrix
model. The analytic continuation presented here is a streamlined version
of that given for the partition function in \cite{Awata:2012jb}.


\subsection{The calculation of the lens space Wilson loop}
\label{lensWL}

We provide computational details of the derivation of (\ref{LensSpaceWilsonLoop}). We first recall the definition of the (un-normalized) Wilson loop in the lens space matrix model:
\begin{align}
{\cal W}_{\rm lens}^{\rm I}(N_1,N_2;n)_k:=
\left\langle\sum_{j=1}^{N_1}e^{n \mu_j}\right\rangle\ ,
\label{WilsonLensapp}
\end{align}
where we have defined the expectation value of ${\cal O}$ by
\be
\langle{\cal O}\rangle :=\cN_{\rm lens}
\int\prod_{i=1}^{N_1}{d\mu_i\over 2\pi}\prod_{a=1}^{N_2}{d\nu_a\over 2\pi}
\Delta_{\rm sh}(\mu)^2\Delta_{\rm sh}(\nu)^2\Delta_{\rm ch}(\mu,\nu)^2
\,{\cal O}\,
e^{-{1\over 2g_s}\left(\sum_{i=1}^{N_1}\mu_i^2+\sum_{a=1}^{N_2}\nu_a^2\right)}\ .
\label{vevapp}
\ee
The integrals (\ref{WilsonLensapp}) are actually Gaussian and can be performed exactly. To see it, it is convenient to shift the eigenvalues as
\be
\mu_i\to \mu_i-{i\pi\over 2}\ ,\qquad\qquad
\nu_a\to \nu_a+{i\pi\over 2}\ .
\ee
These yield
\begin{align}
 \Delta_{\rm sh}(\mu)\Delta_{\rm sh}(\nu)\Delta_{\rm ch}(\mu,\nu)=
 e^{-{i\pi\over 2}N_1N_2-{N-1\over 2}(\sum_j \mu_j+\sum_a \nu_a)}\Delta(\mu,\nu)\ ,
\end{align}
where $N:= N_1+N_2$. The Vandermonde determinant $\Delta(\mu,\nu)$ takes the following form and can be expanded as
\begin{align}
\Delta(\mu,\nu) 
 &:=
 \prod_{j<k} (e^{\mu_j}-e^{\mu_k})
 \prod_{a<b} (e^{\nu_a}-e^{\nu_b})
 \prod_{j,a} (e^{\mu_j}-e^{\nu_a})\notag\\
 &=\sum_{\sigma\in S_{N}} (-1)^\sigma 
 e^{ \sum_{j=1}^{N_1} (\sigma(j)-1)\mu_j
 +\sum_{a=1}^{N_2}(\sigma(N_1+a)-1)\nu_a }\ ,
 \label{Vandermonde}
\end{align}
where $S_N$ is the permutation group of length $N$ and $(-1)^\sigma$ is
the signature of an element $\sigma\in S_N$. Because each term in
\eqref{Vandermonde} is an exponential whose exponent is linear in
$\mu_j,\nu_a$, the integrals \eqref{vevapp} are all Gaussian.
To proceed, we define
\be
{\cal W}(j):=\int\prod_{i=1}^{N_1}{d\mu_i\over 2\pi}\prod_{a=1}^{N_2}{d\nu_a\over 2\pi}
\Delta_{\rm sh}(\mu)^2\Delta_{\rm sh}(\nu)^2\Delta_{\rm ch}(\mu,\nu)^2
e^{n\mu_j}
e^{-{1\over 2g_s}\left(\sum_{i=1}^{N_1}\mu_i^2+\sum_{a=1}^{N_2}\nu_a^2\right)}\ ,
\ee
so that ${\cal W}_{\rm lens}^{\rm I}(N_1,N_2;n)_k=\cN_{\rm lens}\sum_{j=1}^{N_1}{\cal W}(j)$. 
By using the expansion of the Vandermonde determinants, this becomes
\begin{align}
{\cal W}(j)&=i^{-2N_1N_2}\int\prod_{i=1}^{N_1}{d\mu_i\over 2\pi}\prod_{a=1}^{N_2}{d\nu_a\over 2\pi}
\prod_{i=1}^{N_1}\prod_{a=1}^{N_2}e^{n\left(\mu_j-{i\pi\over 2}\right)-(N-1)(\mu_i+\nu_a)-{\pi^2\over 4}N}
\sum_{\sigma, \tau\in S_N}(-1)^{\sigma+\tau}\label{Wj-1}\\
&\times e^{\sum_{A=1}^{N_1}\left(\sigma(A)+\tau(A)-2\right)\mu_A+\sum_{B=1}^{N_2}\left(\sigma(N_1+B)+\tau(N_1+B)-2\right)\nu_B-{1\over 2g_s}\left[\sum_{i=1}^{N_1}\mu_i^2+\sum_{a=1}^{N_2}\nu_a^2-i\pi\left(\sum_{i=1}^{N_1}\mu_i
-\sum_{a=1}^{N_2}\nu_a\right)\right]}\ .\nn
\end{align}
It is then straightforward to perform the Gaussian integrals and find that
\begin{align}
{\cal W}(j)&=(-1)^{N_1N_2}e^{{N\pi^2\over 8g_s}}\left({g_s\over 2\pi}\right)^{N\over 2}
\sum_{\sigma, \tau\in S_N}(-1)^{\sigma+\tau}
e^{\sum_{A=1}^{N_1}{1\over 2g_s}\left(g_s\left(\sigma(A)+\tau(A)+n\delta_{Aj}-N-1\right)+{i\pi\over 2}\right)^2}\nn\\
&\times e^{-{i\pi n\over 2}}e^{\sum_{B=N_1+1}^{N}{1\over 2g_s}\left(g_s\left(\sigma(B)+\tau(B)-N-1\right)-{i\pi\over 2}\right)^2}\nn\\
&=\left({g_s\over 2\pi}\right)^{N\over 2}e^{-{g_s\over 6}N(N+1)(N+2)
+g_s\left({1\over 2}n^2-n-nN\right)}\nn\\
&\times  \sum_{\sigma, \tau\in S_N}(-1)^{\sigma+\tau} e^{g_s\sum_{A=1}^N\sigma(A)\tau(A)
+i\pi\sum_{A=1}^{N_1}\left(\sigma(A)+\tau(A)\right)+g_s n\left(\sigma(j)+\tau(j)\right)}\ ,
\label{Wj0}
\end{align}
where in the second equality we used $i^{-2N_1N_2}e^{-{i\pi\over 2}(N+1)(N_1-N_2+N)}=1$ and 
\begin{align}
\sum_{A=1}^N\sigma(A)=\sum_{A=1}^N\tau(A)=\half N(N+1)\ ,\quad
\sum_{A=1}^N\sigma(A)^2=\sum_{A=1}^N\tau(A)^2={1\over 6}N(N+1)(2N+1)\ .\nn
\end{align}
Notice that the sum over the permutation $\tau$ in (\ref{Wj0}) is the determinant of the Vandermonde matrix
\be
\left(x_A\right)^B:=
\left\{
\begin{array}{cl}
\left(-q^{\sigma(A)+n\delta_{Aj}}\right)^{B} & (A=1,\cdots,j,\cdots, N_1) \\
\left(q^{-\sigma(A)}\right)^{B} & (A=N_1+1,\cdots, N)
\end{array}
\right.\ ,
\ee
where we introduced $q= e^{-g_s}$. Thus (\ref{Wj0}) yields
\begin{align}
{\cal W}(j)&=q^{-{n^2\over 2}+n}\left({g_s\over 2\pi}\right)^{N\over 2}q^{-{1\over 3}N(N^2-1)}
\sum_{\sigma\in S_N}(-1)^{\sigma+\sum_{A=1}^{N_1}\sigma(A)}q^{-n\sigma(j)}\nn\\
&\times  (-1)^{N_1(N_1+1)\over 2}
\prod_{1\le A<B\le N_1}\left(q^{\sigma(A)+n\delta_{Aj}}-q^{\sigma(B)+n\delta_{Bj}}\right)
\label{Wj1}\\
&\times\prod_{N_1+1\le A<B\le N}\left(q^{\sigma(A)}-q^{\sigma(B)}\right)
\prod_{A=1}^{N_1}\prod_{B=N_1+1}^{N}\left(q^{\sigma(B)}+q^{\sigma(A)+n\delta_{Aj}}\right)\ .\nn
\end{align}
We now rewrite the sum over the permutation $\sigma$ as the sum over the ways of partitioning $N$ numbers $\{1,2,\cdots, N\}$ into two groups of ordered numbers $\{\{C_1,C_2,\cdots, C_{N_1}\},\{D_1,D_2,\cdots, D_{N_2}\}\}$ where $C_1\le C_2\le\cdots\le C_{N_1}$ and $D_1\le D_2\le\cdots\le D_{N_2}$. 
In rewriting, we start with the sequence of numbers $\{\sigma(1),\sigma(2),\cdots,\sigma(N)\}$ and then reorder it into 
\be
\{D_1,\cdots,D_{a_1}, C_1,D_{a_1+1}, \cdots, D_{a_2},C_2,D_{a_2+1},\cdots,D_{a_{N_1}},C_{N_1},D_{a_{N_1}+1},\cdots, D_{N_2}\}
\ee
that is just a way of expressing $\{1,2,\cdots, N\}$.
This obviously yields the sign $(-1)^{\sigma}$. We further reorder it into the partition
\be
\{C_1,\cdots, C_{N_1},D_1,\cdots, D_{N_2}\}\ .
\ee
We can find the sign picked up by this reordering as follows. 
We first move $C_1$ farthest to the left. This gives the sign $(-1)^{C_1-1}$. We next move $C_2$ to the next to and the right of $C_1$. This gives the sign $(-1)^{C_2-2}$. In repeating this process, the move of $C_i$ picks up the sign $(-1)^{C_i-i}$. Thus the sign picked up in the end of all the moves is 
\be
(-1)^{\sigma}(-1)^{\sum_{i=1}^{N_1}C_i-\half N_1(N_1+1)}=(-1)^{\sigma+\sum_{A=1}^{N_1}\sigma(A)-\half N_1(N_1+1)}\ .
\ee
This is exactly the same sign factor as in (\ref{Wj1}) and thus canceled out. Hence we obtain
\begin{align}
{\cal W}(j)&=(N_1-1)!N_2!\,q^{-{n^2\over 2}+n}\left({g_s\over 2\pi}\right)^{N\over 2}q^{-{1\over 3}N(N^2-1)}
\sum_{l=1}^{N_1}\sum_{\{N_1,N_2\}}q^{-nC_l}\label{Wilson16II}\\
&\times  
\prod_{1\le i<k\le N_1}\left(q^{C_i+n\delta_{il}}-q^{C_k+n\delta_{kl}}\right)
\prod_{N_1+1\le a<b\le N}\left(q^{D_a}-q^{D_b}\right)
\prod_{i=1}^{N_1}\prod_{a=N_1+1}^{N}\left(q^{C_i+n\delta_{il}}+q^{D_a}\right)\ ,\nn
\end{align}
where $\{N_1,N_2\}$ is the partition that we have just discussed. The factor $(N_1-1)!N_2!$ arises for a fixed $l$ since there are so many numbers of the sequence that yield a given partition. 
Note that, as anticipated from the definition of the Wilson loop, this is $j$-independent and thus yields 
\begin{align}
{\cal W}_{\rm lens}^{\rm I}(N_1,N_2;n)_k&=i^{-{\kappa\over 2}(N_1^2+N_2^2)}\,q^{-{n^2\over 2}+n}\left({g_s\over 2\pi}\right)^{N\over 2}q^{-{1\over 3}N(N^2-1)}
\sum_{l=1}^{N_1}\sum_{\{N_1,N_2\}}q^{-nC_l}\label{Wilson16II}\\
&\times  
\prod_{1\le i<k\le N_1}\left(q^{C_i+n\delta_{il}}-q^{C_k+n\delta_{kl}}\right)
\prod_{N_1+1\le a<b\le N}\left(q^{D_a}-q^{D_b}\right)
\prod_{i=1}^{N_1}\prod_{a=N_1+1}^{N}\left(q^{C_i+n\delta_{il}}+q^{D_a}\right)\ .\nn
\end{align}
This agrees with the partition function in \cite{Awata:2012jb} (up to the difference in normalizations) when the winding $n=0$.
As we have done so for the partition function, using (\ref{qBarnesVande}), we further rewrite the Wilson loop (\ref{Wilson16II}) as a product of the $q$-Barnes $G$-function and a ``generalization of multiple $q$-hypergeometric function'':
\begin{align}
{\cal W}_{\rm lens}^{\rm I}(N_1,N_2;n)_k&=i^{-{\kappa\over 2}(N_1^2+N_2^2)}
q^{-{n^2\over 2}+n}
\left({g_s\over 2\pi}\right)^{N\over 2}q^{-{N(N^2-1)\over 6}}\nn\\
&\qquad\times (1-q)^{N(N-1)\over 2}G_2(N+1;q)S(N_1,N_2;n)_k\ ,\label{lensWilson}
\end{align}
where the special function $S_n(N_1,N_2)$ is defined by
\begin{align}
S(N_1,N_2;n)_k&=\sum_{l=1}^{N_1}\sum_{\{N_1,N_2\}}q^{-nC_l}\prod_{l<k\le N_1}
{q^{C_l+n}-q^{C_k}\over q^{C_l}-q^{C_k}}
\prod_{1\le i<l}{q^{C_i}-q^{C_l+n}\over q^{C_i}-q^{C_l}}\nn\\
&\times \prod_{C_i<D_a}{q^{C_i+n\delta_{il}}+q^{D_a}\over q^{C_i}-q^{D_a}}
\prod_{C_i>D_a}{q^{D_a}+q^{C_i+n\delta_{il}}\over q^{D_a}-q^{C_i}}\ .
\label{Sndef}
\end{align}
We now define the normalized Wilson loop by $W_{\rm lens}^{\rm I}(N_1,N_2;n)_k:={{\cal W}_{\rm lens}^{\rm I}(N_1,N_2;n)_k\over{\cal W}_{\rm lens}^{\rm I}(N_1,N_2;0)_k}$ that takes the following simpler form
\begin{align}
W_{\rm lens}^{\rm I}(N_1,N_2;n)_k=q^{-{n^2\over 2}+n}{S(N_1,N_2;n)_k\over S(N_1,N_2;0)_k}\ .
\end{align}
In the rest of this appendix we massage (\ref{Sndef}) into a more convenient form: Notice first that once the set of $C_i$'s is selected out of the numbers $\{1,2,\cdots, N\}$, $D_a$'s simply fill in the rest of the numbers. Taking this fact into account, we can rewrite the special function (\ref{Sndef}) as
\begin{align}
S(N_1,N_2;n)_k&=\sum_{l=1}^{N_1}\sum_{1\le C_1<\cdots <C_{N_1}\le N}q^{-nC_l}\prod_{l<k\le N_1}
{q^{C_l+n}-q^{C_k}\over q^{C_l}-q^{C_k}}
\prod_{1\le i<l}{q^{C_i}-q^{C_l+n}\over q^{C_i}-q^{C_l}}\nn\\
&\quad\times \prod_{1\le a<C_1}\prod_{j=1}^{N_1}{q^{a}+q^{C_j+n\delta_{jl}}\over q^{a}-q^{C_j}}
\prod_{i=1}^{N_1-1}\left(\prod_{C_i<a<C_{i+1}}\prod_{j=1}^{i}{q^{C_j+n\delta_{jl}}+q^{a}\over q^{C_j}-q^{a}}\right)\nn\\
&\quad\times\left(\prod_{C_i<a<C_{i+1}}\prod_{j=i+1}^{N_1}{q^{a}+q^{C_{j}+n\delta_{jl}}\over q^{a}-q^{C_{j}}}\right)
\prod_{C_{N_1}< a\le N}\prod_{j=1}^{N_1}{q^{C_j+n\delta_{jl}}+q^{a}\over q^{C_j}-q^{a}}
\label{SnRW1}\\
&=\sum_{l=1}^{N_1}\sum_{1\le C_1<\cdots <C_{N_1}\le N}q^{-2nC_l+n(N+l-1)}\prod_{k\ne l}
{1-q^{C_k-C_{l}-n}\over 1-q^{C_k-C_{l}}}\nn\\
&\quad\times\prod_{i=1}^{N_1}\left(\prod_{j=i}^{N_1-1}{\left(-q^{C_j-C_i-n\delta_{il}+1}\right)_{C_{j+1}-C_j-1}
\over \left(q^{C_j-C_i+1}\right)_{C_{j+1}-C_j-1}}
\prod_{j=1}^{i-1}{\left(-q^{C_i-C_{j+1}+n\delta_{il}+1}\right)_{C_{j+1}-C_j-1}
\over \left(q^{C_i-C_{j+1}+1}\right)_{C_{j+1}-C_j-1}}\right)\nn\\
&\quad\times \prod_{i=1}^{N_1}\left({\left(-q^{C_i-C_1+n\delta_{il}+1}\right)_{C_1-1}\left(-q^{C_{N_1}-C_i-n\delta_{il}+1}\right)_{N-C_{N_1}}
\over \left(q^{C_i-C_1+1}\right)_{C_1-1}\left(q^{C_{N_1}-C_i+1}\right)_{N-C_{N_1}}}\right)\ .\label{SnRW2}
\end{align}
In going from (\ref{SnRW1}) to (\ref{SnRW2}), we passed the expression
\begin{align}
S(N_1,N_2;n)_k&=\sum_{l=1}^{N_1}\sum_{1\le C_1<\cdots <C_{N_1}\le N}q^{-nC_l}\prod_{k=l+1}^{N_1}
{1-q^{C_l-C_{k}+n}\over 1-q^{C_l-C_{k}}}
\prod_{k=1}^{l-1}{1-q^{C_l-C_k+n}\over 1-q^{C_l-C_k}}\nn\\
&\times\prod_{i=1}^{N_1}\left\{\left(\prod_{j=i}^{N_1-1}\prod_{b=1}^{C_{j+1}-C_j-1}{q^{C_i-C_j-b+n\delta_{il}}+1\over q^{C_i-C_j-b}-1}\right)
\left(\prod_{j=1}^{i-1}\prod_{b=1}^{C_{j+1}-C_j-1}{1+q^{C_i-C_{j+1}+b+n\delta_{il}}\over 1-q^{C_i-C_{j+1}+b}}\right)\right\}\nn\\
&\times \prod_{j=1}^{N_1}\left(\prod_{b=1}^{C_1-1}{1+q^{C_j-C_1+b+n\delta_{jl}}\over 1-q^{C_j-C_1+b}}
\prod_{b=1}^{N-C_{N_1}}{q^{C_j-C_{N_1}-b+n\delta_{jl}}+1\over q^{C_j-C_{N_1}-b}-1}\right)\ .
\end{align}
Note that we can extend the range of sums from $1\le C_1<\cdots <C_{N_1}\le N$ to the semi-infinite one $1\le C_1<\cdots <C_{N_1}$, since, by using $\left(q^{\alpha}\right)_{-m}=1/\left(q^{\alpha-m}\right)_m$, the factor $1/\left(q^{C_{N_1}-C_j+1}\right)_{N-C_{N_1}}=\left(q^{N-C_j+1}\right)_{C_{N_1}-N}$ for $C_{N_1}>N$ vanishes when $j=N_1$.
By repeatedly applying the formula $(a)_n=(a)_m\left(a q^m\right)_{n-m}$, we can simplify (\ref{SnRW2}) to (\ref{LensSpaceSfunction})
\begin{align}
S(N_1,N_2;n)_k&={1\over N_1!}\sum_{l=1}^{N_1}\sum_{1\le C_1,\cdots, C_{N_1}}q^{-2nC_l+n(N+l-1)}\prod_{k\ne l}
{1-q^{C_k-C_{l}-n}\over 1-q^{C_k-C_{l}}}\label{SnSimple}\\
&\times\prod_{i=1}^{N_1}\left[{\left(-q^{1+n\delta_{il}}\right)_{C_i-1}\left(-q^{1-n\delta_{il}}\right)_{N-C_i}\over(q)_{C_i-1}(q)_{N-C_i}}
\prod_{j=1}^{i-1}{\left(q^{C_i-C_j}\right)_{1}\over\left(-q^{C_i-C_j+n\delta_{il}}\right)_{1}}
\prod_{j=i+1}^{N_1}{\left(q^{C_j-C_i}\right)_{1}\over\left(-q^{C_j-C_i-n\delta_{il}}\right)_{1}}\right]\ .\nn
\end{align}
We note that the summation over $C_i$'s is originally ordered, $\sum_{1\le C_1<\cdots <C_{N_1}}$.
However, the summand in (\ref{SnSimple}) is invariant under permutation of $C_i$'s, as the marked index $l$ is summed over, we have replaced $\sum_{1\le C_1<\cdots <C_{N_1}}$ by the unordered sum ${1\over N_1!} \sum_{1\le C_1,\cdots, C_{N_1}}$.


\subsection{Details of the analytic continuation}
\label{analcontdetail}

Having derived the lens space result, in particular, (\ref{SnSimple}), 
here we provide the details of the analytic continuation, $N_2\to -N_2$, to obtain the ABJ result:
\be
S(N_1,N_2;n)_k\quad \stackrel{N_2\to -N_2}{\longrightarrow}\quad S^{\rm ABJ}(N_1,N_2;n)_k\ .
\ee
What we are going to do is simply to replace $N_2$ in (\ref{SnSimple}) by $-N_2+\epsilon$ and take the $\epsilon\to 0$ limit.
The basic formula to use is 
\begin{align}
(q)_{-|z|+\epsilon}&=\prod_{k=0}^{\infty}{1-q^{k+1}\over 1-q^{\epsilon-|z|+k+1}}
=-{1\over \epsilon \ln q}\prod_{k=0}^{|z|-2}{1\over 1-q^{k-|z|+1}}+{\cal O}\left(\epsilon^0\right)\nn\\
&=-{1\over \epsilon \ln q \left(q^{1-|z|}\right)_{|z|-1}}+{\cal O}\left(\epsilon^0\right)
=-{(-1)^{|z|-1}q^{\half|z|\left(|z|-1\right)}\over \epsilon \ln q (q)_{|z|-1}}+{\cal O}\left(\epsilon^0\right)\ ,
\label{analcontformula}
\end{align}
where we used (\ref{maux20Sep12}) in the first equality. The nontrivial part is the denominator in the factor 
\be
\prod_{i=1}^{N_1}{\left(-q^{1+n\delta_{il}}\right)_{C_i-1}\left(-q^{1-n\delta_{il}}\right)_{N-C_i}\over(q)_{C_i-1}(q)_{N-C_i}}
\label{nontrivialfactor}
\ee
in the second line of (\ref{SnSimple}). However, we need to treat the two cases $N_1\le N_2$ and $N_1\ge N_2$ separately, and the latter turns out to be more involved than the former.

\subsubsection{The $N_1\le N_2$ case}

In this case it is straightforward to apply the formula (\ref{analcontformula}). The denominator of (\ref{nontrivialfactor}) yields 
\be
(q)_{C_i-1}(q)_{N-C_i}\,\,\to\,\,
(q)_{C_i-1}(q)_{-M-C_i+\epsilon}
=-{(-1)^{M+C_i-1}q^{\half\left(M+C_i\right)\left(M+C_i-1\right)}\over\epsilon\ln q\left(q^{C_i}\right)_{M}}\ ,
\ee
where $N=N_1+N_2$ and $M:= |N_2-N_1|=N_2-N_1$ and we used $(q)_{C_i-1}/(q)_{M+C_i-1}=1/(q^{C_i})_M$.
Meanwhile, we can rewrite the numerator of (\ref{nontrivialfactor}) after the analytic continuation  as
\begin{align}
\prod_{i=1}^{N_1}{\left(-q^{1+n\delta_{il}}\right)_{C_i-1}\left(-q^{1-n\delta_{il}}\right)_{-M-C_i}}
&=\prod_{i=1}^{N_1}{q^{\left(C_i+M\right)n\delta_{il}+\half\left(C_i+M\right)\left(C_i+M-1\right)}\over 
\left(1+q^{n\delta_{il}}\right)\left(-q^{C_i+n\delta_{il}}\right)_{M}}\ .
\end{align}
Putting these together yields
\begin{align}
\prod_{i=1}^{N_1}{\left(-q^{1+n\delta_{il}}\right)_{C_i-1}\left(-q^{1-n\delta_{il}}\right)_{N-C_i}\over(q)_{C_i-1}(q)_{N-C_i}}
&\,\,\to\,\, \left(\epsilon\ln q\right)^{N_1}\prod_{i=1}^{N_1}(-1)^{C_i+M}{q^{\left(C_i+M\right)n\delta_{il}}\left(q^{C_i}\right)_M
\over\left(1+q^{n\delta_{il}}\right)\left(-q^{C_i+n\delta_{il}}\right)_{M}}\ .
\label{nontrivialfactorcntd}
\end{align}
We thus find the analytic continuation (\ref{N1lessN2analcont})
\begin{align}
S(N_1,-N_2+\epsilon;n)_k&\sim\left(\epsilon\ln q\right)^{N_1}S^{\rm ABJ}(N_1,N_2;n)_k\ ,
\end{align}
with the special function (\ref{ABJSfunction1}) for the ABJ theory
\begin{align}
&S^{\rm ABJ}(N_1,N_2;n)_k={1\over N_1!}\sum_{l=1}^{N_1}\sum_{1\le C_1,\cdots,C_{N_1}}q^{-2nC_l+n(-M+l-1)}\prod_{k\ne l}
{1-q^{C_k-C_{l}-n}\over 1-q^{C_k-C_{l}}}\label{Soriginal}\\
&\qquad\times\prod_{i=1}^{N_1}\left[(-1)^{C_i+M}{q^{\left(C_i+M\right)n\delta_{il}}\left(q^{C_i}\right)_M
\over\left(1+q^{n\delta_{il}}\right)\left(-q^{C_i+n\delta_{il}}\right)_{M}}\prod_{j=1}^{i-1}{\left(q^{C_i-C_j}\right)_{1}\over\left(-q^{C_i-C_j+n\delta_{il}}\right)_{1}}
\prod_{j=i+1}^{N_1}{\left(q^{C_j-C_i}\right)_{1}\over\left(-q^{C_j-C_i-n\delta_{il}}\right)_{1}}\right]\ .\nn
\end{align}

\subsubsection{The $N_1\ge N_2$ case}

As mentioned, this case is more involved than the previous case. The major difference stems from the fact that the factor $(q)_{N-C_i}$ in the denominator of (\ref{nontrivialfactor}), when analytically continued to $(q)_{N_1-N_2-C_i+\epsilon}$, has the index $N_1-N_2-C_i+\epsilon=M-C_i+\epsilon$ that is not always negative in contrast to the previous case. This index becomes negative when $C_i> M$. When the index is negative, the factor $(q)_{M-C_i+\epsilon}$ is singular and of order $\epsilon^{-1}$ as in (\ref{analcontformula}). This means that the factor (\ref{nontrivialfactor}), when analytically continued, vanishes with some power of $\epsilon$. 
For the purpose of the analytic continuation, we are only concerned with the leading vanishing term.
In the previous case the leading vanishing term was of order $\epsilon^{N_1}$ as in (\ref{nontrivialfactorcntd}).
 
\medskip
To extract the leading vanishing term of the factor (\ref{nontrivialfactor}), we need to find in which case the factor $\prod_{i=1}^{N_1}(q)_{M-C_i+\epsilon}$ is least singular.
Now the summand in (\ref{analcontformula}) is identically zero whenever any of $C_i$'s coincide. In other words, it is only nonzero, when all of $C_i$'s are different from each other. Thus we can focus on the case where none of $C_i$'s are equal. Clearly, the factor $\prod_{i=1}^{N_1}(q)_{M-C_i+\epsilon}$ is least singular when $M$ of $C_i$'s, being all different, take values in $\{1,\cdots, M\}$. Since the order does not matter, without loss of generality, we can choose $\{C_1, C_2, \cdots, C_M\}=\{1,2,\cdots, M\}$ by taking into account the combinatorial factor $\mbox{}_{N_1}\!C_{M}M!=N_1!/N_2!$. This also implies that the leading vanishing term is of order $\epsilon^{N_1-M}=\epsilon^{N_2}$.

\medskip
Having understood this point, the remaining task is to (1) plug $\{C_1, C_2, \cdots, C_M\}=\{1,2,\cdots, M\}$ into (\ref{SnSimple}) and (2)  apply the formula (\ref{analcontformula}) to the factors $(q)_{M-C_i+\epsilon}$ with $C_{i\ge M+1}\ge M+1$, while taking care of the sum over $l$ by splitting the sum $\sum_{l=1}^{N_1}$ into $\sum_{l=1}^M+\sum_{l=M+1}^{N_1}$.

\medskip
We first deal with the factors
\be
\prod_{i=1}^{N_1}\left(
\prod_{j=1}^{i-1}{\left(q^{C_i-C_j}\right)_{1}\over\left(-q^{C_i-C_j+n\delta_{il}}\right)_{1}}
\prod_{j=i+1}^{N_1}{\left(q^{C_j-C_i}\right)_{1}\over\left(-q^{C_j-C_i-n\delta_{il}}\right)_{1}}\right)
\label{facN1geN2}
\ee
in the second line of (\ref{SnSimple}). The factors with all the indices less than or equal to $M$ yield  
\begin{align}
\prod_{i=1}^{M}\left(
\prod_{j=1}^{i-1}{\left(q^{i-j}\right)_{1}\over\left(-q^{i-j+n\delta_{il}}\right)_{1}}
\prod_{j=i+1}^{M}{\left(q^{j-i}\right)_{1}\over\left(-q^{j-i-n\delta_{il}}\right)_{1}}\right)
&=\prod_{i=1}^{M}{(q)_{i-1}(q)_{M-i}\over\left(-q^{1+n\delta_{il}}\right)_{i-1}\left(-q^{1-n\delta_{il}}\right)_{M-i}}\ .
\label{N1geN2factor1}
\end{align}
The factors with one of the indices less than or equal to $M$ and the other one greater than $M$ yield
\bea
&\hspace{-5cm}\left(\prod_{i=M+1}^{N_1}\prod_{j=1}^{M}{\left(q^{C_i-j}\right)_{1}\over\left(-q^{C_i-j+n\delta_{il}}\right)_{1}}\right)
\left(\prod_{i=1}^M\prod_{j=M+1}^{N_1}{\left(q^{C_j-i}\right)_{1}\over\left(-q^{C_j-i-n\delta_{il}}\right)_{1}}\right)\nn\\
\hspace{3cm}&=\left\{\begin{array}{ll}
\prod_{a=1}^{N_2}{\left(q^{D_a}\right)^2_{M}\left(-q^{D_a+c}\right)_{1}\over\left(-q^{D_a}\right)^2_{M}\left(-q^{D_a+c-n}\right)_{1}}
& (l\le M)\\
\prod_{a=1}^{N_2}{\left(q^{D_a}\right)^2_{M}\over\left(-q^{D_a+n\delta_{ad}}\right)_{M}\left(-q^{D_a}\right)_{M}}
& (l> M)
\end{array}
\right.
\eea
where we relabeled $D_a=C_{a+M}-M$ with $a=1,\cdots, N_2$ and $c=M-l$ for $l\le M$ and $d=l-M$ for $l> M$.
The factors with all the indices greater than $M$ are trivially of a similar form as (\ref{facN1geN2}).
We next examine the factor in the first line of (\ref{SnSimple})
\bea
\prod_{k\ne l}{1-q^{C_k-C_l-n}\over 1-q^{C_k-C_l}}
&=&\left\{
\begin{array}{ll}
\prod_{\substack{k=1 \\ k\ne l}}^{M}{1-q^{k-l-n}\over 1-q^{k-l}}\prod_{k=M+1}^{N_1}{1-q^{C_k-l-n}\over 1-q^{C_k-l}}
& (l\le M)\\
\prod_{k=1}^{M}{1-q^{k-C_l-n}\over 1-q^{k-C_l}}\prod_{\substack{k=M+1 \\ k\ne l}}^{N_1}{1-q^{C_k-C_l-n}\over 1-q^{C_k-C_l}} 
& (l>M)
\end{array}
\right.\nn\\
&=&\left\{
\begin{array}{ll}
q^{-n(M-1-c)}{(q^{1-n})_c(q^{1+n})_{M-1-c}\over (q)_c(q)_{M-1-c}}\prod_{a=1}^{N_2}{1-q^{D_a+c-n}\over 1-q^{D_a+c}}
& (l\le M)\\
q^{-nM}{(q^{D_d+n})_M\over (q^{D_d})_M}\prod_{\substack{a=1\\ a\ne d}}^{N_2}{1-q^{D_a-D_{d}-n}\over 1-q^{D_a-D_{d}}} 
& (l>M)
\end{array}
\right.\ .
\eea
Finally, we look into the factor (\ref{nontrivialfactor}).
The factors with $i\le M$ are exactly the inverse of (\ref{N1geN2factor1}) and canceled out, whereas the analytic continuation of the rest of the factors yields
\begin{align}
\prod_{i=M+1}^{N_1}{\left(-q^{1+n\delta_{il}}\right)_{C_i-1}\left(-q^{1-n\delta_{il}}\right)_{M+\epsilon-C_i}\over(q)_{C_i-1}(q)_{M+\epsilon-C_i}}
\sim\left(\epsilon\ln q\right)^{N_2}\prod_{a=1}^{N_2}(-1)^{D_a}
{q^{nD_a\delta_{a+M,l}}\left(-q^{D_a+n\delta_{a+M, l}}\right)_M\over \left(1+q^{n\delta_{a+M, l}}\right)\left(q^{D_a}\right)_M}\ .
\end{align}
Putting all factors together, we find the analytic continuation (\ref{N1geN2Sn})
\be
S(N_1,-N_2+\epsilon;n)_k\sim \left(\epsilon\ln q\right)^{N_2}\left(S^{\rm ABJ}_{(1)}(N_1,N_2;n)_k+S^{\rm ABJ}_{(2)}(N_1,N_2;n)_k\right)\ ,
\ee
where the special functions for the ABJ theory are given by
\begin{align}
S^{\rm ABJ}_{(1)}(N_1,N_2;n)_k&={1\over N_2!}\sum_{c=0}^{M-1}\sum_{1\le D_1,\cdots ,D_{N_2}}
q^{n(2c-M)}{\left(q^{1-n}\right)_c\left(q^{1+n}\right)_{M-1-c}\over (q)_c(q)_{M-1-c}}
\prod_{a=1}^{N_2}{1-q^{D_a+c-n}\over 1-q^{D_a+c}}\label{S1original}\\
&\times \prod_{a=1}^{N_2}\left[{(-1)^{D_a}\left(q^{D_a}\right)_M\left(-q^{D_a+c}\right)_M
\over 2\left(-q^{D_a}\right)_{M}\left(-q^{D_a+c-n}\right)_M}\prod_{b=1}^{a-1}{\left(q^{D_a-D_b}\right)_{1}
\over\left(-q^{D_a-D_b}\right)_{1}}
\prod_{b=a+1}^{N_2}{\left(q^{D_b-D_a}\right)_{1}\over\left(-q^{D_b-D_a}\right)_{1}}\right]\nn
\end{align}
and 
\begin{align}
S^{\rm ABJ}_{(2)}(N_1,N_2;n)_k&={1\over N_2!}
\sum_{d=1}^{N_2}\sum_{1\le D_1,\cdots ,D_{N_2}}q^{-nD_d+n(d-M-1)}
\prod_{\stackrel{a=1}{a\ne d}}^{N_2}{\left(q^{D_a-D_d-n}\right)_1\over \left(q^{D_a-D_d}\right)_1} 
\label{S2original}\\
&\times\prod_{a=1}^{N_2}\left[{(-1)^{D_a}\left(q^{D_a+n\delta_{ad}}\right)_M
\over\left(1+q^{n\delta_{ad}}\right)\left(-q^{D_a}\right)_{M}}\prod_{b=1}^{a-1}{\left(q^{D_a-D_b}\right)_{1}
\over\left(-q^{D_a-D_b+n\delta_{ad}}\right)_{1}}
\prod_{b=a+1}^{N_2}{\left(q^{D_b-D_a}\right)_{1}\over\left(-q^{D_b-D_a-n\delta_{ad}}\right)_{1}}\right]\ .\nn
\end{align}

\section{Wilson loops in general representations}
\label{WL_gen_rep}

In this appendix, we present the expressions for Wilson loops in general
representations.  We will be very brief in explaining how to derive
these results, because it is similar to the one for partition function
(Ref.~\cite{Awata:2012jb}) and for Wilson loops with winding number
$n$ (Appendix \ref{Details}).

\subsection{Lens space Wilson loops}

Let us start with Wilson loops in the $U(N_1)\times U(N_2)$ lens space
matrix model.  In \eqref{WilsonLens}, we considered representations with
winding number $n$, but here we would like to consider general
representations.
For the $U(N_1)$ representation with Young diagram $\lambda$, the Wilson
loop can be computed by inserting
$S_\lambda(e^{\mu_1},\dots,e^{\mu_{N_1}})$ in the matrix integral, where
$S_\lambda(x_1,\dots,x_{N_1})$ is the Schur polynomial for $\lambda$
\cite{Drukker:2009hy}.  Because each term in the polynomial
$S_\lambda(e^{\mu_1},\dots,e^{\mu_{N_1}})$ has the form $e^{\sum_j m_j
\mu_j}$ with $m_j\in \bbZ$, all we have to compute in principle is the
matrix integral with $e^{\sum_j m_j \mu_j}$ inserted.  If the Wilson
line carries a non-trivial representation for $U(N_2)$, we must insert a
similar factor also for $\nu_a$.

Therefore, given $\{m_j\}_{j=1}^{N_1},\{n_a\}_{a=1}^{N_2}$, the object
of our interest here is the following matrix integral:
\begin{align}
 &\cW^{\rm lens}_{\{m\},\{n\}}(N_1,N_2)_k
 :=\ev{e^{m\cdot\mu+n\cdot \nu}}\notag\\
 &:=
 \cN_{\rm lens}
 \int \prod_{j=1}^{N_1} {d\mu_j\over 2\pi}
 \int \prod_{a=1}^{N_2} {d\nu_a\over 2\pi}
 \Delta_{\rm sh}(\mu)^2\Delta_{\rm sh}(\nu)^2\Delta_{\rm ch}(\mu,\nu)^2
 e^{-{1\over 2g_s}(\sum_j \mu_j^2+\sum_a \nu_a^2)}
 e^{m\cdot\mu+n\cdot \nu},
\end{align}
where $m\cdot\mu=\sum_j m_j \mu_j$, $n\cdot \nu=\sum_a n_a \nu_a$, with
$m_j,n_a\in\bbZ$.  By symmetry, it is clear that this is invariant under
$m_j\leftrightarrow m_l$ and $n_a\leftrightarrow n_b$.
If $m=n=0$, this reduces to partition function while,
for the case with a single winding number, e.g.,
$m_j=m\delta_{1j}$, $n_{a}=0$, this reduces to the Wilson loop studied
in the main text, \eqref{WilsonLens}, up to a factor:
\begin{align}
 \cW^{\rm lens}_{\{m,0,\dots\},\{0,\dots\}}(N_1,N_2)_k
 ={1\over N_1}\cW^{\rm I}_{\rm lens}(N_1,N_2;m)_k.
\end{align}

By carrying out the Gauss integration and doing manipulations similar to
\eqref{Wj-1}--\eqref{Sndef}, we arrive at the following simple
combinatorial expression:
\begin{align}
 \cW^{\rm lens}_{\{m\},\{n\}}(N_1,N_2)_k
 &=i^{-{\kappa\over 2}(N_1^2+N_2^2)}\,
 \left({g_s\over 2\pi}\right)^{N\over 2}
 q^{-{1\over 6}N(N^2-1)+\sum_{A=1}^N (-{P_A^2\over 2}+P_A)}
 \notag\\ 
 &\qquad\times
 (1-q)^{\half N(N-1)}
 G_2(N+1,q)
  \,S_{\{m\},\{n\}}(N_1,N_2),\label{dpq11Jun14}
 \\[2ex]
 S_{\{m\},\{n\}}(N_1,N_2)
 &:=
 \sum_{(\cN_1,\cN_2)}
 {1\over N_1!}\sum_{\sigma\in S_{N_1}}
 {1\over N_2!}\sum_{\tau\in S_{N_2}}
 q^{-\sum_{j=1}^{N_1}m_{\sigma(j)}C_j-\sum_{a=1}^{N_2}n_{\sigma(a)}D_a}
 \notag\\
 &\qquad\times
 \prod_{C_j<C_k}{q^{C_j+m_{\sigma(j)}}-q^{C_k+m_{\sigma(k)}}
 \over q^{C_j}-q^{C_k}}
 \prod_{D_a<D_b}{q^{D_a+n_{\tau(a)}}-q^{D_b+n_{\tau(b)}}
 \over q^{D_a}-q^{D_b}}
 \notag\\
 &\qquad\times
 \prod_{C_j<D_a}{q^{C_j+m_{\sigma(j)}}+q^{D_a+n_{\tau(a)}}
 \over q^{C_j}-q^{D_a}}
 \prod_{D_a<C_j}{q^{D_a+n_{\tau(a)}}+q^{C_j+m_{\sigma(j)}}
 \over q^{D_a}-q^{C_j}},\label{fsgt24May13}
\end{align}
where $N:=N_1+N_2$ and $P_A=(m_j,n_a)$, $A=1,\dots,N$.
The symbol $(\cN_1,\,\cN_2)$ denotes the partition of the numbers $(1,
2, \cdots, N)$ into two groups $\cN_1=(C_1, C_2, \cdots, C_{N_1})$ and
$\cN_2=(D_1, D_2, \cdots, D_{N_2})$ where $C_i$'s and $D_a$'s are
ordered as $C_1<\cdots<C_{N_1}$ and $D_1<\cdots<D_{N_2}$.

To proceed, let us focus on the case with $n_a=0$, namely on the
$U(N_1)$ Wilson loop henceforth.  In this case, we can rewrite the
product in $S$ in favor of the $q$-Pochhammer symbol, just as we did in
\eqref{SnRW1}--\eqref{SnSimple}. After using various formulas for the
$q$-Pochhammer symbol in Appendix \ref{qanalogs}, the final result can
be written as
\begin{align}
 &S_{\{m_j\},\{0\}}(N_1,N_2)=
 {1\over N_1!}\sum_{\sigma\in S_{N_1}}
 \prod_{j=1}^{N_1}
 {(-q^{N_1+1-j-m_{\sigma(j)}})_{N_2} \over (q^j)_{N_2}}
 {(-q^{j+m_{\sigma(j)}})_{-N_1-N_2}\over (q^j)_{-N_1-N_2}}
 \prod_{j\neq k}^{N_1}{1+q^{k-j-m_{\sigma(j)}}\over 1-q^{k-j}}
 \notag\\
 &\qquad\times
 \sum_{1\le C_1<\dots<C_{N_1}\le N_1+N_2}
 (-1)^{\sum_{j=1}^{N_1}(C_j-1)}q^{\sum_{j=1}^{N_1}(N_1+N_2-1-C_j)m_{\sigma(j)}}
 \notag\\
 &\qquad\times
 \prod_{j=1}^{N_1}
 {(q^{C_j})_{-N_1-N_2}\over (-q^{C_j+m_{\sigma(j)}})_{-N_1-N_2}}
 \prod_{j<k}^{N_1}
 {(q^{C_k-C_j+m_{\sigma(k)}-m_{\sigma(j)}})_1 (q^{C_k-C_j})_1
 \over 
 (-q^{C_k-C_j-m_{\sigma(j)}})_1  (-q^{C_k-C_j+m_{\sigma(k)}})_1 
 }
 \label{gwts28May13}
\end{align}
In the above, we treated $N_2$ as a continuous variable.  To obtain an
expression for integral $N_2$, let us shift $N_2\to N_2+\epsilon$ with
$N_2\in\bbZ$ where it is understood that $\epsilon$ will be taken to
zero at the end of computation.  By extracting powers of $\epsilon$ from
the $q$-Pochhammer symbols in the first line of \eqref{gwts28May13}, we
find, to leading order in small $\epsilon$ expansion,
\begin{align}
 &S_{\{m_j\},\{0\}}(N_1,N_2+\epsilon)=
 {(-1)^{N_1 N_2}({\epsilon \log q})^{N_1}
  \over \prod_{j=1}^{N_1}(1+q^{m_j})}  \notag\\
 &\qquad\times {1\over N_1!}
 \sum_{\sigma\in S_{N_1}}
 \sum_{1\le C_1<\dots<C_{N_1}}
 (-1)^{\sum_{j=1}^{N_1}(C_j-1)}q^{-\sum_{j=1}^{N_1}C_j m_{\sigma(j)}}
 \notag\\
 &\qquad\times
 \prod_{j=1}^{N_1}
 {(q^{C_j})_{-N_1-N_2-\epsilon}\over (-q^{C_j+m_{\sigma(j)}})_{-N_1-N_2-\epsilon}}
 \prod_{j<k}^{N_1}
 {(q^{C_k-C_j+m_{\sigma(k)}-m_{\sigma(j)}})_1 (q^{C_k-C_j})_1
 \over 
 (-q^{C_k-C_j-m_{\sigma(j)}})_1  (-q^{C_k-C_j+m_{\sigma(k)}})_1 
 }.
 \label{jwkt28May13}
\end{align}
We will drop subleading terms in small $\epsilon$ expansion henceforth.
For $N_2\in\bbZ_{>0}$, of course, the factor $\epsilon^{N_1}$ in front
of \eqref{jwkt28May13} must be canceled by the $q$-Pochhammer symbol
$(q^{C_j})_{-N_1-N_2-\epsilon}$, because the original expression
\eqref{fsgt24May13} was finite to begin with.  This $q$-Pochhammer
symbol can be rewritten as $(q^{C_j})_{-N_1-N_2-\epsilon}$
$=1/(q^{C_j-N_1-N_2-\epsilon})_{N_1+N_2+\epsilon}$ $\approx
1/(q^{C_j-N_1-N_2-\epsilon})_{N_1+N_2}$ $=
1/[(1-q^{C_j-N_1-N_2-\epsilon})\cdots (1-q^{C_j-1-\epsilon})]$, where
``$\approx$'' means up to subleading terms in powers of $\epsilon$.
Indeed, for $C_j\le N_1+N_2$, this always contains the factor
$1/(1-q^{-\epsilon})=1/(\epsilon\log q)$ and, collecting contributions
from $C_1,\dots,C_{N_1}$, we see that this completely cancels the
$\epsilon^{N_1}$.  Actually, this also means that, if $C_j>N_1+N_2$, we
have less powers of $\epsilon$ in the denominator and, as a result, the
summand vanishes as $\epsilon\to 0$.  Therefore, we are free to remove
the upper bound in the $C_j$-sum, as we have already done in
\eqref{jwkt28May13}.

In \eqref{jwkt28May13}, summing over permutations $\sigma\in
S_{N_1}$ acting on $m_j$ is the same as summing over permutations acting
on $C_j$ (it is easy to show that, if we set
$j'=\sigma(j),k'=\sigma(k)$, summing over $j,k,\sigma$ is the same as
summing over $j',k',\sigma'=\sigma^{-1}$, with $C_j,C_k$ replaced by
$C_{\sigma'(j')},C_{\sigma'(k')}$).  Therefore, we can relax the
ordering constraint on $C_j$ and forget about the summation over
permutations:
 \begin{align}
 S_{\{m_j\},\{0\}}(N_1,N_2+\epsilon)&=
 {(-1)^{N_1 N_2}({\epsilon \log q})^{N_1}
   \over \prod_{j=1}^{N_1}(1+q^{m_j})}
  \times
  {1\over N_1!}
 \sum_{C_1,\dots,C_{N_1}=1}^\infty 
 (-1)^{\sum_{j=1}^{N_1}(C_j-1)}q^{-\sum_{j=1}^{N_1}C_j m_{j}}
  \notag\\
&\qquad\times
 \prod_{j=1}^{N_1}
 {(q^{C_j})_{-N}\over (-q^{C_j+m_{j}})_{-N}}
 \prod_{j<k}^{N_1}
 {(q^{C_k-C_j+m_{k}-m_{j}})_1 (q^{C_k-C_j})_1
 \over 
 (-q^{C_k-C_j-m_{j}})_1  (-q^{C_k-C_j+m_{k}})_1 
 }.
\label{hrty24Apr14}
 \end{align}
Note that the sum over $C_1,\dots,C_{N_1}$ is completely unconstrained,
because the summand in \eqref{jwkt28May13} vanishes if $C_j=C_k$ with
$j\neq k$.  We can easily see that, upon setting $m_j=0$, this reduces
to the $S$ function for the partition function \cite{Awata:2012jb}. We
will refer to this as the unordered formula.  The formula
\eqref{jwkt28May13} is referred to as the ordered formula.

\subsection{Explicit expression for ABJ Wilson Loop ($m_j\neq 0,n_a=0$)}

The ABJ Wilson loop is obtained by setting $N_2\to -N_2'$ with
$N_2'\in\bbZ_{>0}$ in the above expressions.  More precisely, according
\cite{Awata:2012jb}, the formula for analytic continuation
is
\begin{align}
 \cW^{\rm ABJ}_{\{m\},\{0\}}(N_1,N_2')_k
 =\lim_{\epsilon\to 0}\left[(2\pi)^{-N_2'}{G_2(N_2'+1)\over G_2(-N_2'+1+\epsilon)}
 \cW^{\rm lens}_{\{m\},\{0\}}(N_1,-N_2'+\epsilon)_k\right].\label{vn11Jun14}
\end{align}

First, consider the case with $N_1\le N_2'$.  Using the unordered
expression \eqref{hrty24Apr14}, we straightforwardly obtain
\begin{multline}
 S_{\{m_j\},\{0\}}(N_1,-N_2'+\epsilon)
 =
 {(-1)^{N_1N_2'}(\epsilon \log q)^{N_1}\over  
 \prod_{j=1}^{N_1} (1+q^{m_j})}
 \times{1\over N_1!}
 \sum_{C_j\ge 1}
 (-1)^{\sum_j (C_j-1)}
 q^{-\sum_j C_j m_j}
 \\
 \times
 \prod_{j=1}^{N_1} {(q^{C_j})_{N_2'-N_1} \over (-q^{C_j+m_j})_{N_2'-N_1}}
 \prod_{j<k}^{N_1}
 {(q^{C_k-C_j+m_k-m_j})_1 (q^{C_k-C_j})_1
 \over 
 (-q^{C_k-C_j-m_j})_1 (-q^{C_k-C_j+m_k})_1}
 \label{maeb26May13}
\end{multline}
We have the $\epsilon^{N_1}$ as a prefactor, which remains uncanceled by
the $q$-Pochhammer.  It is not difficult to show that, when
$m_j=m\delta_{1j}$, this reduces to the formula \eqref{Soriginal}, up to
normalization; namely,
$S_{\{m,0,\dots\},\{0\}}(N_1,-N_2'+\epsilon)=(1/N_1)S^{\rm
ABJ}(N_1,N_2',m)_k$. The integral representation is
\begin{multline}
 S_{\{m_j\},\{0\}}(N_1,-N_2'+\epsilon)
 =
 {(-1)^{N_1N_2'}(\epsilon \log q)^{N_1}\over  \prod_j (1+q^{m_j})}
 \times {1\over N_1!}
 \left[\prod_{j=1}^{N_1} {-1\over 2\pi i}\int {\pi ds_j\over \sin(\pi s_j)}\right]
 q^{-\sum_j (s_j+1) m_j}
 \\
 \times 
 \prod_{j=1}^{N_1} {(q^{s_j+1})_M \over (-q^{s_j+1+m_j})_M}
 \prod_{j<k}^{N_1}
 {(q^{s_k-s_j+m_k-m_j})_1 (q^{s_k-s_j})_1
 \over 
 (-q^{s_k-s_j-m_j})_1 (-q^{s_k-s_j+m_k})_1},
\end{multline}
where $C_j\leftrightarrow s_j+1$ and $M:=N_2'-N_1$.
Using the formula \eqref{vn11Jun14},
the full expression including the prefactors in \eqref{dpq11Jun14} is
\begin{align}
 &\cW^{\rm ABJ}_{\{m_j\},\{0\}}(N_1,N_2')_k
 =
 i^{-{\kappa\over 2}(N_1^2+N_2'^2)}
 (-1)^{{1\over 2}N_1(N_1-1)}
 \left({g_s\over 2\pi}\right)^{N_1+N_2'\over 2}
 \notag\\ 
 &\qquad\times
 (1-q)^{\half M(M-1)}\,
 q^{\sum_{j=1}^{N_1} (-{1\over 2}\mu_j^2+\mu_j)}\,
 G_2(M+1,q)\,
 {S_{\{m_j\},\{0\}}(N_1,-N_2'+\epsilon)\over (-1)^{N_1N_2'}\left(\epsilon \ln q\right)^{N_1} }.
\end{align}

The case with $N_1>N_2'$ is more nontrivial as is the case for partition
function \cite{Awata:2012jb}.  In this case, there are some powers of
$\epsilon$ coming from
$(q^{C_j})_{-(N_1-N_2')}=(q^{C_j})_{-M'}=1/(q^{C_j-M'})_{M'}$ in the
summand, where we set $M':=N_1-N_2'>0$.  One choice to get the most
singular contribution is
\begin{align}
 \begin{split}
 C_1&=1,\quad C_2=2,\quad  \dots,\quad C_{M'}=M',\\
 C_{M'+1}&=M'+C_1',\quad  \dots, \quad C_{N_1}=M'+C_{N_2}'.
 \end{split}\label{mcry26May13}
\end{align}
For this particular choice, by extracting powers of $\epsilon$ from the
$q$-Pochhammer symbol, we can show that \eqref{hrty24Apr14} can be
rewritten as
\begin{align}
 &
 {(-\epsilon \ln q)^{N_2'}\over N_1!\prod_{j=1}^{N_2'}(1+q^{m_{M'+j}})}
 \sum_{C_j'\ge 1}
 {(-1)^{\sum_{j=1}^{N_2'} (C_j'-1)} 
 q^{\sum_{j=1}^{M'} (-2j+M')m_j-\sum_{j=1}^{N_2'}(C_j'+M')m_{M'+j}}}
 \notag\\
 &\qquad
 \times {\prod_{j=1}^{N_2'}(-q^{C_j'+m_{M'+j}})_M'\over \prod_{j=1}^{M'-1}(q)_j}
 \prod_{1\le j<k\le M'}(q^{k-j+m_k-m_j})_1
 \prod_{j=1}^{M'} \prod_{k=1}^{N_2'}
 {(q^{M'+C_k'-j+m_{M'+k}-m_j})_1\over
 (-q^{M'+C_k'-j-m_j})_1(-q^{M'+C_k'-j+m_{M'+k}})_1}
 \notag\\
 &\qquad\times
 \prod_{1\le j<k\le N_2'}{(q^{C_k'-C_j'+m_{M'+k}-m_{M'+j}})_1 (q^{C_k'-C_j'})_1
 \over (-q^{C_k'-C_j'-m_{M'+j}})_1(-q^{C_k'-C_j'+m_{M'+k}})_1}.
\end{align}
The choice \eqref{mcry26May13} is only one possibility and there are
more; there are ${N_1\choose M'}$ ways to choose $M'$ special $C_j$'s
out of $N_1$.  Furthermore, there are $M'!$ ways to permute those $M'$
special $C_j$'s.  So, in total, we have $N_1!\over N_2'!$ ways.  We
should sum over all these choices.  Or equivalently, as we have seen in
going between \eqref{jwkt28May13} and \eqref{hrty24Apr14}, we can fix
the order of $C_j$'s to be $C_1<C_2<\cdots<C_{N_1}$, consider only
\eqref{mcry26May13}, and sum over the permultations of $m_j$.  So,
\eqref{gwts28May13} can be expressed as
\begin{align}
 &S_{\{m_j\},\{0\}}(N_1,-N_2'+\epsilon)\notag\\
 &
 =
 {(-\epsilon \ln q)^{N_2'}\over N_1!\prod_{j=1}^{M'-1}(q)_j}
 \sum_{\sigma\in S_{N_1}}
 {q^{\sum_{j=1}^{M'} (-2j+M')m_{\sigma(j)}} \prod_{1\le j<k\le M'}(q^{k-j+m_{\sigma(k)}-m_{\sigma(j)}})_1
\over \prod_{j=1}^{N_2'}(1+q^{m_{\sigma(M'+j)}})}
 \notag\\
 &\qquad\times
 \sum_{1\le C_1'<\cdots<C_{N_2'}'}
 {(-1)^{\sum_{j=1}^{N_2'} (C_j'-1)} 
 q^{-\sum_{j=1}^{N_2'}(C_j'+M')m_{\sigma(M'+j)}}}
 \notag\\
 &\qquad\times 
 {\prod_{j=1}^{N_2'}(-q^{C_j'+m_{\sigma(M'+j)}})_M'}
 \prod_{j=1}^{M'} \prod_{k=1}^{N_2'}
 {(q^{M'+C_k'-j+m_{\sigma(M'+k)}-m_{\sigma(j)}})_1\over
 (-q^{M'+C_k'-j-m_{\sigma(j)}})_1(-q^{M'+C_k'-j+m_{\sigma(M'+k)}})_1
 }\notag\\
 &\qquad\times
 \prod_{1\le j<k\le N_2'}{(q^{C_k'-C_j'+m_{\sigma(M'+k)}-m_{\sigma(M'+j)}})_1 (q^{C_k'-C_j'})_1
 \over (-q^{C_k'-C_j'-m_{\sigma(M'+j)}})_1(-q^{C_k'-C_j'+m_{\sigma(M'+k)}})_1}.
 \label{kkxj4Jun13}
\end{align}
If we want to relax the ordering constraint and let $C_j'$'s run over all
positive integers, then the summation will be over $S_{N_1}/S_{N_2'}$,
meaning that two permutations $\sigma_1,\sigma_2\in S_{N_1}$ are
identified if $\{\sigma_1(M'+1),\dots,\sigma_1(N_1)_{N_2'}\}$ and
$\{\sigma_2(M'+1),\dots,\sigma_2(N_1)_{N_2'}\}$ are permutations of each other.
One can show that, when $m_j=m\delta_{1j}$, Eq.\ \eqref{kkxj4Jun13}
reduces to  \eqref{S1original} plus \eqref{S2original}, up to
normalization. Namely,
$S_{\{m,0,\dots\},\{0\}}(N_1,-N_2'+\epsilon)=(1/N_1)[S^{\rm
ABJ}_{(1)}(N_1,N_2',m)_k+S^{\rm ABJ}_{(2)}(N_1,N_2',m)_k]$.  The
integral representation is
\begin{align}
 &S_{\{m_j\},\{0\}}(N_1,-N_2'+\epsilon)\notag\\
 &
 =
 {(-\epsilon \ln q)^{N_2'}\over N_1!\prod_{j=1}^{M'-1}(q)_j}
 \sum_{\sigma\in S_{N_1}}
 {q^{\sum_{j=1}^{M'} (-2j+M')m_{\sigma(j)}} \prod_{1\le j<k\le M'}(q^{k-j+m_{\sigma(k)}-m_{\sigma(j)}})_1
\over \prod_{j=1}^{N_2'}(1+q^{m_{\sigma(M'+j)}})}
 \notag\\
 &\qquad\times
 \left[\prod_{j=1}^{N_2'} {-1\over 2\pi i}\int {\pi ds_j\over \sin(\pi s_j)}\right]
 q^{-\sum_{j=1}^{N_2'}(s_j+1+M')m_{\sigma(M'+j)}}
 \notag\\
 &\qquad\times 
 {\prod_{j=1}^{N_2'}(-q^{s_j+1+m_{\sigma(M'+j)}})_{M'}}
 \prod_{j=1}^{M'} \prod_{k=1}^{N_2'}
 {(q^{M'+s_k+1-j+m_{\sigma(M'+k)}-m_{\sigma(j)}})_1\over
 (-q^{M'+s_k+1-j-m_{\sigma(j)}})_1(-q^{M'+s_k+1-j+m_{\sigma(M'+k)}})_1
 }\notag\\
 &\qquad\times
 \prod_{1\le j<k\le N_2'}{(q^{s_k-s_j+m_{\sigma(M'+k)}-m_{\sigma(M'+j)}})_1 (q^{s_k-s_j})_1
 \over (-q^{s_k-s_j-m_{\sigma(M'+j)}})_1(-q^{s_k-s_j+m_{\sigma(M'+k)}})_1}.
\end{align}
The full expression including the prefactors in \eqref{dpq11Jun14} is
\begin{align}
 &\cW^{\rm ABJ}_{\{m_j\},\{0\}}(N_1,N_2')_k
 =
 i^{-{\kappa\over 2}(N_1^2+N_2'^2)}
 (-1)^{{1\over 2}(N_2'-1)N_2'}
 \left({g_s\over 2\pi}\right)^{{N_1+N_2'\over 2}}
 \notag\\
 &\qquad\times
 q^{-{1\over 6}M'(M'^2-1)+\sum_{j=1}^{N_1}(-{1\over 2}{\mu_j^2}+\mu_j)}
 (1-q)^{{1\over 2}M'(M'-1)}G_2(M'+1,q)
 { S_{\{m_j\},\{0\}}(N_1,-N_2'+\epsilon)\over \left(-\epsilon \ln q\right)^{N_2'}}.
\end{align}

\section{The cancellation of residues at $s_a=-1,-2,\cdots, -n$}
\label{Cancellation}

In this appendix we show that, in our expression (\ref{1/6WilsonI2}) for ${1\over 6}$-BPS Wilson loops in the $N_1\ge N_2$ case, the contributions from the P poles at $s_a=-1,\cdots,-n$ are absent. Namely, these contributions are canceled between those from $I^{(1)}(N_1,N_2;n)_k$ and $I^{(2)}(N_1,N_2;n)_k$. This is necessary, in particular, for the ${1\over 6}$-BPS Wilson loops (\ref{1/6WilsonI2}) to correctly reproduce  the perturbative expansion in $g_s$ and also fills the gap of a proof that the two expressions (\ref{1/6WilsonI1}) and (\ref{1/6WilsonI2}) agree in the ABJM limit $N_1=N_2$.  

\medskip
We first recall the expressions of our interest:
\begin{align}
I^{(1)}(N_1,N_2;n)_k&:={1\over N_2!}\sum_{c=0}^{n-1}
\prod_{a=1}^{N_2}\left[{-1\over 2\pi i}\int_{C_1\![c]}{\pi ds_a\over\sin(\pi s_a)}\right]
q^{n(2c-M)}{\left(q^{1-n}\right)_c\left(q^{1+n}\right)_{M-1-c}\over (q)_c(q)_{M-1-c}}
\prod_{a=1}^{N_2}{\left(q^{s_a+1}\right)_M\over 2\left(-q^{s_a+1}\right)_{M}}\nn\\
&\times \prod_{a=1}^{N_2}\left[{\left(-q^{s_a+1+c}\right)_1\left(q^{s_a+1+c-n}\right)_1
\over \left(q^{s_a+1+c}\right)_1\left(-q^{s_a+1+c-n}\right)_1}\prod_{b=1}^{a-1}{\left(q^{s_a-s_b}\right)_{1}
\over\left(-q^{s_a-s_b}\right)_{1}}
\prod_{b=a+1}^{N_2}{\left(q^{s_b-s_a}\right)_{1}\over\left(-q^{s_b-s_a}\right)_{1}}\right]\ ,
\label{IntegralII1App}
\end{align}
and
\begin{align}
I^{(2)}(N_1,N_2;n)_k &:={1\over N_2!}\sum_{d=1}^{N_2}
\prod_{a=1}^{N_2}\left[{-1\over 2\pi i}\int_{C_2}{\pi ds_a\over\sin(\pi s_a)}\right]
q^{-ns_d+n(d-M-2)}
\prod_{\substack{a=1 \\ a\ne d}}^{N_2}{\left(q^{s_a-s_{d}-n}\right)_1\over \left(q^{s_a-s_{d}}\right)_1} \label{IntegralII2App}\\
&\times\prod_{a=1}^{N_2}\left[{\left(q^{s_a+1+n\delta_{ad}}\right)_M
\over\left(1+q^{n\delta_{ad}}\right)\left(-q^{s_a+1}\right)_{M}}
\prod_{b=1}^{a-1}{\left(q^{s_a-s_b}\right)_{1}
\over\left(-q^{s_a-s_b+n\delta_{ad}}\right)_{1}}
\prod_{b=a+1}^{N_2}{\left(q^{s_b-s_a}\right)_{1}\over\left(-q^{s_b-s_a-n\delta_{ad}}\right)_{1}}\right]\ .\nn
\end{align}
In Section \ref{IntRep} we discussed the pole structures of the integrands and the integration contours $C_1$ and $C_2$ in detail. The contour $C_1$ for a given $c$ is placed to the left of $s_a={\rm min}(-1-c, {k\over 2}-M)$ and the right of $s_a={\rm max}(-M-1, -{k\over 2}+n-1-c)$,\footnote{A remark similar to footnote \ref{footnoteC1} applies.} whereas the contour $C_2$ is placed, for $a\ne d$, to the left of $s_a={\rm min}(0, {k\over 2}-M)$ and the right of $s_a={\rm max}(-M-1, -{k\over 2}-1)$ and, for $a=d$, to the left of $s_d={\rm min}(-n, {k\over 2}-M)$ and the right of $s_d={\rm max}(-M-1, -{k\over 2}-1)$. In particular, there are residues from the P poles at $s_d=-1-c$, ($c=0,\dots,n-1$), for (\ref{IntegralII1App}) and $s_d=-1,\cdots,-n$ for (\ref{IntegralII2App}). Recall (\ref{limitresidue}) for generic (including non-integral) $M$
\bea
\lim_{s_d\to -1-c}{(q^{1-n})_c(q^{1+n})_{M-1-c}\over (q)_c(q)_{M-1-c}}{(q^{s_d+1})_M\over (q^{s_d+1+c})_1}
&=&{q^{-nc}(q^{n-c})_M\over (q^n)_1(q^{-c})_c(q)_{M-1-c}}\lim_{s_d\to -1-c}{(q^{s_d+1})_M\over  (q^{s_d+1+c})_1}\nn\\
&=&{q^{-nc}(q^{n-c})_M\over (q^n)_1}
\label{limitresidue2}
\eea
where we used $(q^{1-n})_c(q^{1+n})_{M-1-c}=(-1)^cq^{-nc+\half c(c+1)}(q^{n-c})_M/(q^n)_1$, $\lim_{\epsilon\to 0}(q^{\epsilon-c})_M/(q^{\epsilon})_1=(q^{-c})_c(q)_{M-1-c}$ and (\ref{maux20Sep12}). 
As remarked below \eqref{limitresidue} in Section \ref{ABJMlimit}, 
implicit in this calculation is the $\epsilon$-prescription that always enables us to regard the P pole at $s_d=-1-c$ as a simple pole. Namely, the integer $M$ is shifted to a non-integral value $M+\epsilon$ and the contour $C_1[c]$ is placed between $s_d=-1-c$ and $-1-c-\epsilon$ so as to avoid the latter pole.

This yields\footnote{With an abuse of notation, we denote the integrands of $I^{(1)}(N_1,N_2;n)_k$ and $I^{(2)}(N_1,N_2;n)_k$ by the same symbols.} 
\begin{align}
-2\pi i{\rm Res}_{s_d=-1-c}\left[I^{(1)}(N_1,N_2;n)_k\right]&={(-1)^c\over N_2!}
\prod_{\substack{a=1 \\ a\ne d}}^{N_2}\left[{-1\over 2\pi i}\int_{C_1\![c]}{\pi ds_a\over\sin(\pi s_a)}\right]
{q^{n(c-M)}(q^{n-c})_M\over (-q^{-c})_M(-q^{n})_1}
\nn\\
&\times \prod_{\substack{a=1 \\ a\ne d}}^{N_2}\Biggl[
{\left(q^{s_a+1}\right)_M\left(q^{s_a+1+c}\right)_1\left(q^{s_a+1+c-n}\right)_1
\over 2\left(-q^{s_a+1}\right)_{M}\left(-q^{s_a+1+c}\right)_1\left(-q^{s_a+1+c-n}\right)_1}
\label{I1residue}\\
&\times 
\prod_{\substack{b=1 \\ b\ne d}}^{a-1}{\left(q^{s_a-s_b}\right)_{1}
\over\left(-q^{s_a-s_b}\right)_{1}}
\prod_{\substack{b=a+1\\ b\ne d}}^{N_2}{\left(q^{s_b-s_a}\right)_{1}\over\left(-q^{s_b-s_a}\right)_{1}}\Biggr]\ .\nn
\end{align}
Similarly, it is straightforward to find that
\begin{align}
-2\pi i{\rm Res}_{s_d=-1-c}\left[I^{(2)}(N_1,N_2;n)_k\right]&={(-1)^{c+1}\over N_2!}
\prod_{\substack{a=1 \\ a\ne d}}^{N_2}\left[{-1\over 2\pi i}\int_{C_2}{\pi ds_a\over\sin(\pi s_a)}\right]{q^{n(c-M)}\left(q^{n-c}\right)_M
\over\left(-q^n\right)_1\left(-q^{-c}\right)_{M}}
\nn\\
&\times
\prod_{\substack{a=1 \\ a\ne d}}^{N_2}\Biggl[{\left(q^{s_a+1}\right)_M\left(q^{s_a+1+c}\right)_{1}\left(q^{s_a+1+c-n}\right)_{1}
\over 2\left(-q^{s_a+1}\right)_{M}\left(-q^{s_a+1+c}\right)_{1}\left(-q^{s_a+1+c-n}\right)_{1}}\label{I2residue}\\
&\times
\prod_{\substack{b=1 \\ b\ne d}}^{a-1}{\left(q^{s_a-s_b}\right)_{1}
\over\left(-q^{s_a-s_b}\right)_{1}}
\prod_{\substack{b=a+1\\ b\ne d}}^{N_2}{\left(q^{s_b-s_a}\right)_{1}\over\left(-q^{s_b-s_a}\right)_{1}}\Biggr]\ .\nn
\end{align}
Hence the sum of the two residues ${\rm Res}_{s_d=-1-c}\left[I^{(1)}(N_1,N_2;n)_k+I^{(2)}(N_1,N_2;n)_k\right]$ exactly cancels out. As $c$ runs from $0$ to $n-1$, this implies the cancellation of the residues in (\ref{1/6WilsonI2}) at the P poles $s_d=-1,\cdots,-n$, $(d=1,\cdots, N_2)$.
More precisely, the cancellation requires the equivalence of the contours $C_1$ and $C_2$. In this regard, note that the integrand of the residue (\ref{I1residue}), in particular, does not have a pole at $s_a=-1-c$ and the contour $C_1$ can thus be shifted,  past $s_a=-1-c$, to the left of $s_a={\rm min}(0,{k\over 2}-M)$ so that $C_1$ becomes identical to $C_2$.

\medskip
We would like to address the subtlety remarked below (\ref{S1}) and (\ref{S2}) concerning the range, in particular, of the sum over $c$ in (\ref{S1}) and (\ref{IntegralII1}). In its original form, the sum over $c$ ran from $0$ to $M-1$. When $n$ is less than $M$, the sum simply terminates at $c=n$, as the factors (\ref{limitresidue2}) vanish when $c\ge n$. When $n$ is greater than $M$, however, in order to replace the upper limit $M-1$ by $n-1$, we need to show that the contribution from $c=M$ to $n-1$ is absent in (\ref{1/6WilsonI2}) and (\ref{formalABJ1/6WilsonLoop}). Note that when $c\ge M$ in the first line of (\ref{limitresidue2}) the factor $(q)_{M-1-c}$ in the denominator diverges. Thus, for $c\ge M$, (\ref{limitresidue2}) might appear to vanish. However, there are nonvanishing contributions coming from the poles at $s_d=-1-c$ as the second line of (\ref{limitresidue2}) indicates. This implies that the cancellation we have shown above is all we  need to ensure the absence of the contribution from $c=M$ to $n-1$ in (\ref{1/6WilsonI2}) and (\ref{formalABJ1/6WilsonLoop}). The cancellation of the residues at $s_a=-1,\dots,-n$ also justifies the extension of the lower limits of the sum over $D_a$'s in (\ref{S1}) and (\ref{S2}).

\section{The $U(1)_k \times U(1)_{-k}$ ABJM theory}
\label{U1U1check}
As a simplest check of our prescriptions that lack first principle derivations, we compare the result from the integral representation \eqref{1/6WilsonI1}--\eqref{IntegralI} with that from the direct calculation in the case of the $U(1)_k \times U(1)_{-k}$ ABJM theory.

The direct integral of \eqref{1/6wilson0} yields
\begin{align}
\mathcal{W}^{\mathrm{I}}_{\frac{1}{6}} (1,1;n)_k &=\int \frac{d \mu}{2\pi}\frac{d\nu}{2 \pi} \frac{\mathrm{e}^{n\mu} \mathrm{e}^{-\frac{1}{2g_s}(\mu^2 -\nu^2) } }{(2 \cosh \left( \frac{\mu-\nu}{2} \right))^2} =\frac{1}{|k|} \frac{q^{-\frac{1}{2}n^2+n}}{(1+q^n)^2} ~~~\mbox{for}~~|n|<\frac{k}{2}\ ,
\end{align}
where we used the Fourier transform of $1/\cosh x$
\begin{align}
\frac{1}{2\cosh \frac{x}{2}} =\int \frac{dp}{2 \pi} \frac{\mathrm{e}^{\frac{ixp}{2 \pi}}}{2 \cosh \frac{p}{2}}\ .
\end{align}
As commented in Section \ref{WilsonLoopResults}, the restriction on winding number, $|n| <\frac{k}{2}$, is necessary for the convergence of integrals. Hence the direct calculation gives the normalized $\frac{1}{6}$-BPS Wilson loop
\begin{align}
W_{\frac{1}{6}}^{\mathrm{I}} (1,1;n)_k &=\frac{4q^{-\frac{1}{2}n^2 +n}}{(1+q^n)^2}\ . \label{W11}
\end{align}
On the other hand, as calculated in  Section \ref{U1UNexample}, the result from our integral representation yields\footnote{For even $k$, since $(-q^{-n})^k=1$, the sum $\sum_{s=0}^{k-1}(-q^{-n})^s=0$ for an integer $n$. Meanwhile, this sum equals ${1-q^{-nk}\over 1+q^{-n}}=: f(n)$ for a non-integral $n$, To find the sum $\sum_{s=0}^{k-1}(-q^{-n})^s s$ for even $k$, we differentiate $f(n)$ w.r.t. $n$ and then send $n$ to an integral value.}
\begin{align}
I(1,1;n)_k&=\frac{q^{-n}}{1+q^n}\begin{cases}  
\frac{1}{2}\sum_{s=0}^{k-1} (-1)^s q^{-ns} ~~~~~~~~~~~~~~\mbox{for~odd}~k\\
-\frac{1}{2k} \sum_{s=0}^{k-1}(-1)^s q^{-ns} (s+a)~~~\mbox{for~even}~k
\end{cases} \nonumber \\
&=\frac{1}{(1+q^n)^2}\ . 
\end{align}
Thus the normalized $\frac{1}{6}$-BPS Wilson loop \eqref{1/6WilsonI1} indeed agrees with \eqref{W11} providing evidence for our prescriptions.


\section{An alternative derivation of Seiberg duality}
\label{KapustinWillett}

In this appendix we provide an alternative derivation of the duality transformations of ABJ Wilson loops by following Kapustin and Willett~\cite{Kapustin:2013hpk}. In their approach Seiberg duality is understood as an isomorphism of the algebras that BPS Wilson loops generate. 


Before discussing the Wilson loops in the ABJ theory, 
we first give a brief sketch of their construction of Wilson loop algebras and derivation of  duality transformations for generic three-dimensional supersymmetric gauge theories that admit localization. 
In order to construct the Wilson loop algebras, we first define
\begin{align}
\Phi (t) &= \prod_{j=1}^{N} (1+t x_j)  \nonumber \\
&=1+t \left( \sum_{j=1}^N x_j \right)+t^2 \left( \sum_{i<j}x_i x_j \right) +\cdots+t^N \prod_{j=1} ^N x_j \nonumber \\
&=1+t \,{\tiny\yng(1)} +t^2 \, {\tiny\yng(1,1)} +t^3 \, {\tiny\yng(1,1,1)} +\cdots +t^N \nonumber \\
&=: \sum_{i=0}^N t^i \phi_i 
\end{align}
and
\begin{align}
\Psi(t) &= \prod_{j=1}(1-t x_j)^{-1}  \nonumber \\
&= 1+t \sum_{j=1}^N x_j +t^2 \left( \sum_{j=1}^{N} x_j^2 + \sum_{i < j} x_i x_j \right) 
+\cdots \nonumber \\
&=1+ t\, {\tiny\yng(1)}  +t^2\, {\tiny\yng(2)} +t^3   \,  {\tiny\yng(3)}  +\cdots \nonumber \\
&=: \sum_{i=0} t^i \psi_i\ ,
\end{align}
where we denoted symmetric polynomials by the corresponding Young diagrams. Each Young diagram corresponds to a specific representation of the Wilson loop. The variables $x_i\, (i=1,\cdots,N)$ will be  identified with the integration variables of the matrix models derived from localization.  The $\phi_i$'s and $\psi_i$'s generate the ring of symmetric polynomials. Notice, however, that $\Phi (-t) \Psi(t) =1$ by definition
and thus the $\phi_i$'s and $\psi_i$'s are not independent. 

Next, to find the duality transformations for BPS Wilson loops, we construct an algebra  for a given theory from the ring of symmetric polynomials that \emph{quantum} Wilson loops generate. 
As it turns out,  there is an isomorphism between the algebras for the original and dual theories that can be regarded as the duality transformations.
Here we only outline the derivation of the duality transformations:
\begin{enumerate}
\item Using the matrix model obtained by localization, the invariance under the shift of an integration variable yields the following identity: 
\begin{align}
 \left\langle p(x) \right\rangle = \left\langle x^M +a_1 x^{M-1} + \cdots \right\rangle =0\ , \label{quantum}
\end{align}
where $x$ is an integration variable and $\langle \cdots \rangle$ means vev of the matrix model.
The polynomial $p(x)$ is at most of the $M$-th order in $x$, where $M$ is a constant   determined by the rank and level of the gauge theory. This is a \emph{quantum} constraint on the BPS Wilson loops that the Wilson loop algebra is endowed with.
\item From the polynomial $p(x)$ we construct the following quantities $\tilde{p}(t)$ and $\Psi_p (t)$:
\begin{align}
\tilde{p}(t) &:= t^M p(t^{-1}) 
\end{align}
and
\begin{align}
\Psi_p (t) &:=   \left. \tilde{p}(t) \Psi (t) \right|_{\text{truncated\,at}\,\cO(t^{M-N})} \nonumber\\
&=\sum_{i=0}^{M-N} t^i \psi_{pi}  \label{truncation} \\
&=1+(\,{\tiny\yng(1)}+a_1)t +\cO(t^2)\ .   \nonumber
\end{align}
In the algebra that Kapustin and Willett identify with a certain quotient of the ring, the classical constraint $\Phi(-t)\Psi(t)=1$ is deformed to 
\begin{align}
\Phi(-t) \Psi_p(t) =\tilde{p}(t) \label{constraint}
\end{align}
that the elements, $\phi_i$'s and $\psi_{pi}$'s, of the algebra obey in quantum theory. The quantum constraint \eqref{quantum} is crucial, and it is important that the left hand side of the constraint \eqref{constraint} is truncated to a finite polynomial in $t$ in contrast to the classical constraint $\Phi(-t)\Psi(t)=1$.

\item 
Owing to the truncation in \eqref{truncation} and thus in \eqref{constraint}, there exists an isomorphism under the following transformations:
\begin{align}
N &\leftrightarrow M-N\ , \\
\phi_i &\leftrightarrow (-1)^i \psi_{pi}\ . \label{Wilsonduality}
\end{align}
That is, the quantum constraint \eqref{constraint} is invariant under these transformations. 
The first transformation can be identified with the map of the ranks of the two gauge groups in a dual pair and the second transformations with the maps of the BPS Wilson loops.
Thus the duality transformations of the Wilson loops can be extracted order by order in $t$ from \eqref{Wilsonduality}. 
%
%
%
%
At $\cO(t)$, for instance, we find for the fundamental representation
\begin{align}
{\tiny\yng(1)} \rightarrow -\,\widetilde{\tiny\yng(1)}-a_1,
\end{align}
where the tilde indicates that the Wilson loop is that in the dual theory. 
\end{enumerate} 

\subsection{ABJ Wilson loop duality}
We now apply the above method to ABJ Wilson loops.
The original theory is the $U(N_1)_k \times U(N_2)_{-k}$ ABJ theory and the dual theory is the $U(\widetilde{N}_2)_k \times U(N_1)_{-k}$ theory with the dual gauge group $\widetilde{N}_2 =k+2N_1-N_2$. In implementing the above procedure it is useful to regard the original theory as the $U(N_2)_{-k}$ Chern-Simons matter theory and the $U(N_1)$ part as flavor.
Let us first recall that the $\frac{1}{6}$-BPS Wilson loop on the gauge group $U(N_2)$ with winding $n$ is given by the eigenvalue integrals
\begin{align}
W^{\mathrm{II}}_{\frac{1}{6}}(N_1,N_2;n)_k \propto 
 \int\prod_{i=1}^{N_1}{d\mu_i\over 2\pi}\prod_{a=1}^{N_2}{d\nu_a\over 2\pi}
 {\Delta_{\rm sh}(\mu)^2\Delta_{\rm sh}(\nu)^2\over
 \Delta_{\rm ch}(\mu,\nu)^2}\,
 e^{-{1\over 2g_s}\left(\sum_{i=1}^{N_1}\mu_i^2-\sum_{a=1}^{N_2}\nu_a^2\right) } \sum_{a=1}^{N_2} e^{n\nu_a}\ ,
\end{align}
where we omitted the normalization factor as it is not relevant in the following discussion.
For later convenience, we introduce the following notations:
\begin{align}
\langle \,\dots \rangle_{\mu} &:= \int \prod_{j=1}^{N_1} \frac{d \mu_j}{2 \pi} 
{\Delta_{\rm sh}(\mu)^2 \over \Delta_{\rm ch}(\mu,\nu)^2}
 \, e^{-{1\over 2g_s}  \sum_{i=1}^{N_1}\mu_i^2  }(\,\dots)\ ,  \label{muint}\\
\langle \,\dots \rangle_{\nu} &:= \int \prod \frac{d \nu_a}{2 \pi}  {\Delta_{\rm sh}(\nu)^2\over
 \Delta_{\rm ch}(\mu,\nu)^2} \, e^{+{1\over 2g_s}\sum_{a=1}^{N_2}\nu_a^2  }(\,\dots)\ , \label{nuint} \\
 \langle \,\dots \rangle_{\mu ,\nu}&:= \int \prod_{j=1}^{N_1} \frac{d \mu_j}{2 \pi}  \prod_{a=1}^{N_2}\frac{d \nu_a}{2\pi} 
{\Delta_{\rm sh}(\mu)^2\Delta_{\rm sh}(\nu)^2\over \Delta_{\rm ch}(\mu,\nu)^2}
 \, e^{-{1\over 2g_s}\left(\sum_{i=1}^{N_1}\mu_i^2-\sum_{a=1}^{N_2}\nu_a^2\right) }(\,\dots)\ . 
\end{align}
Note that in \eqref{muint} only $\mu_i$'s are integrated and $\nu_a$'s are regarded as  parameters of the theory.  In \eqref{nuint} the roles of $\mu_i$'s and $\nu_a$'s are interchanged. 
In this notation the $\frac{1}{6}$-BPS Wilson loop on the $U(N_2)$ is expressed as
\begin{align}
W^{\mathrm{II}}_{\frac{1}{6}}(N_1,N_2;n)_k  \propto \left\langle \sum_{a=1}^{N_2}\mathrm{e}^{n \nu_a} \right\rangle_{\mu ,\nu}.
\end{align}
In the expression \eqref{nuint} we can regard the $U(N_1)_k \times U(N_2)_{-k}$ ABJ theory as the $U(N_2)$ Chern-Simons matter theory at level $-k$ with $2N_1$ hypermultiplets without  Fayet-Iliopoulos terms. From this viewpoint the flavor group is $U(2N_1)$ instead of $U(N_1)$. In \eqref{nuint}, $\mu_i\,(i=1,\cdots,N_1)$ can be thought of as mass parameters for the hypermultiplets. The Seiberg dual is then the $U(\widetilde{N}_2)$ Chern-Simons matter theory at level $k$ with $2N_1$ hypermultiplets of masses $\nu_a \, (a=1,\cdots,N_1)$ without Fayet-Iliopoulos terms \cite{Giveon:2008zn}.\footnote{Precisely speaking, the ABJ theory has a superpotential, whereas the class of theories considered in the Giveon-Kutasov duality \cite{Giveon:2008zn} do not. Furthermore, the dual theories in the latter case have singlet meson fields that together with dual quarks generate a superpotential.
However, this difference can be safely disregarded since the superpotential has no effect on the results of localization and only affects R-charges of the matter fields.}
These parameters are to be integrated in the end.

In Step 1, to find \eqref{quantum} for the ABJ theory, we consider the following quantity:
\begin{align}
\label{KWQ1}
\left\langle \mathrm{e}^{n\nu_a} \prod_{i=1}^{N_1} \left(  e^{-(\mu_i -\nu_a) } +1\right)^2   \right\rangle_\nu\ .
\end{align}
Integrals are invariant under shifts of integration variables. For our purpose, we consider, in particular, the shift
\begin{align}
\nu_a \rightarrow \nu_a+ 2 \pi i\ ,
\end{align}
where $a$ is a selected index but can be chosen arbitrarily.
Note that the inserted factor $\prod_{a=1}^{N_2} \left(  e^{-(\mu_l -\nu_a) } +1\right)^2 $ exactly cancels the poles that appear in the integrand \eqref{nuint} and allows us to make the above shift keeping the range of integration intact. Under this shift \eqref{KWQ1} becomes
\begin{align}
\left\langle \mathrm{e}^{k \pi i +k \nu_a}   \mathrm{e}^{n\nu_a} \prod_{l=1}^{N_1} \left(  e^{-(\mu_l -\nu_a) } +1\right)^2    \right\rangle_\nu.
\end{align}
Since this is identical to \eqref{KWQ1}, we obtain the identity:
\begin{align}
\left\langle (1-\mathrm{e}^{k \pi i +k \nu_a}  )\, \mathrm{e}^{n\nu_a} \prod_{l=1}^{N_1} \left(  e^{-(\mu_l -\nu_a) } +1\right)^2    \right\rangle_\nu =0\ .
\end{align}
This equation must hold for any representation of BPS Wilson loops and then yields the operator identity
\begin{align}
(1-(-1)^k x^k) \prod_{l=1}^{N_1} \left( x \mathrm{e}^{-\mu_l} +1\right)^2 =0\ ,
\end{align}
where we introduced $x=\mathrm{e}^{\nu_a}$. For later convenience, we rewrite this as 
\begin{align}
0=p(x) &:=\left( x^k -(-1)^k \right) \prod_{l=1}^{N_1} \left( x+ \mathrm{e}^{\mu_l} \right)^2 \\
&= x^{k+2N_1} +\left( 2 \sum_{i=1}^{N_1} \mathrm{e}^{\mu_i }\right) x^{k+2N_1-1} +\left( \sum_{i=1}^{N_1} \mathrm{e}^{2 \mu_i} \right) x^{k+2N_1-2}  +\cdots\ .
\end{align}
Note that this polynomial is at most of order $\cO(x^{k+2N_1})$. 

In Step 2, we introduce the following quantities 
\begin{align}
\tilde{p} (t) &= t^{k+2N_1} p(t^{-1})  \nonumber \\ 
&=\left(1+(-1)^{k+1} t^k  \right)  \left[  1+t \left( 2 \sum_{i=1}^{N_1} \mathrm{e}^{\mu_i}  \right) +t^2 \left( \sum_{i=1}^{N_1} \mathrm{e}^{2 \mu_i} +4 \sum_{i<j} \mathrm{e}^{\mu_i +\mu_j}  \right) +\cO(t^3) \right] 
\end{align}
and
\begin{align}
\Psi_p (t) & := \tilde{p}(t) \Psi(t) = \sum_{i=1} t^i \psi_{pi} \nonumber \\
&=1+t\left( {\tiny\yng(1)}+2\sum_{i=1}^{N_1} \mathrm{e}^{\mu_i} \right) +t^2 \left(  {\tiny\yng(2)} +2\,{\tiny\yng(1)}\sum_{i=1}^{N_1} \mathrm{e}^{\mu_i}+\sum_{i=1}^{N_1}\mathrm{e}^{2\mu_i} +4 \sum_{i<j} \mathrm{e}^{\mu_i +\mu_j} \right) +\cO(t^3)  \nonumber \\
&\qquad+(-1)^{k+1} t^k \left(1 +t \left(2 \sum \mathrm{e}^{\mu_i}  +{\tiny\yng(1)} \right)+\cO(t^2) \right)\ . \label{psip} 
\end{align}
%
The Wilson loop algebra is generated by $\phi_i$ and $\psi_{pi}$ that are constrained by \eqref{constraint}. 

In Step 3, Seiberg duality of the Wilson loops is extracted from the transformation \eqref{Wilsonduality}.
At $\cO(t)$ we obtain
\begin{align}
{\tiny\yng(1)}_{\,\mathrm{II}} \rightarrow -\,\widetilde{\tiny\yng(1)}_{\,\mathrm{I}} -2\sum_i  \mathrm{e}^{\mu_i}\ , \label{ordertdual}
\end{align}
where the subscripts $\mathrm{I}$ and $\mathrm{II}$ indicate that the Wilson loops are on the first and second gauge group, respectively.
In this example, the BPS Wilson loop of the original $U(N_2)_{-k}$ theory in the fundamental representation maps to minus that of the dual $U(\widetilde{N}_2)_{k}$ theory in the fundamental representation shifted by a singlet that depends on the masses  $\mu_i$'s in the dual theory.\footnote{ 
In the $k=1$ case it might look naively that there is an additional identity operator coming from the $\cO(t^k)$ term that contributes to the duality map. However, the $\frac{1}{6}$-BPS Wilson loop with winding $n$ is only well-defined for $n<{k\over 2}$ and thus the $n=1$ fundamental Wilson loop is ill-defined for $k=1$. This method is therefore not applicable to the ABJ theory with $k=1$. For the same reason the $\cO(t^{k/2})$ terms in \eqref{psip} are meaningless because the $\frac{1}{6}$-BPS Wilson loops with winding $n \ge  \frac{k}{2}$ are involved at this order.}  

In order to find the duality maps for the ABJ theory, we need to integrate over the mass parameters $\mu_i\,(i=1,\cdots,N_1)$. 
The map \eqref{ordertdual} then becomes
%
\begin{align}
{\tiny\yng(1)}_{\,\mathrm{II}} \rightarrow -\,\widetilde{\tiny\yng(1)}_{\,\mathrm{I}} -2\,\widetilde{\tiny\yng(1)}_{\,\mathrm{II}}.
\end{align}
In our notation this reads
\begin{align}
W^{\mathrm{II}} _{\frac{1}{6}} (N_1,N_2;1)_{k} \rightarrow -W^{\mathrm{I}}_{\frac{1}{6}}(\widetilde{N}_2 ,N_1;1)_{k}  -2W^{\mathrm{I}}_{\frac{1}{6}}(\widetilde{N}_2,N_1;1)_{k}\ . \label{Wgauge}
\end{align}
This indeed agrees with \eqref{1/6duality}. 

We can run the same procedure for the flavor Wilson loops. 
Since the flavor group is unchanged under Seiberg duality, we simply obtain the following map:
\begin{align}
{\tiny\yng(1)}_{\,\mathrm{I}} \rightarrow \widetilde{\tiny\yng(1)}_{\,\mathrm{II}}
\end{align}
which reads
\begin{align}
W^{\mathrm{I}} _{\frac{1}{6}} (N_1,N_2;1)_{k} \rightarrow  W^{\mathrm{II}} _{\frac{1}{6}} (\widetilde{N}_2,N_1;1)_{k}\ . \label{Wflavor}
\end{align}
This agrees with \eqref{flavorduality}. 
Finally, the maps \eqref{Wgauge} and \eqref{Wflavor} imply that
\begin{align}
W_{\frac{1}{2}} (N_1,N_2;1)_{k} \rightarrow -W_{\frac{1}{2} } (\widetilde{N}_2,N_1;1)_{k}\ .
\end{align}
This agrees with \eqref{1/2duality}. 
We have thus succeeded to reproduce Seiberg duality for the ABJ Wilson loops with the winding $n=1$.

\subsection{Wilson loop duality in more general representations}

From the isomorphism of the Wilson loop algebras, we can also extract the duality transformations for the Wilson loops in higher dimensional representations in the ABJ theory. Here we consider the Wilson loops that involve two boxes in Young diagrams corresponding to  the $\cO(t^2)$ terms in \eqref{psip}:
%
%
%
\begin{align}
\psi_{p2} &={\tiny\yng(2)} +2\,{\tiny\yng(1)} \, \sum_i \mathrm{e}^{\mu_i} +\sum_i \mathrm{e}^{2 \mu_i} +4 \sum_{i <j} \mathrm{e}^{\mu_i+\mu_j}\ .
\end{align}
Integrating over the mass parameters $\mu_i$, we find the duality transformation of the antisymmetric Wilson loop:
\begin{align}
W_{\bullet~ {\tiny\yng(1,1)} } \rightarrow 
 \widetilde{W}_{{\tiny\yng(2)}~\bullet } 
 +2 \widetilde{W}_{{\tiny\yng(1)}\,{\tiny\yng(1)}}+\widetilde{W}_{\bullet~{\tiny\yng(1)}}^{n=2} +4 \widetilde{W}_{\bullet~{\tiny\yng(1,1)}}\ ,\label{antisym}
\end{align}
where $\widetilde{W}_{\bullet~{\tiny\yng(1)}}^{n=2} $ is a fundamental Wilson loop with winding $n=2$ and decomposed into
\begin{align}
\widetilde{W}_{\bullet~{\tiny\yng(1)}}^{n=2}  =\widetilde{W}_{\bullet ~ {\tiny\yng(2)}} -\widetilde{W}_{\bullet ~ {\tiny\yng(1,1)}}\ .
\end{align}
Using this relation, \eqref{antisym} yields
\begin{align}
W_{\bullet ~ {\tiny\yng(1,1)} } \rightarrow \widetilde{W}_{{\tiny\yng(2)} ~\bullet } +\widetilde{W}_{\bullet ~ {\tiny\yng(2)}}+ 3 \widetilde{W}_{\bullet ~{\tiny\yng(1,1)}} +2 \widetilde{W}_{{\tiny\yng(1)}~{\tiny\yng(1)}}\ . \label{Wdotbb}
\end{align}
This agrees with the result from the heuristic derivation in the brane picture in Section \ref{branepicturesection}.
For the flavor Wilson loops, we simply have
\begin{align}
W_{{\tiny\yng(1,1)} ~ \bullet} &  \rightarrow \widetilde{W}_{\bullet ~ {\tiny\yng(1,1)} }\ , \\
W_{{\tiny\yng(2)} ~ \bullet}  &   \rightarrow \widetilde{W}_{\bullet ~{\tiny\yng(2)} }\ .  \label{Wbbdot}
\end{align}
Lastly, we consider the Wilson loop that is on both gauge groups
\begin{align}
W_{{\tiny\yng(1)} \, {\tiny\yng(1)}}\ .
\end{align}
The map of this Wilson loop is obtained from the duality transformation for the fundamental Wilson loops:
\begin{align}
W_{{\tiny\yng(1)} \, {\tiny\yng(1)}} ={\tiny\yng(1)}_{\,\mathrm{I}}\, {\tiny\yng(1)}_{\,\mathrm{II}}  & \rightarrow \widetilde{{\tiny\yng(1)}}_{\,\mathrm{II}} \, \left( -\,\widetilde{{\tiny\yng(1)}}_{\,\mathrm{I}} -2 \,\widetilde{{\tiny\yng(1)}}_{\,\mathrm{II}} \right) \nonumber \\
   & = -\widetilde{W}_{{\tiny\yng(1)} \,{\tiny\yng(1)}} -2 \Bigl( \sum_a \mathrm{e}^{\nu_a} \Bigr)^2  \nonumber \\
   &=-\widetilde{W}_{{\tiny\yng(1)} \, {\tiny\yng(1)}} -2 \widetilde{W}_{\bullet~ {\tiny\yng(2)} } -2 \widetilde{W}_{ \bullet ~ {\tiny\yng(1,1)}}\ , \label{Wbb}
\end{align}
where integration over $\nu_a$'s is implied. 
Combining \eqref{Wdotbb}, \eqref{Wbbdot} and \eqref{Wbb}, we find the map of the $\frac{1}{2}$-BPS Wilson loop
\begin{align}
W^{1/2}_{{\tiny\yng(2)}  } \rightarrow \widetilde{ W}^{1/2}_{{\tiny\yng(2)} }\ . 
\end{align}
%
This again agrees with the result in Section \ref{branepicturesection}.
Although it becomes increasingly cumbersome, it is straightforward to generalize this procedure to more general higher dimensional representations.

\end{document}